\documentclass[longauth]{aa} 
\usepackage{graphicx}
\usepackage[usenames, dvipsnames]{xcolor}
\usepackage{txfonts}
\usepackage{amsmath}
\usepackage{amssymb}
\usepackage{mathtools}
\usepackage{multirow}
\usepackage{epstopdf}
\usepackage{textcomp,gensymb}
\usepackage[normalem]{ulem}
\usepackage{natbib}
\usepackage[draft]{hyperref}
\usepackage{enumitem}

\usepackage{listings}
\usepackage{color}

\definecolor{dkgreen}{rgb}{0,0.6,0}
\definecolor{gray}{rgb}{0.5,0.5,0.5}
\definecolor{mauve}{rgb}{0.58,0,0.82}
\definecolor{golden}{rgb}{0.86,0.65,0.01}

\newcommand{\orcit}[1]{\protect\href{https://orcid.org/#1}{\protect\includegraphics[width=8pt]{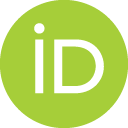}}}

\def\L#1{{\textcolor{LimeGreen}{#1}}}

\newcommand\Gaiaf{\textit{Gaia}}
\newcommand\Gaia{\textit{Gaia }}

\newcommand{\esagaia}{{\sl Gaia}}
\newcommand\gaia{\textit{Gaia }}

\newcommand\gdrtwo{\gaia~DR2}

\newcommand\secref[1]{Sect.~\ref{#1}}

\newcommand\figref[1]{Fig.~\ref{#1}}

\newcommand\figrefalt[1]{Figure~\ref{#1}}

\newcommand\equref[1]{Eq.~\eqref{#1}}

\newcommand{\kms}{{\,\mathrm{km\,s^{-1}}}}
\newcommand{\kpc}{{\,\mathrm{kpc}}}
\newcommand{\Gyr}{{\,\mathrm{Gyr}}}
\providecommand{\masyr}{\,mas\,yr$^{-1}$}
\providecommand{\mas}{\,\mathrm{mas}}
\providecommand{\muas}{\,\mu\mathrm{as}}
\providecommand{\muasyr}{\,\mu\mathrm{as}\,\mathrm{yr}^{-1}}
\providecommand{\yr}{\,\mathrm{yr}}
\newcommand{\kmskpc}{{\,\mathrm{km\,s^{-1}\,kpc}}}

\newcommand{\gbp}{{G_\mathrm{BP}}}
\newcommand{\grp}{{G_\mathrm{RP}}}

\newcommand{\piepi}{\varpi/\sigma_\varpi}

\newcommand{\vl}{V_\ell}
\newcommand{\vb}{V_b}
\newcommand{\Lz}{L_z^*}

\newcommand{\Vp}{V_\phi^*}
\newcommand{\VZ}{V_Z^*}

\newcommand{\Vt}{V_t}

\newcommand{\absmag}{$M_{\lambda}$}
\newcommand{\twomass}{{\sl 2MASS}}
\newcommand{\panstars}{{\sl Pan-STARRS}}
\newcommand{\wise}{{\sl WISE}}
\newcommand{\bayestar}{{\sl Bayestar}}

\newcommand{\lucey}{{L20}}

\newcommand{\feh}{\ensuremath{[\mathrm{Fe/H}]}}

\newcommand{\logg}{\mbox{$\log g$}}
\newcommand{\kiel}{{\sl Kiel}}
 
\newcommand{\AC}{{AC20-$\piepi>3$ }}
\newcommand{\ACf}{{AC20-$\piepi>3$}}

\makeatletter
\renewcommand*\maketitle{%
  \thispagestyle{firstpage}
\begingroup
    \if@wideboxfn
    \setlength\bibindent{1.4\parindent}
    \else
    \setlength\bibindent{\parindent}
    \fi
    \renewcommand*\thefootnote{\@fnsymbol\c@footnote}%
    \renewcommand\@makefntext[1]{%
    \ifaa@longfn\hsize\textwidth\fi
    \noindent
    \hb@xt@\bibindent{\hss\@makefnmark\enspace}##1}
  \ifaa@twocolumn
  \begingroup
    \begin{aa@strip}
          \aa@maketitle
    \end{aa@strip}
    \@thanks            
  \endgroup
  \else
    \begingroup
      \let\thanks\footnote
      \aa@maketitle
    \endgroup
  \fi
\endgroup
  \setcounter{footnote}{0}%
}
\makeatother

\begin{document}

   \title{Gaia Early Data Release 3: The Galactic anticentre}

\author{
{\it Gaia} Collaboration
\and T.        ~Antoja                        \orcit{0000-0003-2595-5148}\inst{\ref{inst:0001}}
\and P.J.      ~McMillan                      \orcit{0000-0002-8861-2620}\inst{\ref{inst:0002}}
\and G.        ~Kordopatis                    \orcit{0000-0002-9035-3920}\inst{\ref{inst:0003}}
\and P.        ~Ramos                         \orcit{0000-0002-5080-7027}\inst{\ref{inst:0001}}
\and A.        ~Helmi                         \orcit{0000-0003-3937-7641}\inst{\ref{inst:0005}}
\and E.        ~Balbinot                      \orcit{0000-0002-1322-3153}\inst{\ref{inst:0005}}
\and T.        ~Cantat-Gaudin                 \orcit{0000-0001-8726-2588}\inst{\ref{inst:0001}}
\and L.        ~Chemin                        \orcit{0000-0002-3834-7937}\inst{\ref{inst:0008}}
\and F.        ~Figueras                      \orcit{0000-0002-3393-0007}\inst{\ref{inst:0001}}
\and C.        ~Jordi                         \orcit{0000-0001-5495-9602}\inst{\ref{inst:0001}}
\and S.        ~Khanna                        \orcit{0000-0002-2604-4277}\inst{\ref{inst:0005}}
\and M.        ~Romero-G\'{o}mez              \orcit{0000-0003-3936-1025}\inst{\ref{inst:0001}}
\and G.M.      ~Seabroke                      \inst{\ref{inst:0013}}
\and A.G.A.    ~Brown                         \orcit{0000-0002-7419-9679}\inst{\ref{inst:0014}}
\and A.        ~Vallenari                     \orcit{0000-0003-0014-519X}\inst{\ref{inst:0015}}
\and T.        ~Prusti                        \orcit{0000-0003-3120-7867}\inst{\ref{inst:0016}}
\and J.H.J.    ~de Bruijne                    \orcit{0000-0001-6459-8599}\inst{\ref{inst:0016}}
\and C.        ~Babusiaux                     \orcit{0000-0002-7631-348X}\inst{\ref{inst:0018},\ref{inst:0019}}
\and M.        ~Biermann                      \inst{\ref{inst:0020}}
\and O.L.      ~Creevey                       \orcit{0000-0003-1853-6631}\inst{\ref{inst:0003}}
\and D.W.      ~Evans                         \orcit{0000-0002-6685-5998}\inst{\ref{inst:0022}}
\and L.        ~Eyer                          \orcit{0000-0002-0182-8040}\inst{\ref{inst:0023}}
\and A.        ~Hutton                        \inst{\ref{inst:0024}}
\and F.        ~Jansen                        \inst{\ref{inst:0016}}
\and S.A.      ~Klioner                       \orcit{0000-0003-4682-7831}\inst{\ref{inst:0026}}
\and U.        ~Lammers                       \orcit{0000-0001-8309-3801}\inst{\ref{inst:0027}}
\and L.        ~Lindegren                     \orcit{0000-0002-5443-3026}\inst{\ref{inst:0002}}
\and X.        ~Luri                          \orcit{0000-0001-5428-9397}\inst{\ref{inst:0001}}
\and F.        ~Mignard                       \inst{\ref{inst:0003}}
\and C.        ~Panem                         \inst{\ref{inst:0031}}
\and D.        ~Pourbaix                      \orcit{0000-0002-3020-1837}\inst{\ref{inst:0032},\ref{inst:0033}}
\and S.        ~Randich                       \orcit{0000-0003-2438-0899}\inst{\ref{inst:0034}}
\and P.        ~Sartoretti                    \inst{\ref{inst:0019}}
\and C.        ~Soubiran                      \orcit{0000-0003-3304-8134}\inst{\ref{inst:0036}}
\and N.A.      ~Walton                        \orcit{0000-0003-3983-8778}\inst{\ref{inst:0022}}
\and F.        ~Arenou                        \orcit{0000-0003-2837-3899}\inst{\ref{inst:0019}}
\and C.A.L.    ~Bailer-Jones                  \inst{\ref{inst:0039}}
\and U.        ~Bastian                       \orcit{0000-0002-8667-1715}\inst{\ref{inst:0020}}
\and M.        ~Cropper                       \orcit{0000-0003-4571-9468}\inst{\ref{inst:0013}}
\and R.        ~Drimmel                       \orcit{0000-0002-1777-5502}\inst{\ref{inst:0042}}
\and D.        ~Katz                          \orcit{0000-0001-7986-3164}\inst{\ref{inst:0019}}
\and M.G.      ~Lattanzi                      \orcit{0000-0003-0429-7748}\inst{\ref{inst:0042},\ref{inst:0045}}
\and F.        ~van Leeuwen                   \inst{\ref{inst:0022}}
\and J.        ~Bakker                        \inst{\ref{inst:0027}}
\and J.        ~Casta\~{n}eda                 \orcit{0000-0001-7820-946X}\inst{\ref{inst:0048}}
\and F.        ~De Angeli                     \inst{\ref{inst:0022}}
\and C.        ~Ducourant                     \orcit{0000-0003-4843-8979}\inst{\ref{inst:0036}}
\and C.        ~Fabricius                     \orcit{0000-0003-2639-1372}\inst{\ref{inst:0001}}
\and M.        ~Fouesneau                     \orcit{0000-0001-9256-5516}\inst{\ref{inst:0039}}
\and Y.        ~Fr\'{e}mat                    \orcit{0000-0002-4645-6017}\inst{\ref{inst:0053}}
\and R.        ~Guerra                        \orcit{0000-0002-9850-8982}\inst{\ref{inst:0027}}
\and A.        ~Guerrier                      \inst{\ref{inst:0031}}
\and J.        ~Guiraud                       \inst{\ref{inst:0031}}
\and A.        ~Jean-Antoine Piccolo          \inst{\ref{inst:0031}}
\and E.        ~Masana                        \orcit{0000-0002-4819-329X}\inst{\ref{inst:0001}}
\and R.        ~Messineo                      \inst{\ref{inst:0059}}
\and N.        ~Mowlavi                       \inst{\ref{inst:0023}}
\and C.        ~Nicolas                       \inst{\ref{inst:0031}}
\and K.        ~Nienartowicz                  \orcit{0000-0001-5415-0547}\inst{\ref{inst:0062},\ref{inst:0063}}
\and F.        ~Pailler                       \inst{\ref{inst:0031}}
\and P.        ~Panuzzo                       \orcit{0000-0002-0016-8271}\inst{\ref{inst:0019}}
\and F.        ~Riclet                        \inst{\ref{inst:0031}}
\and W.        ~Roux                          \inst{\ref{inst:0031}}
\and R.        ~Sordo                         \orcit{0000-0003-4979-0659}\inst{\ref{inst:0015}}
\and P.        ~Tanga                         \orcit{0000-0002-2718-997X}\inst{\ref{inst:0003}}
\and F.        ~Th\'{e}venin                  \inst{\ref{inst:0003}}
\and G.        ~Gracia-Abril                  \inst{\ref{inst:0071},\ref{inst:0020}}
\and J.        ~Portell                       \orcit{0000-0002-8886-8925}\inst{\ref{inst:0001}}
\and D.        ~Teyssier                      \orcit{0000-0002-6261-5292}\inst{\ref{inst:0074}}
\and M.        ~Altmann                       \orcit{0000-0002-0530-0913}\inst{\ref{inst:0020},\ref{inst:0076}}
\and R.        ~Andrae                        \inst{\ref{inst:0039}}
\and I.        ~Bellas-Velidis                \inst{\ref{inst:0078}}
\and K.        ~Benson                        \inst{\ref{inst:0013}}
\and J.        ~Berthier                      \orcit{0000-0003-1846-6485}\inst{\ref{inst:0080}}
\and R.        ~Blomme                        \orcit{0000-0002-2526-346X}\inst{\ref{inst:0053}}
\and E.        ~Brugaletta                    \orcit{0000-0003-2598-6737}\inst{\ref{inst:0082}}
\and P.W.      ~Burgess                       \inst{\ref{inst:0022}}
\and G.        ~Busso                         \orcit{0000-0003-0937-9849}\inst{\ref{inst:0022}}
\and B.        ~Carry                         \orcit{0000-0001-5242-3089}\inst{\ref{inst:0003}}
\and A.        ~Cellino                       \orcit{0000-0002-6645-334X}\inst{\ref{inst:0042}}
\and N.        ~Cheek                         \inst{\ref{inst:0087}}
\and G.        ~Clementini                    \orcit{0000-0001-9206-9723}\inst{\ref{inst:0088}}
\and Y.        ~Damerdji                      \inst{\ref{inst:0089},\ref{inst:0090}}
\and M.        ~Davidson                      \inst{\ref{inst:0091}}
\and L.        ~Delchambre                    \inst{\ref{inst:0089}}
\and A.        ~Dell'Oro                      \orcit{0000-0003-1561-9685}\inst{\ref{inst:0034}}
\and J.        ~Fern\'{a}ndez-Hern\'{a}ndez   \inst{\ref{inst:0094}}
\and L.        ~Galluccio                     \orcit{0000-0002-8541-0476}\inst{\ref{inst:0003}}
\and P.        ~Garc\'{i}a-Lario              \inst{\ref{inst:0027}}
\and M.        ~Garcia-Reinaldos              \inst{\ref{inst:0027}}
\and J.        ~Gonz\'{a}lez-N\'{u}\~{n}ez    \orcit{0000-0001-5311-5555}\inst{\ref{inst:0087},\ref{inst:0099}}
\and E.        ~Gosset                        \inst{\ref{inst:0089},\ref{inst:0033}}
\and R.        ~Haigron                       \inst{\ref{inst:0019}}
\and J.-L.     ~Halbwachs                     \orcit{0000-0003-2968-6395}\inst{\ref{inst:0103}}
\and N.C.      ~Hambly                        \orcit{0000-0002-9901-9064}\inst{\ref{inst:0091}}
\and D.L.      ~Harrison                      \orcit{0000-0001-8687-6588}\inst{\ref{inst:0022},\ref{inst:0106}}
\and D.        ~Hatzidimitriou                \orcit{0000-0002-5415-0464}\inst{\ref{inst:0107}}
\and U.        ~Heiter                        \orcit{0000-0001-6825-1066}\inst{\ref{inst:0108}}
\and J.        ~Hern\'{a}ndez                 \inst{\ref{inst:0027}}
\and D.        ~Hestroffer                    \orcit{0000-0003-0472-9459}\inst{\ref{inst:0080}}
\and S.T.      ~Hodgkin                       \inst{\ref{inst:0022}}
\and B.        ~Holl                          \orcit{0000-0001-6220-3266}\inst{\ref{inst:0023},\ref{inst:0062}}
\and K.        ~Jan{\ss}en                    \inst{\ref{inst:0114}}
\and G.        ~Jevardat de Fombelle          \inst{\ref{inst:0023}}
\and S.        ~Jordan                        \orcit{0000-0001-6316-6831}\inst{\ref{inst:0020}}
\and A.        ~Krone-Martins                 \orcit{0000-0002-2308-6623}\inst{\ref{inst:0117},\ref{inst:0118}}
\and A.C.      ~Lanzafame                     \orcit{0000-0002-2697-3607}\inst{\ref{inst:0082},\ref{inst:0120}}
\and W.        ~L\"{ o}ffler                  \inst{\ref{inst:0020}}
\and A.        ~Lorca                         \inst{\ref{inst:0024}}
\and M.        ~Manteiga                      \orcit{0000-0002-7711-5581}\inst{\ref{inst:0123}}
\and O.        ~Marchal                       \inst{\ref{inst:0103}}
\and P.M.      ~Marrese                       \inst{\ref{inst:0125},\ref{inst:0126}}
\and A.        ~Moitinho                      \orcit{0000-0003-0822-5995}\inst{\ref{inst:0117}}
\and A.        ~Mora                          \inst{\ref{inst:0024}}
\and K.        ~Muinonen                      \orcit{0000-0001-8058-2642}\inst{\ref{inst:0129},\ref{inst:0130}}
\and P.        ~Osborne                       \inst{\ref{inst:0022}}
\and E.        ~Pancino                       \orcit{0000-0003-0788-5879}\inst{\ref{inst:0034},\ref{inst:0126}}
\and T.        ~Pauwels                       \inst{\ref{inst:0053}}
\and A.        ~Recio-Blanco                  \inst{\ref{inst:0003}}
\and P.J.      ~Richards                      \inst{\ref{inst:0136}}
\and M.        ~Riello                        \orcit{0000-0002-3134-0935}\inst{\ref{inst:0022}}
\and L.        ~Rimoldini                     \orcit{0000-0002-0306-585X}\inst{\ref{inst:0062}}
\and A.C.      ~Robin                         \orcit{0000-0001-8654-9499}\inst{\ref{inst:0139}}
\and T.        ~Roegiers                      \inst{\ref{inst:0140}}
\and J.        ~Rybizki                       \orcit{0000-0002-0993-6089}\inst{\ref{inst:0039}}
\and L.M.      ~Sarro                         \orcit{0000-0002-5622-5191}\inst{\ref{inst:0142}}
\and C.        ~Siopis                        \inst{\ref{inst:0032}}
\and M.        ~Smith                         \inst{\ref{inst:0013}}
\and A.        ~Sozzetti                      \orcit{0000-0002-7504-365X}\inst{\ref{inst:0042}}
\and A.        ~Ulla                          \inst{\ref{inst:0146}}
\and E.        ~Utrilla                       \inst{\ref{inst:0024}}
\and M.        ~van Leeuwen                   \inst{\ref{inst:0022}}
\and W.        ~van Reeven                    \inst{\ref{inst:0024}}
\and U.        ~Abbas                         \orcit{0000-0002-5076-766X}\inst{\ref{inst:0042}}
\and A.        ~Abreu Aramburu                \inst{\ref{inst:0094}}
\and S.        ~Accart                        \inst{\ref{inst:0152}}
\and C.        ~Aerts                         \orcit{0000-0003-1822-7126}\inst{\ref{inst:0153},\ref{inst:0154},\ref{inst:0039}}
\and J.J.      ~Aguado                        \inst{\ref{inst:0142}}
\and M.        ~Ajaj                          \inst{\ref{inst:0019}}
\and G.        ~Altavilla                     \orcit{0000-0002-9934-1352}\inst{\ref{inst:0125},\ref{inst:0126}}
\and M.A.      ~\'{A}lvarez                   \orcit{0000-0002-6786-2620}\inst{\ref{inst:0160}}
\and J.        ~\'{A}lvarez Cid-Fuentes       \orcit{0000-0001-7153-4649}\inst{\ref{inst:0161}}
\and J.        ~Alves                         \orcit{0000-0002-4355-0921}\inst{\ref{inst:0162}}
\and R.I.      ~Anderson                      \orcit{0000-0001-8089-4419}\inst{\ref{inst:0163}}
\and E.        ~Anglada Varela                \orcit{0000-0001-7563-0689}\inst{\ref{inst:0094}}
\and M.        ~Audard                        \orcit{0000-0003-4721-034X}\inst{\ref{inst:0062}}
\and D.        ~Baines                        \orcit{0000-0002-6923-3756}\inst{\ref{inst:0074}}
\and S.G.      ~Baker                         \orcit{0000-0002-6436-1257}\inst{\ref{inst:0013}}
\and L.        ~Balaguer-N\'{u}\~{n}ez        \orcit{0000-0001-9789-7069}\inst{\ref{inst:0001}}
\and Z.        ~Balog                         \orcit{0000-0003-1748-2926}\inst{\ref{inst:0020},\ref{inst:0039}}
\and C.        ~Barache                       \inst{\ref{inst:0076}}
\and D.        ~Barbato                       \inst{\ref{inst:0023},\ref{inst:0042}}
\and M.        ~Barros                        \orcit{0000-0002-9728-9618}\inst{\ref{inst:0117}}
\and M.A.      ~Barstow                       \orcit{0000-0002-7116-3259}\inst{\ref{inst:0175}}
\and S.        ~Bartolom\'{e}                 \orcit{0000-0002-6290-6030}\inst{\ref{inst:0001}}
\and J.-L.     ~Bassilana                     \inst{\ref{inst:0152}}
\and N.        ~Bauchet                       \inst{\ref{inst:0080}}
\and A.        ~Baudesson-Stella              \inst{\ref{inst:0152}}
\and U.        ~Becciani                      \orcit{0000-0002-4389-8688}\inst{\ref{inst:0082}}
\and M.        ~Bellazzini                    \orcit{0000-0001-8200-810X}\inst{\ref{inst:0088}}
\and M.        ~Bernet                        \inst{\ref{inst:0001}}
\and S.        ~Bertone                       \orcit{0000-0001-9885-8440}\inst{\ref{inst:0183},\ref{inst:0184},\ref{inst:0042}}
\and L.        ~Bianchi                       \inst{\ref{inst:0186}}
\and S.        ~Blanco-Cuaresma               \orcit{0000-0002-1584-0171}\inst{\ref{inst:0187}}
\and T.        ~Boch                          \orcit{0000-0001-5818-2781}\inst{\ref{inst:0103}}
\and A.        ~Bombrun                       \inst{\ref{inst:0189}}
\and D.        ~Bossini                       \orcit{0000-0002-9480-8400}\inst{\ref{inst:0190}}
\and S.        ~Bouquillon                    \inst{\ref{inst:0076}}
\and A.        ~Bragaglia                     \orcit{0000-0002-0338-7883}\inst{\ref{inst:0088}}
\and L.        ~Bramante                      \inst{\ref{inst:0059}}
\and E.        ~Breedt                        \orcit{0000-0001-6180-3438}\inst{\ref{inst:0022}}
\and A.        ~Bressan                       \orcit{0000-0002-7922-8440}\inst{\ref{inst:0195}}
\and N.        ~Brouillet                     \inst{\ref{inst:0036}}
\and B.        ~Bucciarelli                   \orcit{0000-0002-5303-0268}\inst{\ref{inst:0042}}
\and A.        ~Burlacu                       \inst{\ref{inst:0198}}
\and D.        ~Busonero                      \orcit{0000-0002-3903-7076}\inst{\ref{inst:0042}}
\and A.G.      ~Butkevich                     \inst{\ref{inst:0042}}
\and R.        ~Buzzi                         \orcit{0000-0001-9389-5701}\inst{\ref{inst:0042}}
\and E.        ~Caffau                        \orcit{0000-0001-6011-6134}\inst{\ref{inst:0019}}
\and R.        ~Cancelliere                   \orcit{0000-0002-9120-3799}\inst{\ref{inst:0203}}
\and H.        ~C\'{a}novas                   \orcit{0000-0001-7668-8022}\inst{\ref{inst:0024}}
\and R.        ~Carballo                      \inst{\ref{inst:0205}}
\and T.        ~Carlucci                      \inst{\ref{inst:0076}}
\and M.I       ~Carnerero                     \orcit{0000-0001-5843-5515}\inst{\ref{inst:0042}}
\and J.M.      ~Carrasco                      \orcit{0000-0002-3029-5853}\inst{\ref{inst:0001}}
\and L.        ~Casamiquela                   \orcit{0000-0001-5238-8674}\inst{\ref{inst:0036}}
\and M.        ~Castellani                    \orcit{0000-0002-7650-7428}\inst{\ref{inst:0125}}
\and A.        ~Castro-Ginard                 \orcit{0000-0002-9419-3725}\inst{\ref{inst:0001}}
\and P.        ~Castro Sampol                 \inst{\ref{inst:0001}}
\and L.        ~Chaoul                        \inst{\ref{inst:0031}}
\and P.        ~Charlot                       \inst{\ref{inst:0036}}
\and A.        ~Chiavassa                     \orcit{0000-0003-3891-7554}\inst{\ref{inst:0003}}
\and M.-R. L.  ~Cioni                         \orcit{0000-0002-6797-696x}\inst{\ref{inst:0114}}
\and G.        ~Comoretto                     \inst{\ref{inst:0217}}
\and W.J.      ~Cooper                        \orcit{0000-0003-3501-8967}\inst{\ref{inst:0218},\ref{inst:0042}}
\and T.        ~Cornez                        \inst{\ref{inst:0152}}
\and S.        ~Cowell                        \inst{\ref{inst:0022}}
\and F.        ~Crifo                         \inst{\ref{inst:0019}}
\and M.        ~Crosta                        \orcit{0000-0003-4369-3786}\inst{\ref{inst:0042}}
\and C.        ~Crowley                       \inst{\ref{inst:0189}}
\and C.        ~Dafonte                       \orcit{0000-0003-4693-7555}\inst{\ref{inst:0160}}
\and A.        ~Dapergolas                    \inst{\ref{inst:0078}}
\and M.        ~David                         \orcit{0000-0002-4172-3112}\inst{\ref{inst:0227}}
\and P.        ~David                         \inst{\ref{inst:0080}}
\and P.        ~de Laverny                    \inst{\ref{inst:0003}}
\and F.        ~De Luise                      \orcit{0000-0002-6570-8208}\inst{\ref{inst:0230}}
\and R.        ~De March                      \orcit{0000-0003-0567-842X}\inst{\ref{inst:0059}}
\and J.        ~De Ridder                     \orcit{0000-0001-6726-2863}\inst{\ref{inst:0153}}
\and R.        ~de Souza                      \inst{\ref{inst:0233}}
\and P.        ~de Teodoro                    \inst{\ref{inst:0027}}
\and A.        ~de Torres                     \inst{\ref{inst:0189}}
\and E.F.      ~del Peloso                    \inst{\ref{inst:0020}}
\and E.        ~del Pozo                      \inst{\ref{inst:0024}}
\and A.        ~Delgado                       \inst{\ref{inst:0022}}
\and H.E.      ~Delgado                       \orcit{0000-0003-1409-4282}\inst{\ref{inst:0142}}
\and J.-B.     ~Delisle                       \orcit{0000-0001-5844-9888}\inst{\ref{inst:0023}}
\and P.        ~Di Matteo                     \inst{\ref{inst:0019}}
\and S.        ~Diakite                       \inst{\ref{inst:0242}}
\and C.        ~Diener                        \inst{\ref{inst:0022}}
\and E.        ~Distefano                     \orcit{0000-0002-2448-2513}\inst{\ref{inst:0082}}
\and C.        ~Dolding                       \inst{\ref{inst:0013}}
\and D.        ~Eappachen                     \inst{\ref{inst:0246},\ref{inst:0154}}
\and H.        ~Enke                          \orcit{0000-0002-2366-8316}\inst{\ref{inst:0114}}
\and P.        ~Esquej                        \orcit{0000-0001-8195-628X}\inst{\ref{inst:0250}}
\and C.        ~Fabre                         \inst{\ref{inst:0251}}
\and M.        ~Fabrizio                      \orcit{0000-0001-5829-111X}\inst{\ref{inst:0125},\ref{inst:0126}}
\and S.        ~Faigler                       \inst{\ref{inst:0254}}
\and G.        ~Fedorets                      \inst{\ref{inst:0129},\ref{inst:0256}}
\and P.        ~Fernique                      \orcit{0000-0002-3304-2923}\inst{\ref{inst:0103},\ref{inst:0258}}
\and A.        ~Fienga                        \orcit{0000-0002-4755-7637}\inst{\ref{inst:0259},\ref{inst:0080}}
\and C.        ~Fouron                        \inst{\ref{inst:0198}}
\and F.        ~Fragkoudi                     \inst{\ref{inst:0262}}
\and E.        ~Fraile                        \inst{\ref{inst:0250}}
\and F.        ~Franke                        \inst{\ref{inst:0264}}
\and M.        ~Gai                           \orcit{0000-0001-9008-134X}\inst{\ref{inst:0042}}
\and D.        ~Garabato                      \orcit{0000-0002-7133-6623}\inst{\ref{inst:0160}}
\and A.        ~Garcia-Gutierrez              \inst{\ref{inst:0001}}
\and M.        ~Garc\'{i}a-Torres             \orcit{0000-0002-6867-7080}\inst{\ref{inst:0268}}
\and A.        ~Garofalo                      \orcit{0000-0002-5907-0375}\inst{\ref{inst:0088}}
\and P.        ~Gavras                        \orcit{0000-0002-4383-4836}\inst{\ref{inst:0250}}
\and E.        ~Gerlach                       \orcit{0000-0002-9533-2168}\inst{\ref{inst:0026}}
\and R.        ~Geyer                         \orcit{0000-0001-6967-8707}\inst{\ref{inst:0026}}
\and P.        ~Giacobbe                      \inst{\ref{inst:0042}}
\and G.        ~Gilmore                       \orcit{0000-0003-4632-0213}\inst{\ref{inst:0022}}
\and S.        ~Girona                        \orcit{0000-0002-1975-1918}\inst{\ref{inst:0161}}
\and G.        ~Giuffrida                     \inst{\ref{inst:0125}}
\and A.        ~Gomez                         \orcit{0000-0002-3796-3690}\inst{\ref{inst:0160}}
\and I.        ~Gonzalez-Santamaria           \orcit{0000-0002-8537-9384}\inst{\ref{inst:0160}}
\and J.J.      ~Gonz\'{a}lez-Vidal            \inst{\ref{inst:0001}}
\and M.        ~Granvik                       \orcit{0000-0002-5624-1888}\inst{\ref{inst:0129},\ref{inst:0281}}
\and R.        ~Guti\'{e}rrez-S\'{a}nchez     \inst{\ref{inst:0074}}
\and L.P.      ~Guy                           \orcit{0000-0003-0800-8755}\inst{\ref{inst:0062},\ref{inst:0217}}
\and M.        ~Hauser                        \inst{\ref{inst:0039},\ref{inst:0286}}
\and M.        ~Haywood                       \orcit{0000-0003-0434-0400}\inst{\ref{inst:0019}}
\and S.L.      ~Hidalgo                       \orcit{0000-0002-0002-9298}\inst{\ref{inst:0288},\ref{inst:0289}}
\and T.        ~Hilger                        \orcit{0000-0003-1646-0063}\inst{\ref{inst:0026}}
\and N.        ~H\l{}adczuk                   \inst{\ref{inst:0027}}
\and D.        ~Hobbs                         \orcit{0000-0002-2696-1366}\inst{\ref{inst:0002}}
\and G.        ~Holland                       \inst{\ref{inst:0022}}
\and H.E.      ~Huckle                        \inst{\ref{inst:0013}}
\and G.        ~Jasniewicz                    \inst{\ref{inst:0295}}
\and P.G.      ~Jonker                        \orcit{0000-0001-5679-0695}\inst{\ref{inst:0154},\ref{inst:0246}}
\and J.        ~Juaristi Campillo             \inst{\ref{inst:0020}}
\and F.        ~Julbe                         \inst{\ref{inst:0001}}
\and L.        ~Karbevska                     \inst{\ref{inst:0023}}
\and P.        ~Kervella                      \orcit{0000-0003-0626-1749}\inst{\ref{inst:0301}}
\and A.        ~Kochoska                      \orcit{0000-0002-9739-8371}\inst{\ref{inst:0302}}
\and M.        ~Kontizas                      \orcit{0000-0001-7177-0158}\inst{\ref{inst:0107}}
\and A.J.      ~Korn                          \orcit{0000-0002-3881-6756}\inst{\ref{inst:0108}}
\and Z.        ~Kostrzewa-Rutkowska           \inst{\ref{inst:0014},\ref{inst:0246}}
\and K.        ~Kruszy\'{n}ska                \orcit{0000-0002-2729-5369}\inst{\ref{inst:0307}}
\and S.        ~Lambert                       \orcit{0000-0001-6759-5502}\inst{\ref{inst:0076}}
\and A.F.      ~Lanza                         \orcit{0000-0001-5928-7251}\inst{\ref{inst:0082}}
\and Y.        ~Lasne                         \inst{\ref{inst:0152}}
\and J.-F.     ~Le Campion                    \inst{\ref{inst:0311}}
\and Y.        ~Le Fustec                     \inst{\ref{inst:0198}}
\and Y.        ~Lebreton                      \orcit{0000-0002-4834-2144}\inst{\ref{inst:0301},\ref{inst:0314}}
\and T.        ~Lebzelter                     \orcit{0000-0002-0702-7551}\inst{\ref{inst:0162}}
\and S.        ~Leccia                        \orcit{0000-0001-5685-6930}\inst{\ref{inst:0316}}
\and N.        ~Leclerc                       \inst{\ref{inst:0019}}
\and I.        ~Lecoeur-Taibi                 \orcit{0000-0003-0029-8575}\inst{\ref{inst:0062}}
\and S.        ~Liao                          \inst{\ref{inst:0042}}
\and E.        ~Licata                        \orcit{0000-0002-5203-0135}\inst{\ref{inst:0042}}
\and H.E.P.    ~Lindstr{\o}m                  \inst{\ref{inst:0042},\ref{inst:0322}}
\and T.A.      ~Lister                        \orcit{0000-0002-3818-7769}\inst{\ref{inst:0323}}
\and E.        ~Livanou                       \inst{\ref{inst:0107}}
\and A.        ~Lobel                         \inst{\ref{inst:0053}}
\and P.        ~Madrero Pardo                 \inst{\ref{inst:0001}}
\and S.        ~Managau                       \inst{\ref{inst:0152}}
\and R.G.      ~Mann                          \orcit{0000-0002-0194-325X}\inst{\ref{inst:0091}}
\and J.M.      ~Marchant                      \inst{\ref{inst:0329}}
\and M.        ~Marconi                       \orcit{0000-0002-1330-2927}\inst{\ref{inst:0316}}
\and M.M.S.    ~Marcos Santos                 \inst{\ref{inst:0087}}
\and S.        ~Marinoni                      \orcit{0000-0001-7990-6849}\inst{\ref{inst:0125},\ref{inst:0126}}
\and F.        ~Marocco                       \orcit{0000-0001-7519-1700}\inst{\ref{inst:0334},\ref{inst:0335}}
\and D.J.      ~Marshall                      \inst{\ref{inst:0336}}
\and L.        ~Martin Polo                   \inst{\ref{inst:0087}}
\and J.M.      ~Mart\'{i}n-Fleitas            \orcit{0000-0002-8594-569X}\inst{\ref{inst:0024}}
\and A.        ~Masip                         \inst{\ref{inst:0001}}
\and D.        ~Massari                       \orcit{0000-0001-8892-4301}\inst{\ref{inst:0088}}
\and A.        ~Mastrobuono-Battisti          \orcit{0000-0002-2386-9142}\inst{\ref{inst:0002}}
\and T.        ~Mazeh                         \orcit{0000-0002-3569-3391}\inst{\ref{inst:0254}}
\and S.        ~Messina                       \orcit{0000-0002-2851-2468}\inst{\ref{inst:0082}}
\and D.        ~Michalik                      \orcit{0000-0002-7618-6556}\inst{\ref{inst:0016}}
\and N.R.      ~Millar                        \inst{\ref{inst:0022}}
\and A.        ~Mints                         \orcit{0000-0002-8440-1455}\inst{\ref{inst:0114}}
\and D.        ~Molina                        \orcit{0000-0003-4814-0275}\inst{\ref{inst:0001}}
\and R.        ~Molinaro                      \orcit{0000-0003-3055-6002}\inst{\ref{inst:0316}}
\and L.        ~Moln\'{a}r                    \orcit{0000-0002-8159-1599}\inst{\ref{inst:0349},\ref{inst:0350},\ref{inst:0351}}
\and P.        ~Montegriffo                   \inst{\ref{inst:0088}}
\and R.        ~Mor                           \orcit{0000-0002-8179-6527}\inst{\ref{inst:0001}}
\and R.        ~Morbidelli                    \orcit{0000-0001-7627-4946}\inst{\ref{inst:0042}}
\and T.        ~Morel                         \inst{\ref{inst:0089}}
\and D.        ~Morris                        \inst{\ref{inst:0091}}
\and A.F.      ~Mulone                        \inst{\ref{inst:0059}}
\and D.        ~Munoz                         \inst{\ref{inst:0152}}
\and T.        ~Muraveva                      \orcit{0000-0002-0969-1915}\inst{\ref{inst:0088}}
\and C.P.      ~Murphy                        \inst{\ref{inst:0027}}
\and I.        ~Musella                       \orcit{0000-0001-5909-6615}\inst{\ref{inst:0316}}
\and L.        ~Noval                         \inst{\ref{inst:0152}}
\and C.        ~Ord\'{e}novic                 \inst{\ref{inst:0003}}
\and G.        ~Orr\`{u}                      \inst{\ref{inst:0059}}
\and J.        ~Osinde                        \inst{\ref{inst:0250}}
\and C.        ~Pagani                        \inst{\ref{inst:0175}}
\and I.        ~Pagano                        \orcit{0000-0001-9573-4928}\inst{\ref{inst:0082}}
\and L.        ~Palaversa                     \inst{\ref{inst:0368},\ref{inst:0022}}
\and P.A.      ~Palicio                       \orcit{0000-0002-7432-8709}\inst{\ref{inst:0003}}
\and A.        ~Panahi                        \orcit{0000-0001-5850-4373}\inst{\ref{inst:0254}}
\and M.        ~Pawlak                        \orcit{0000-0002-5632-9433}\inst{\ref{inst:0372},\ref{inst:0307}}
\and X.        ~Pe\~{n}alosa Esteller         \inst{\ref{inst:0001}}
\and A.        ~Penttil\"{ a}                 \orcit{0000-0001-7403-1721}\inst{\ref{inst:0129}}
\and A.M.      ~Piersimoni                    \orcit{0000-0002-8019-3708}\inst{\ref{inst:0230}}
\and F.-X.     ~Pineau                        \orcit{0000-0002-2335-4499}\inst{\ref{inst:0103}}
\and E.        ~Plachy                        \orcit{0000-0002-5481-3352}\inst{\ref{inst:0349},\ref{inst:0350},\ref{inst:0351}}
\and G.        ~Plum                          \inst{\ref{inst:0019}}
\and E.        ~Poggio                        \orcit{0000-0003-3793-8505}\inst{\ref{inst:0042}}
\and E.        ~Poretti                       \orcit{0000-0003-1200-0473}\inst{\ref{inst:0383}}
\and E.        ~Poujoulet                     \inst{\ref{inst:0384}}
\and A.        ~Pr\v{s}a                      \orcit{0000-0002-1913-0281}\inst{\ref{inst:0302}}
\and L.        ~Pulone                        \orcit{0000-0002-5285-998X}\inst{\ref{inst:0125}}
\and E.        ~Racero                        \inst{\ref{inst:0087},\ref{inst:0388}}
\and S.        ~Ragaini                       \inst{\ref{inst:0088}}
\and M.        ~Rainer                        \orcit{0000-0002-8786-2572}\inst{\ref{inst:0034}}
\and C.M.      ~Raiteri                       \orcit{0000-0003-1784-2784}\inst{\ref{inst:0042}}
\and N.        ~Rambaux                       \inst{\ref{inst:0080}}
\and M.        ~Ramos-Lerate                  \inst{\ref{inst:0393}}
\and P.        ~Re Fiorentin                  \orcit{0000-0002-4995-0475}\inst{\ref{inst:0042}}
\and S.        ~Regibo                        \inst{\ref{inst:0153}}
\and C.        ~Reyl\'{e}                     \inst{\ref{inst:0139}}
\and V.        ~Ripepi                        \orcit{0000-0003-1801-426X}\inst{\ref{inst:0316}}
\and A.        ~Riva                          \orcit{0000-0002-6928-8589}\inst{\ref{inst:0042}}
\and G.        ~Rixon                         \inst{\ref{inst:0022}}
\and N.        ~Robichon                      \orcit{0000-0003-4545-7517}\inst{\ref{inst:0019}}
\and C.        ~Robin                         \inst{\ref{inst:0152}}
\and M.        ~Roelens                       \orcit{0000-0003-0876-4673}\inst{\ref{inst:0023}}
\and L.        ~Rohrbasser                    \inst{\ref{inst:0062}}
\and N.        ~Rowell                        \inst{\ref{inst:0091}}
\and F.        ~Royer                         \orcit{0000-0002-9374-8645}\inst{\ref{inst:0019}}
\and K.A.      ~Rybicki                       \orcit{0000-0002-9326-9329}\inst{\ref{inst:0307}}
\and G.        ~Sadowski                      \inst{\ref{inst:0032}}
\and A.        ~Sagrist\`{a} Sell\'{e}s       \orcit{0000-0001-6191-2028}\inst{\ref{inst:0020}}
\and J.        ~Sahlmann                      \orcit{0000-0001-9525-3673}\inst{\ref{inst:0250}}
\and J.        ~Salgado                       \orcit{0000-0002-3680-4364}\inst{\ref{inst:0074}}
\and E.        ~Salguero                      \inst{\ref{inst:0094}}
\and N.        ~Samaras                       \orcit{0000-0001-8375-6652}\inst{\ref{inst:0053}}
\and V.        ~Sanchez Gimenez               \inst{\ref{inst:0001}}
\and N.        ~Sanna                         \inst{\ref{inst:0034}}
\and R.        ~Santove\~{n}a                 \orcit{0000-0002-9257-2131}\inst{\ref{inst:0160}}
\and M.        ~Sarasso                       \orcit{0000-0001-5121-0727}\inst{\ref{inst:0042}}
\and M.        ~Schultheis                    \orcit{0000-0002-6590-1657}\inst{\ref{inst:0003}}
\and E.        ~Sciacca                       \orcit{0000-0002-5574-2787}\inst{\ref{inst:0082}}
\and M.        ~Segol                         \inst{\ref{inst:0264}}
\and J.C.      ~Segovia                       \inst{\ref{inst:0087}}
\and D.        ~S\'{e}gransan                 \orcit{0000-0003-2355-8034}\inst{\ref{inst:0023}}
\and D.        ~Semeux                        \inst{\ref{inst:0251}}
\and H.I.      ~Siddiqui                      \orcit{0000-0003-1853-6033}\inst{\ref{inst:0423}}
\and A.        ~Siebert                       \orcit{0000-0001-8059-2840}\inst{\ref{inst:0103},\ref{inst:0258}}
\and L.        ~Siltala                       \orcit{0000-0002-6938-794X}\inst{\ref{inst:0129}}
\and E.        ~Slezak                        \inst{\ref{inst:0003}}
\and R.L.      ~Smart                         \orcit{0000-0002-4424-4766}\inst{\ref{inst:0042}}
\and E.        ~Solano                        \inst{\ref{inst:0429}}
\and F.        ~Solitro                       \inst{\ref{inst:0059}}
\and D.        ~Souami                        \orcit{0000-0003-4058-0815}\inst{\ref{inst:0301},\ref{inst:0432}}
\and J.        ~Souchay                       \inst{\ref{inst:0076}}
\and A.        ~Spagna                        \orcit{0000-0003-1732-2412}\inst{\ref{inst:0042}}
\and F.        ~Spoto                         \orcit{0000-0001-7319-5847}\inst{\ref{inst:0187}}
\and I.A.      ~Steele                        \orcit{0000-0001-8397-5759}\inst{\ref{inst:0329}}
\and H.        ~Steidelm\"{ u}ller            \inst{\ref{inst:0026}}
\and C.A.      ~Stephenson                    \inst{\ref{inst:0074}}
\and M.        ~S\"{ u}veges                  \inst{\ref{inst:0062},\ref{inst:0440},\ref{inst:0039}}
\and L.        ~Szabados                      \orcit{0000-0002-2046-4131}\inst{\ref{inst:0349}}
\and E.        ~Szegedi-Elek                  \orcit{0000-0001-7807-6644}\inst{\ref{inst:0349}}
\and F.        ~Taris                         \inst{\ref{inst:0076}}
\and G.        ~Tauran                        \inst{\ref{inst:0152}}
\and M.B.      ~Taylor                        \orcit{0000-0002-4209-1479}\inst{\ref{inst:0446}}
\and R.        ~Teixeira                      \orcit{0000-0002-6806-6626}\inst{\ref{inst:0233}}
\and W.        ~Thuillot                      \inst{\ref{inst:0080}}
\and N.        ~Tonello                       \orcit{0000-0003-0550-1667}\inst{\ref{inst:0161}}
\and F.        ~Torra                         \orcit{0000-0002-8429-299X}\inst{\ref{inst:0048}}
\and J.        ~Torra$^\dagger$               \inst{\ref{inst:0001}}
\and C.        ~Turon                         \orcit{0000-0003-1236-5157}\inst{\ref{inst:0019}}
\and N.        ~Unger                         \orcit{0000-0003-3993-7127}\inst{\ref{inst:0023}}
\and M.        ~Vaillant                      \inst{\ref{inst:0152}}
\and E.        ~van Dillen                    \inst{\ref{inst:0264}}
\and O.        ~Vanel                         \inst{\ref{inst:0019}}
\and A.        ~Vecchiato                     \orcit{0000-0003-1399-5556}\inst{\ref{inst:0042}}
\and Y.        ~Viala                         \inst{\ref{inst:0019}}
\and D.        ~Vicente                       \inst{\ref{inst:0161}}
\and S.        ~Voutsinas                     \inst{\ref{inst:0091}}
\and M.        ~Weiler                        \inst{\ref{inst:0001}}
\and T.        ~Wevers                        \orcit{0000-0002-4043-9400}\inst{\ref{inst:0022}}
\and \L{}.     ~Wyrzykowski                   \orcit{0000-0002-9658-6151}\inst{\ref{inst:0307}}
\and A.        ~Yoldas                        \inst{\ref{inst:0022}}
\and P.        ~Yvard                         \inst{\ref{inst:0264}}
\and H.        ~Zhao                          \orcit{0000-0003-2645-6869}\inst{\ref{inst:0003}}
\and J.        ~Zorec                         \inst{\ref{inst:0467}}
\and S.        ~Zucker                        \orcit{0000-0003-3173-3138}\inst{\ref{inst:0468}}
\and C.        ~Zurbach                       \inst{\ref{inst:0469}}
\and T.        ~Zwitter                       \orcit{0000-0002-2325-8763}\inst{\ref{inst:0470}}
}
\institute{
     Institut de Ci\`{e}ncies del Cosmos (ICCUB), Universitat  de  Barcelona  (IEEC-UB), Mart\'{i} i  Franqu\`{e}s  1, 08028 Barcelona, Spain\relax                                                                                                                                                              \label{inst:0001}
\and Lund Observatory, Department of Astronomy and Theoretical Physics, Lund University, Box 43, 22100 Lund, Sweden\relax                                                                                                                                                                                        \label{inst:0002}
\and Universit\'{e} C\^{o}te d'Azur, Observatoire de la C\^{o}te d'Azur, CNRS, Laboratoire Lagrange, Bd de l'Observatoire, CS 34229, 06304 Nice Cedex 4, France\relax                                                                                                                                            \label{inst:0003}
\and Kapteyn Astronomical Institute, University of Groningen, Landleven 12, 9747 AD Groningen, The Netherlands\relax                                                                                                                                                                                             \label{inst:0005}
\and Centro de Astronom\'{i}a - CITEVA, Universidad de Antofagasta, Avenida Angamos 601, Antofagasta 1270300, Chile\relax                                                                                                                                                                                        \label{inst:0008}
\and Mullard Space Science Laboratory, University College London, Holmbury St Mary, Dorking, Surrey RH5 6NT, United Kingdom\relax                                                                                                                                                                                \label{inst:0013}
\and Leiden Observatory, Leiden University, Niels Bohrweg 2, 2333 CA Leiden, The Netherlands\relax                                                                                                                                                                                                               \label{inst:0014}
\and INAF - Osservatorio astronomico di Padova, Vicolo Osservatorio 5, 35122 Padova, Italy\relax                                                                                                                                                                                                                 \label{inst:0015}
\and European Space Agency (ESA), European Space Research and Technology Centre (ESTEC), Keplerlaan 1, 2201AZ, Noordwijk, The Netherlands\relax                                                                                                                                                                  \label{inst:0016}
\and Univ. Grenoble Alpes, CNRS, IPAG, 38000 Grenoble, France\relax                                                                                                                                                                                                                                              \label{inst:0018}
\and GEPI, Observatoire de Paris, Universit\'{e} PSL, CNRS, 5 Place Jules Janssen, 92190 Meudon, France\relax                                                                                                                                                                                                    \label{inst:0019}
\and Astronomisches Rechen-Institut, Zentrum f\"{ u}r Astronomie der Universit\"{ a}t Heidelberg, M\"{ o}nchhofstr. 12-14, 69120 Heidelberg, Germany\relax                                                                                                                                                       \label{inst:0020}
\and Institute of Astronomy, University of Cambridge, Madingley Road, Cambridge CB3 0HA, United Kingdom\relax                                                                                                                                                                                                    \label{inst:0022}
\and Department of Astronomy, University of Geneva, Chemin des Maillettes 51, 1290 Versoix, Switzerland\relax                                                                                                                                                                                                    \label{inst:0023}
\and Aurora Technology for European Space Agency (ESA), Camino bajo del Castillo, s/n, Urbanizacion Villafranca del Castillo, Villanueva de la Ca\~{n}ada, 28692 Madrid, Spain\relax                                                                                                                             \label{inst:0024}
\and Lohrmann Observatory, Technische Universit\"{ a}t Dresden, Mommsenstra{\ss}e 13, 01062 Dresden, Germany\relax                                                                                                                                                                                               \label{inst:0026}
\and European Space Agency (ESA), European Space Astronomy Centre (ESAC), Camino bajo del Castillo, s/n, Urbanizacion Villafranca del Castillo, Villanueva de la Ca\~{n}ada, 28692 Madrid, Spain\relax                                                                                                           \label{inst:0027}
\and CNES Centre Spatial de Toulouse, 18 avenue Edouard Belin, 31401 Toulouse Cedex 9, France\relax                                                                                                                                                                                                              \label{inst:0031}
\and Institut d'Astronomie et d'Astrophysique, Universit\'{e} Libre de Bruxelles CP 226, Boulevard du Triomphe, 1050 Brussels, Belgium\relax                                                                                                                                                                     \label{inst:0032}
\and F.R.S.-FNRS, Rue d'Egmont 5, 1000 Brussels, Belgium\relax                                                                                                                                                                                                                                                   \label{inst:0033}
\and INAF - Osservatorio Astrofisico di Arcetri, Largo Enrico Fermi 5, 50125 Firenze, Italy\relax                                                                                                                                                                                                                \label{inst:0034}
\and Laboratoire d'astrophysique de Bordeaux, Univ. Bordeaux, CNRS, B18N, all{\'e}e Geoffroy Saint-Hilaire, 33615 Pessac, France\relax                                                                                                                                                                           \label{inst:0036}
\and Max Planck Institute for Astronomy, K\"{ o}nigstuhl 17, 69117 Heidelberg, Germany\relax                                                                                                                                                                                                                     \label{inst:0039}
\and INAF - Osservatorio Astrofisico di Torino, via Osservatorio 20, 10025 Pino Torinese (TO), Italy\relax                                                                                                                                                                                                       \label{inst:0042}
\and University of Turin, Department of Physics, Via Pietro Giuria 1, 10125 Torino, Italy\relax                                                                                                                                                                                                                  \label{inst:0045}
\and DAPCOM for Institut de Ci\`{e}ncies del Cosmos (ICCUB), Universitat  de  Barcelona  (IEEC-UB), Mart\'{i} i  Franqu\`{e}s  1, 08028 Barcelona, Spain\relax                                                                                                                                                   \label{inst:0048}
\and Royal Observatory of Belgium, Ringlaan 3, 1180 Brussels, Belgium\relax                                                                                                                                                                                                                                      \label{inst:0053}
\and ALTEC S.p.a, Corso Marche, 79,10146 Torino, Italy\relax                                                                                                                                                                                                                                                     \label{inst:0059}
\and Department of Astronomy, University of Geneva, Chemin d'Ecogia 16, 1290 Versoix, Switzerland\relax                                                                                                                                                                                                          \label{inst:0062}
\and Sednai S\`{a}rl, Geneva, Switzerland\relax                                                                                                                                                                                                                                                                  \label{inst:0063}
\and Gaia DPAC Project Office, ESAC, Camino bajo del Castillo, s/n, Urbanizacion Villafranca del Castillo, Villanueva de la Ca\~{n}ada, 28692 Madrid, Spain\relax                                                                                                                                                \label{inst:0071}
\and Telespazio Vega UK Ltd for European Space Agency (ESA), Camino bajo del Castillo, s/n, Urbanizacion Villafranca del Castillo, Villanueva de la Ca\~{n}ada, 28692 Madrid, Spain\relax                                                                                                                        \label{inst:0074}
\and SYRTE, Observatoire de Paris, Universit\'{e} PSL, CNRS,  Sorbonne Universit\'{e}, LNE, 61 avenue de l’Observatoire 75014 Paris, France\relax                                                                                                                                                              \label{inst:0076}
\and National Observatory of Athens, I. Metaxa and Vas. Pavlou, Palaia Penteli, 15236 Athens, Greece\relax                                                                                                                                                                                                       \label{inst:0078}
\and IMCCE, Observatoire de Paris, Universit\'{e} PSL, CNRS, Sorbonne Universit{\'e}, Univ. Lille, 77 av. Denfert-Rochereau, 75014 Paris, France\relax                                                                                                                                                           \label{inst:0080}
\and INAF - Osservatorio Astrofisico di Catania, via S. Sofia 78, 95123 Catania, Italy\relax                                                                                                                                                                                                                     \label{inst:0082}
\and Serco Gesti\'{o}n de Negocios for European Space Agency (ESA), Camino bajo del Castillo, s/n, Urbanizacion Villafranca del Castillo, Villanueva de la Ca\~{n}ada, 28692 Madrid, Spain\relax                                                                                                                 \label{inst:0087}
\and INAF - Osservatorio di Astrofisica e Scienza dello Spazio di Bologna, via Piero Gobetti 93/3, 40129 Bologna, Italy\relax                                                                                                                                                                                    \label{inst:0088}
\and Institut d'Astrophysique et de G\'{e}ophysique, Universit\'{e} de Li\`{e}ge, 19c, All\'{e}e du 6 Ao\^{u}t, B-4000 Li\`{e}ge, Belgium\relax                                                                                                                                                                  \label{inst:0089}
\and CRAAG - Centre de Recherche en Astronomie, Astrophysique et G\'{e}ophysique, Route de l'Observatoire Bp 63 Bouzareah 16340 Algiers, Algeria\relax                                                                                                                                                           \label{inst:0090}
\and Institute for Astronomy, University of Edinburgh, Royal Observatory, Blackford Hill, Edinburgh EH9 3HJ, United Kingdom\relax                                                                                                                                                                                \label{inst:0091}
\and ATG Europe for European Space Agency (ESA), Camino bajo del Castillo, s/n, Urbanizacion Villafranca del Castillo, Villanueva de la Ca\~{n}ada, 28692 Madrid, Spain\relax                                                                                                                                    \label{inst:0094}
\and ETSE Telecomunicaci\'{o}n, Universidade de Vigo, Campus Lagoas-Marcosende, 36310 Vigo, Galicia, Spain\relax                                                                                                                                                                                                 \label{inst:0099}
\and Universit\'{e} de Strasbourg, CNRS, Observatoire astronomique de Strasbourg, UMR 7550,  11 rue de l'Universit\'{e}, 67000 Strasbourg, France\relax                                                                                                                                                          \label{inst:0103}
\and Kavli Institute for Cosmology Cambridge, Institute of Astronomy, Madingley Road, Cambridge, CB3 0HA\relax                                                                                                                                                                                                   \label{inst:0106}
\and Department of Astrophysics, Astronomy and Mechanics, National and Kapodistrian University of Athens, Panepistimiopolis, Zografos, 15783 Athens, Greece\relax                                                                                                                                                \label{inst:0107}
\and Observational Astrophysics, Division of Astronomy and Space Physics, Department of Physics and Astronomy, Uppsala University, Box 516, 751 20 Uppsala, Sweden\relax                                                                                                                                         \label{inst:0108}
\and Leibniz Institute for Astrophysics Potsdam (AIP), An der Sternwarte 16, 14482 Potsdam, Germany\relax                                                                                                                                                                                                        \label{inst:0114}
\and CENTRA, Faculdade de Ci\^{e}ncias, Universidade de Lisboa, Edif. C8, Campo Grande, 1749-016 Lisboa, Portugal\relax                                                                                                                                                                                          \label{inst:0117}
\and Department of Informatics, Donald Bren School of Information and Computer Sciences, University of California, 5019 Donald Bren Hall, 92697-3440 CA Irvine, United States\relax                                                                                                                              \label{inst:0118}
\and Dipartimento di Fisica e Astronomia ""Ettore Majorana"", Universit\`{a} di Catania, Via S. Sofia 64, 95123 Catania, Italy\relax                                                                                                                                                                             \label{inst:0120}
\and CITIC, Department of Nautical Sciences and Marine Engineering, University of A Coru\~{n}a, Campus de Elvi\~{n}a S/N, 15071, A Coru\~{n}a, Spain\relax                                                                                                                                                       \label{inst:0123}
\vfill
\and INAF - Osservatorio Astronomico di Roma, Via Frascati 33, 00078 Monte Porzio Catone (Roma), Italy\relax                                                                                                                                                                                                     \label{inst:0125}
\and Space Science Data Center - ASI, Via del Politecnico SNC, 00133 Roma, Italy\relax                                                                                                                                                                                                                           \label{inst:0126}
\and Department of Physics, University of Helsinki, P.O. Box 64, 00014 Helsinki, Finland\relax                                                                                                                                                                                                                   \label{inst:0129}
\and Finnish Geospatial Research Institute FGI, Geodeetinrinne 2, 02430 Masala, Finland\relax                                                                                                                                                                                                                    \label{inst:0130}
\and STFC, Rutherford Appleton Laboratory, Harwell, Didcot, OX11 0QX, United Kingdom\relax                                                                                                                                                                                                                       \label{inst:0136}
\and Institut UTINAM CNRS UMR6213, Universit\'{e} Bourgogne Franche-Comt\'{e}, OSU THETA Franche-Comt\'{e} Bourgogne, Observatoire de Besan\c{c}on, BP1615, 25010 Besan\c{c}on Cedex, France\relax                                                                                                               \label{inst:0139}
\and HE Space Operations BV for European Space Agency (ESA), Keplerlaan 1, 2201AZ, Noordwijk, The Netherlands\relax                                                                                                                                                                                              \label{inst:0140}
\and Dpto. de Inteligencia Artificial, UNED, c/ Juan del Rosal 16, 28040 Madrid, Spain\relax                                                                                                                                                                                                                     \label{inst:0142}
\and Applied Physics Department, Universidade de Vigo, 36310 Vigo, Spain\relax                                                                                                                                                                                                                                   \label{inst:0146}
\and Thales Services for CNES Centre Spatial de Toulouse, 18 avenue Edouard Belin, 31401 Toulouse Cedex 9, France\relax                                                                                                                                                                                          \label{inst:0152}
\and Instituut voor Sterrenkunde, KU Leuven, Celestijnenlaan 200D, 3001 Leuven, Belgium\relax                                                                                                                                                                                                                    \label{inst:0153}
\and Department of Astrophysics/IMAPP, Radboud University, P.O.Box 9010, 6500 GL Nijmegen, The Netherlands\relax                                                                                                                                                                                                 \label{inst:0154}
\and CITIC - Department of Computer Science and Information Technologies, University of A Coru\~{n}a, Campus de Elvi\~{n}a S/N, 15071, A Coru\~{n}a, Spain\relax                                                                                                                                                 \label{inst:0160}
\and Barcelona Supercomputing Center (BSC) - Centro Nacional de Supercomputaci\'{o}n, c/ Jordi Girona 29, Ed. Nexus II, 08034 Barcelona, Spain\relax                                                                                                                                                             \label{inst:0161}
\and University of Vienna, Department of Astrophysics, T\"{ u}rkenschanzstra{\ss}e 17, A1180 Vienna, Austria\relax                                                                                                                                                                                               \label{inst:0162}
\and European Southern Observatory, Karl-Schwarzschild-Str. 2, 85748 Garching, Germany\relax                                                                                                                                                                                                                     \label{inst:0163}
\and School of Physics and Astronomy, University of Leicester, University Road, Leicester LE1 7RH, United Kingdom\relax                                                                                                                                                                                          \label{inst:0175}
\and Center for Research and Exploration in Space Science and Technology, University of Maryland Baltimore County, 1000 Hilltop Circle, Baltimore MD, USA\relax                                                                                                                                                  \label{inst:0183}
\and GSFC - Goddard Space Flight Center, Code 698, 8800 Greenbelt Rd, 20771 MD Greenbelt, United States\relax                                                                                                                                                                                                    \label{inst:0184}
\and EURIX S.r.l., Corso Vittorio Emanuele II 61, 10128, Torino, Italy\relax                                                                                                                                                                                                                                     \label{inst:0186}
\and Harvard-Smithsonian Center for Astrophysics, 60 Garden St., MS 15, Cambridge, MA 02138, USA\relax                                                                                                                                                                                                           \label{inst:0187}
\and HE Space Operations BV for European Space Agency (ESA), Camino bajo del Castillo, s/n, Urbanizacion Villafranca del Castillo, Villanueva de la Ca\~{n}ada, 28692 Madrid, Spain\relax                                                                                                                        \label{inst:0189}
\and CAUP - Centro de Astrofisica da Universidade do Porto, Rua das Estrelas, Porto, Portugal\relax                                                                                                                                                                                                              \label{inst:0190}
\and SISSA - Scuola Internazionale Superiore di Studi Avanzati, via Bonomea 265, 34136 Trieste, Italy\relax                                                                                                                                                                                                      \label{inst:0195}
\and Telespazio for CNES Centre Spatial de Toulouse, 18 avenue Edouard Belin, 31401 Toulouse Cedex 9, France\relax                                                                                                                                                                                               \label{inst:0198}
\and University of Turin, Department of Computer Sciences, Corso Svizzera 185, 10149 Torino, Italy\relax                                                                                                                                                                                                         \label{inst:0203}
\and Dpto. de Matem\'{a}tica Aplicada y Ciencias de la Computaci\'{o}n, Univ. de Cantabria, ETS Ingenieros de Caminos, Canales y Puertos, Avda. de los Castros s/n, 39005 Santander, Spain\relax                                                                                                                 \label{inst:0205}
\and Vera C Rubin Observatory,  950 N. Cherry Avenue, Tucson, AZ 85719, USA\relax                                                                                                                                                                                                                                \label{inst:0217}
\and Centre for Astrophysics Research, University of Hertfordshire, College Lane, AL10 9AB, Hatfield, United Kingdom\relax                                                                                                                                                                                       \label{inst:0218}
\and University of Antwerp, Onderzoeksgroep Toegepaste Wiskunde, Middelheimlaan 1, 2020 Antwerp, Belgium\relax                                                                                                                                                                                                   \label{inst:0227}
\and INAF - Osservatorio Astronomico d'Abruzzo, Via Mentore Maggini, 64100 Teramo, Italy\relax                                                                                                                                                                                                                   \label{inst:0230}
\and Instituto de Astronomia, Geof\`{i}sica e Ci\^{e}ncias Atmosf\'{e}ricas, Universidade de S\~{a}o Paulo, Rua do Mat\~{a}o, 1226, Cidade Universitaria, 05508-900 S\~{a}o Paulo, SP, Brazil\relax                                                                                                              \label{inst:0233}
\and M\'{e}socentre de calcul de Franche-Comt\'{e}, Universit\'{e} de Franche-Comt\'{e}, 16 route de Gray, 25030 Besan\c{c}on Cedex, France\relax                                                                                                                                                                \label{inst:0242}
\and SRON, Netherlands Institute for Space Research, Sorbonnelaan 2, 3584CA, Utrecht, The Netherlands\relax                                                                                                                                                                                                      \label{inst:0246}
\and RHEA for European Space Agency (ESA), Camino bajo del Castillo, s/n, Urbanizacion Villafranca del Castillo, Villanueva de la Ca\~{n}ada, 28692 Madrid, Spain\relax                                                                                                                                          \label{inst:0250}
\and ATOS for CNES Centre Spatial de Toulouse, 18 avenue Edouard Belin, 31401 Toulouse Cedex 9, France\relax                                                                                                                                                                                                     \label{inst:0251}
\and School of Physics and Astronomy, Tel Aviv University, Tel Aviv 6997801, Israel\relax                                                                                                                                                                                                                        \label{inst:0254}
\and Astrophysics Research Centre, School of Mathematics and Physics, Queen's University Belfast, Belfast BT7 1NN, UK\relax                                                                                                                                                                                      \label{inst:0256}
\and Centre de Donn\'{e}es Astronomique de Strasbourg, Strasbourg, France\relax                                                                                                                                                                                                                                  \label{inst:0258}
\and Universit\'{e} C\^{o}te d'Azur, Observatoire de la C\^{o}te d'Azur, CNRS, Laboratoire G\'{e}oazur, Bd de l'Observatoire, CS 34229, 06304 Nice Cedex 4, France\relax                                                                                                                                         \label{inst:0259}
\and Max-Planck-Institut f\"{ u}r Astrophysik, Karl-Schwarzschild-Stra{\ss}e 1, 85748 Garching, Germany\relax                                                                                                                                                                                                    \label{inst:0262}
\and APAVE SUDEUROPE SAS for CNES Centre Spatial de Toulouse, 18 avenue Edouard Belin, 31401 Toulouse Cedex 9, France\relax                                                                                                                                                                                      \label{inst:0264}
\and \'{A}rea de Lenguajes y Sistemas Inform\'{a}ticos, Universidad Pablo de Olavide, Ctra. de Utrera, km 1. 41013, Sevilla, Spain\relax                                                                                                                                                                         \label{inst:0268}
\and Onboard Space Systems, Lule\aa{} University of Technology, Box 848, S-981 28 Kiruna, Sweden\relax                                                                                                                                                                                                           \label{inst:0281}
\and TRUMPF Photonic Components GmbH, Lise-Meitner-Stra{\ss}e 13,  89081 Ulm, Germany\relax                                                                                                                                                                                                                      \label{inst:0286}
\and IAC - Instituto de Astrofisica de Canarias, Via L\'{a}ctea s/n, 38200 La Laguna S.C., Tenerife, Spain\relax                                                                                                                                                                                                 \label{inst:0288}
\and Department of Astrophysics, University of La Laguna, Via L\'{a}ctea s/n, 38200 La Laguna S.C., Tenerife, Spain\relax                                                                                                                                                                                        \label{inst:0289}
\and Laboratoire Univers et Particules de Montpellier, CNRS Universit\'{e} Montpellier, Place Eug\`{e}ne Bataillon, CC72, 34095 Montpellier Cedex 05, France\relax                                                                                                                                               \label{inst:0295}
\and LESIA, Observatoire de Paris, Universit\'{e} PSL, CNRS, Sorbonne Universit\'{e}, Universit\'{e} de Paris, 5 Place Jules Janssen, 92190 Meudon, France\relax                                                                                                                                                 \label{inst:0301}
\and Villanova University, Department of Astrophysics and Planetary Science, 800 E Lancaster Avenue, Villanova PA 19085, USA\relax                                                                                                                                                                               \label{inst:0302}
\and Astronomical Observatory, University of Warsaw,  Al. Ujazdowskie 4, 00-478 Warszawa, Poland\relax                                                                                                                                                                                                           \label{inst:0307}
\and Laboratoire d'astrophysique de Bordeaux, Univ. Bordeaux, CNRS, B18N, all\'{e}e Geoffroy Saint-Hilaire, 33615 Pessac, France\relax                                                                                                                                                                           \label{inst:0311}
\and Universit\'{e} Rennes, CNRS, IPR (Institut de Physique de Rennes) - UMR 6251, 35000 Rennes, France\relax                                                                                                                                                                                                    \label{inst:0314}
\and INAF - Osservatorio Astronomico di Capodimonte, Via Moiariello 16, 80131, Napoli, Italy\relax                                                                                                                                                                                                               \label{inst:0316}
\and Niels Bohr Institute, University of Copenhagen, Juliane Maries Vej 30, 2100 Copenhagen {\O}, Denmark\relax                                                                                                                                                                                                  \label{inst:0322}
\and Las Cumbres Observatory, 6740 Cortona Drive Suite 102, Goleta, CA 93117, USA\relax                                                                                                                                                                                                                          \label{inst:0323}
\and Astrophysics Research Institute, Liverpool John Moores University, 146 Brownlow Hill, Liverpool L3 5RF, United Kingdom\relax                                                                                                                                                                                \label{inst:0329}
\and IPAC, Mail Code 100-22, California Institute of Technology, 1200 E. California Blvd., Pasadena, CA 91125, USA\relax                                                                                                                                                                                         \label{inst:0334}
\and Jet Propulsion Laboratory, California Institute of Technology, 4800 Oak Grove Drive, M/S 169-327, Pasadena, CA 91109, USA\relax                                                                                                                                                                             \label{inst:0335}
\vfill
\and IRAP, Universit\'{e} de Toulouse, CNRS, UPS, CNES, 9 Av. colonel Roche, BP 44346, 31028 Toulouse Cedex 4, France\relax                                                                                                                                                                                      \label{inst:0336}
\and Konkoly Observatory, Research Centre for Astronomy and Earth Sciences, MTA Centre of Excellence, Konkoly Thege Mikl\'{o}s \'{u}t 15-17, 1121 Budapest, Hungary\relax                                                                                                                                        \label{inst:0349}
\and MTA CSFK Lend\"{ u}let Near-Field Cosmology Research Group\relax                                                                                                                                                                                                                                            \label{inst:0350}
\and ELTE E\"{ o}tv\"{ o}s Lor\'{a}nd University, Institute of Physics, 1117, P\'{a}zm\'{a}ny P\'{e}ter s\'{e}t\'{a}ny 1A, Budapest, Hungary\relax                                                                                                                                                               \label{inst:0351}
\and Ru{\dj}er Bo\v{s}kovi\'{c} Institute, Bijeni\v{c}ka cesta 54, 10000 Zagreb, Croatia\relax                                                                                                                                                                                                                   \label{inst:0368}
\and Institute of Theoretical Physics, Faculty of Mathematics and Physics, Charles University in Prague, Czech Republic\relax                                                                                                                                                                                    \label{inst:0372}
\and INAF - Osservatorio Astronomico di Brera, via E. Bianchi 46, 23807 Merate (LC), Italy\relax                                                                                                                                                                                                                 \label{inst:0383}
\and AKKA for CNES Centre Spatial de Toulouse, 18 avenue Edouard Belin, 31401 Toulouse Cedex 9, France\relax                                                                                                                                                                                                     \label{inst:0384}
\and Departmento de F\'{i}sica de la Tierra y Astrof\'{i}sica, Universidad Complutense de Madrid, 28040 Madrid, Spain\relax                                                                                                                                                                                      \label{inst:0388}
\and Vitrociset Belgium for European Space Agency (ESA), Camino bajo del Castillo, s/n, Urbanizacion Villafranca del Castillo, Villanueva de la Ca\~{n}ada, 28692 Madrid, Spain\relax                                                                                                                            \label{inst:0393}
\and Department of Astrophysical Sciences, 4 Ivy Lane, Princeton University, Princeton NJ 08544, USA\relax                                                                                                                                                                                                       \label{inst:0423}
\and Departamento de Astrof\'{i}sica, Centro de Astrobiolog\'{i}a (CSIC-INTA), ESA-ESAC. Camino Bajo del Castillo s/n. 28692 Villanueva de la Ca\~{n}ada, Madrid, Spain\relax                                                                                                                                    \label{inst:0429}
\and naXys, University of Namur, Rempart de la Vierge, 5000 Namur, Belgium\relax                                                                                                                                                                                                                                 \label{inst:0432}
\and EPFL - Ecole Polytechnique f\'{e}d\'{e}rale de Lausanne, Institute of Mathematics, Station 8 EPFL SB MATH SDS, Lausanne, Switzerland\relax                                                                                                                                                                  \label{inst:0440}
\and H H Wills Physics Laboratory, University of Bristol, Tyndall Avenue, Bristol BS8 1TL, United Kingdom\relax                                                                                                                                                                                                  \label{inst:0446}
\and Sorbonne Universit\'{e}, CNRS, UMR7095, Institut d'Astrophysique de Paris, 98bis bd. Arago, 75014 Paris, France\relax                                                                                                                                                                                       \label{inst:0467}
\and Porter School of the Environment and Earth Sciences, Tel Aviv University, Tel Aviv 6997801, Israel\relax                                                                                                                                                                                                    \label{inst:0468}
\and Laboratoire Univers et Particules de Montpellier, Universit\'{e} Montpellier, Place Eug\`{e}ne Bataillon, CC72, 34095 Montpellier Cedex 05, France\relax                                                                                                                                                    \label{inst:0469}
\and Faculty of Mathematics and Physics, University of Ljubljana, Jadranska ulica 19, 1000 Ljubljana, Slovenia\relax                                                                                                                                                                                             \label{inst:0470}
}


   \date{Received September 15, 1996; accepted March 16, 1997}

 
  \abstract
   {}
   {We aim to demonstrate the scientific potential of the \Gaia Early Data Release 3 (EDR3) for the study of different aspects of the Milky Way structure and evolution and we provide, at the same time, a description of several practical aspects of the data and examples of their usage.}
   {We used astrometric positions, proper motions, parallaxes, and photometry from EDR3 to select different populations and components and to calculate the distances and velocities in the direction of the anticentre. In this direction, the \Gaia astrometric data alone enable the calculation of the vertical and azimuthal velocities; also, the extinction is relatively low compared to other directions in the Galactic plane. We then explore the {disturbances} of the current disc, the spatial and kinematical distributions of early accreted versus in situ stars, the structures in the outer parts of the disc, and the orbits of open clusters Berkeley~29 and Saurer~1. }
   {With the improved astrometry and photometry of EDR3, we find that: i) the dynamics of the Galactic disc are very complex with oscillations in the median rotation and vertical velocities as a function of radius, vertical asymmetries, and new correlations, including a bimodality with disc stars with large angular momentum moving vertically upwards from below the plane, and disc stars with slightly lower angular momentum moving preferentially downwards; ii) we resolve the kinematic substructure (diagonal ridges) in the outer parts of the disc for the first time; iii) {the red sequence that has been associated with} the proto-Galactic disc that was present at the time of the merger with Gaia-Enceladus-Sausage is currently radially concentrated up to around 14 kpc, while  {the blue sequence that has been associated with} debris of the satellite extends beyond that; iv) there are density structures in the outer disc, both above and below the plane, most probably related to Monoceros, the Anticentre Stream, and TriAnd, for which the \Gaia data allow an exhaustive selection of candidate member stars and dynamical study; and v) the open clusters Berkeley~29 and Saurer~1, despite being located at large distances from the Galactic centre, are on nearly circular disc-like orbits.}
   {Even with our simple preliminary exploration of the \Gaia EDR3, we demonstrate how, once again, these data from the European Space Agency are crucial for our understanding of the different pieces of our Galaxy and their connection to its global structure and history.}

   \keywords{Galaxy: disc -- Galaxy: halo -- open clusters and associations: individual --  Galaxy: formation  -- Galaxy: kinematics and dynamics -- Stars: distances} 
   \maketitle

   \titlerunning{EDR3 anticentre}
   \authorrunning{Gaia Collaboration}
%
\section{Introduction}\label{sect_introduction}

As for previous releases, the Early Third Data Release  (EDR3, \citealt{Brown2020}) of the \Gaia mission \citep{Prusti2016} of the European Space Agency is accompanied with performance verification articles showing the quality of the data, the improvements with respect to previous releases, and the scientific potential for multiple research areas in astrophysics \citep[see also][]{Luri2020,Smart2020,Klioner2020}. 
In the present study we focus on a specific area in the sky that allows us to explore different elements of the Milky Way's {(MW)} structure and history: the Galactic anticentre. This region of the Galaxy has the advantage that from astrometric measurements alone (proper motions and parallaxes), one can calculate the vertical and azimuthal (rotation) motion of the stars with a negligible contribution of the 
 line-of-sight velocity. Also, the anticentre has relatively low extinction compared to other directions of the Galactic disc.
 
 More importantly, the anticentre is a meeting point of several distinct components of the Galaxy (the disc, the halo) and possibly hosts ancient and recently disrupted stellar systems of extragalactic origin. The anticentre is also an excellent window to the dynamics and the past of the {MW} {since,} due to the lower gravitational potential, any perturbation on the disc would cause more significant deformations than in the inner disc, and, due to the longer dynamical timescales, these could still be observable today \citep[e.g.][]{Bible08,Antoja2018,Laporte2019}.
 
In this paper we focus on several aspects of the Galaxy that coexist in the anticentre and that will help us towards answering a single question: {how the Galaxy appears today and how it became like this.}  Thanks to a combination of models and measurements, in which \Gaia DR2 \citep{Brown2018} played one of the most relevant roles, we have already uncovered part of the  {MW} structure and history. The major accretion event of the so-called Gaia-Enceladus-Sausage around 10 Gyr ago \citep{Helmi2018,Belokurov2018}, together with the ongoing {accretion} of the Sagittarius dwarf galaxy (\citealt{Mateo1996,Majewski2003} and a recent detection with \Gaia data in \citealt{Antoja2020}), and internal structures such as the bar (\citealt{deVaucouleurs1964}; {\citealt{Binney1991};} \citealt{Blitz1991,Weiland1994}) and the spiral arms \citep{Drimmel2001,Reid2009} are among the most important phenomena that have shaped our Galaxy throughout its evolution \citep[see also][]{Bland-Hawthorn2016}. The footprints of these phenomena can still be observed today and that is what we investigate here.

First, we {look into the kinematic disturbances of the disc }
 {that EDR3 allows us to inspect } in its outermost parts {with more detail}. {Already known disturbances include} vertical asymmetries in the number counts linked to vertical bending and breathing waves \citep[e.g.][]{Widrow2012,Bennett2019}, {substructure in the in-plane velocities} (e.g. \citealt{Dehnen1998,Famaey2005,Antoja2008,Antoja2018,Kawata2018,Khanna2019}; {\citealt{fragkoudi19}}) large scale velocity patterns in the disc \citep{Siebert2011,Williams2013,Carlin2013,Antoja2017} and other phase space correlations (e.g. \citealt{Schonrich2018,Friske2019}; {\citealt{Huang2018,Cheng2020}}). {While some of these could be related to the structures such as the bar, the poorly-known spiral arms and the warp,} more recently, 
 the \Gaia vertical phase spiral \citep{Antoja2018,Bland-Hawthorn2019} suggests a phase mixing event after the perturbation of Sagittarius \citep{Binney2018,Laporte2019}.  All of these have proven to be difficult to understand, and also to disentangle or relate. 
Here we look at the rotation and vertical velocities of the outer disc, {showing the power that \Gaia data can have in our understanding of this complexity and the role that recent and past, internal and external, perturbations have had in the  {MW}.}

Second, we go from the current disc of the Galaxy to {more} ancient {components}. Two distinct populations were clearly apparent in the Hertzsprung-Russell (HR) diagram of stars with large tangential velocities near the Sun (i.e. representing the {stellar} halo) by \citet{Babusiaux2018} using \gdrtwo: a blue and a red sequence. It has been {proposed} that {these} two  {populations correspond to} an accreted one stemming largely from the galaxy Gaia-Enceladus-Sausage, and an in situ heated (thick) disc, {different from the canonical thick disc,} that was present at the time of this merger \citep{Helmi2018,Belokurov2018,DiMatteo2019,Gallart2019,Belokurov2020}.  Here we {analyse the spatial distribution and kinematics of stars from each of the HR sequences to} 
investigate out to which distance the debris of Gaia-Enceladus-Sausage may be found, and constrain the extent of the {suggested} proto-Galactic disc present at that time.

Thirdly, we explore   
{the density structures that appear in the edge of the disc in the anticentre direction.}
\citet{Newberg2002} {using} the deep Sloan Digital Sky Survey \citep[SDSS][]{York2000} discovered the existence of a $\sim$100$^\circ$ structure in their A and F star count maps. Now known as Monoceros, later studies have confirmed its existence and large extension on the sky \citep[e.g.][]{Crane2003,Yanny2003,Ibata2003}. Together with the Anticentre stream (ACS, \citealt{Grillmair2006}), both at a  distance $\sim$10\,kpc from the Sun, 
and the Triangulum-Andromeda overdensities (TriAnd, \citealt{Majewski2004,Rocha-Pinto2004,Martin2007}), they are part of a complex and substructured outer disc. The initial interpretation that these could be the remains of an accreted dwarf galaxy \citep[e.g,][]{Martin2004,Penarrubia2005} has become less plausible (although not completely ruled out) since: i) there is no known progenitor to the stream  \citep[the candidate Canis Major has been discarded --][]{Momany2004,Momany2006,Rocha-Pinto2006}, {and} ii) the kinematics of the structures {\citep[e.g.][]{deBoer2018}, their metallicities and their ratio of RR Lyrae to Giants \citep{Li2012,Sheffield2014,Price-Whelan2015,Sheffield2018,Bergemann2018,Laporte2020} are compatible with that of the disc}. Here we explore how these structures look in \Gaia EDR3 and coexist with other structures such as the Sagittarius stream.

Finally, we explore the open clusters Berkeley~29 \citep{Kaluzny94,Lata02,Tosi04,Bragaglia06} and Saurer~1 \citep{Frinchaboy02,Carraro03}  in the anticentre direction that, with ages of several Gyr, are among the oldest Galactic clusters known. Their unusual location at Galactocentric distances of $\sim$20\,kpc and more than 1\,kpc above the Galactic mid-plane is a puzzle that has led several authors to question whether they are associated with the disc, and to propose a possible extragalactic origin \citep[e.g.][]{Frinchaboy04}. Attempts to characterise the orbits of these two objects have returned widely discrepant results \citep{Carraro07,VandePutte10}, mainly due to their poorly-constrained proper motions since at such large distances, small proper motion errors translate into large uncertainties in physical velocities. An additional issue has been the uncertain membership status of individual stars. Here, thanks to \Gaia EDR3, we perform a robust analysis of the membership of these clusters and derive their orbits with high confidence.

To investigate all these aspects, the main \Gaia data products that we use here are the astrometric measurements. For EDR3 these show a substantial decrease of uncertainties resulting from the use of 34 months of data (12 more than for DR2). Apart from a higher completeness at the faint end, there is a significantly larger number of stars at a given parallax precision. The combination of all these improvements essentially means that we can now explore distant regions of the Galaxy in the direction of the anticentre, even reaching around 16 kpc from the Galactic centre and beyond (see Sect.~\ref{sect_data} for details), and thus, the very outskirts of the disc, for a sample with positions and velocities of excellent quality. Moreover, important improvements in the pipelines of the \Gaia photometry have resulted in photometric bands with significantly less systematic error, from which, combined with the improved parallaxes, cleaner  HR diagrams can be built and used to select different populations and components. 

The paper illustrates how, once more, the new \Gaia data {are} set to revolutionise our knowledge of the Galaxy and its past. Additionally, we describe practical aspects of the data and examples of its {use} that might be of interest for the community, such as queries in the \Gaia Archive, quality cuts, derivation of distances (Bayesian inference, considerations on the parallax zero point), etc. We also complement our analysis with the use of simulated data from the Gaia Object Generator (GOG, \citealt{Luri2014}) \& the Gaia Universe Model Snapshot (GUMS, \citealt{Robin2012}), now available directly in the Gaia Archive, to evaluate the effects of selection, errors and extinction.

We start our paper by explaining the different datasets used and demonstrating the different improvements (but also limitations) of the EDR3 data in the anticentre direction (Sect.~\ref{sect_data}). We continue by explaining how the distances and phase space coordinates are derived (Sect.~\ref{sect_phasespace}). 
The results sections {follow,} organised {into} the explorations of the disc dynamics (Sect.~\ref{sect_disc}), halo, thick disc and outer disc structures (Sect.~\ref{sect_haloouter}), and distant open clusters (Sect.~\ref{sect_clusters}). We present our discussion and conclude in Sect.~\ref{sect_conclusions}.

\section{Data}\label{sect_data}


\subsection{Main datasets}\label{sect_selection}

\begin{figure}
\centering
\includegraphics[width=\hsize]{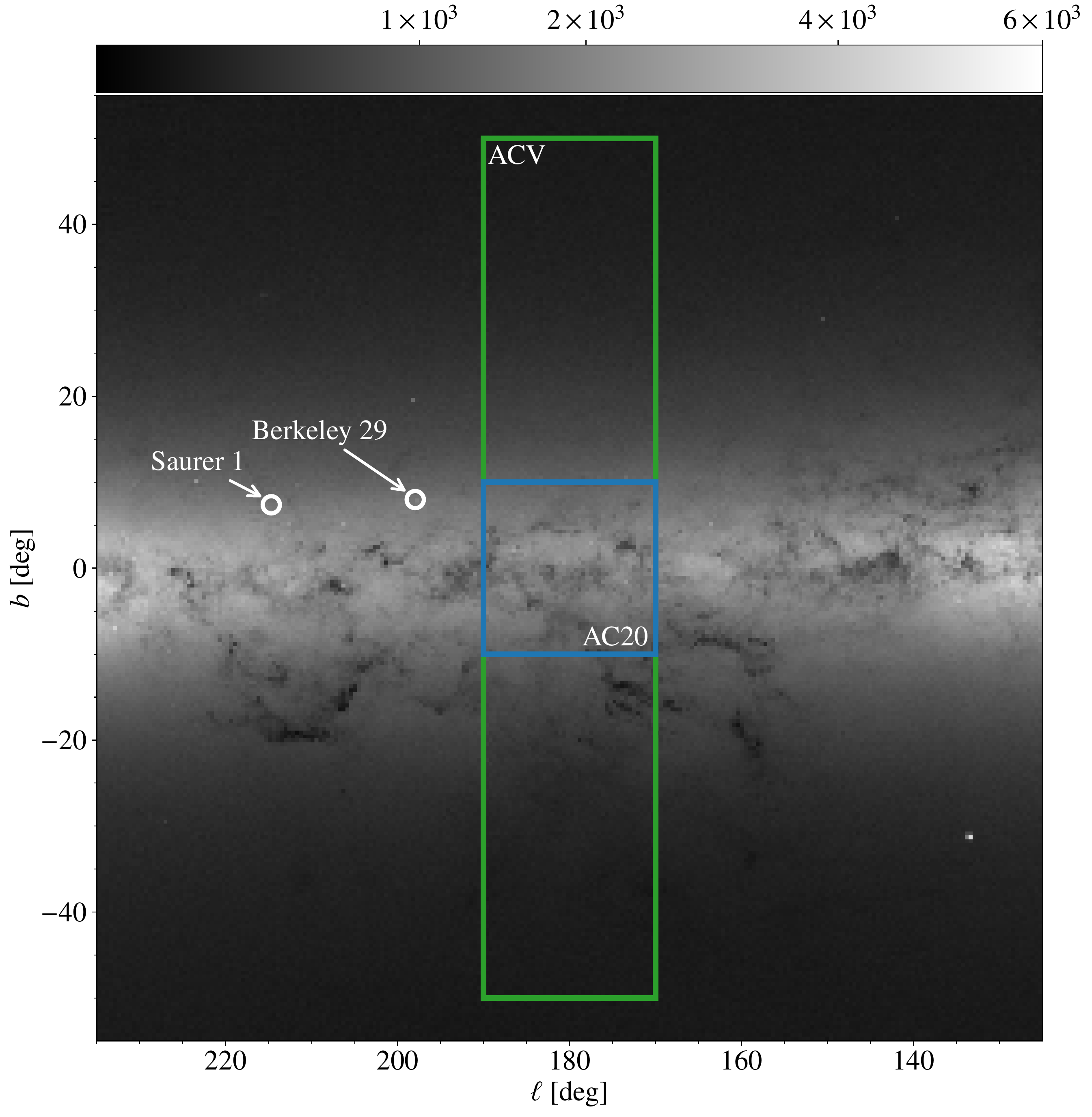}
	\caption{EDR3 star counts in the anticentre region with the different data selections used. The HEALpix map is obtained by querying the \Gaia archive the counts of stars within each HEALpix of level 8 (query \ref{q2} in Appendix~\ref{queries}). The size of the circles to indicate the position of the clusters does not correspond to the size used for the selection which is much smaller. Several other clusters can be seen in the figure, and also the Triangulum Galaxy (M33, bottom right corner).}
\label{fig_AC}
\end{figure}
 
  \begin{table}
\caption{Number of stars in the different samples and comparison with DR2. The numbers are given for the different data {samples} described in Sect.~\ref{sect_selection} and different sub-selections. {The numbers in the first two numerical columns are for samples without the {\tt excess\_flux} and {\tt RUWE} selections since these are not equally defined in the different releases.} \tablefoottext{a}{See footnote \ref{56p}.}}             
\label{tab_data}      
\centering          
\begin{tabular}{l r r r}\hline\hline       
                                      & DR2& \multicolumn{1}{c}{EDR3}& \multicolumn{1}{c}{{EDR3+filters}}\\  \hline                    
1. AC20                                                  &   $13\,307\,312$  &  $14\,120\,029$& {${11\,949\,642}$}\\
\hspace{0.4cm}   5p-6p\tablefootmark{a}  &    $10\,750\,864$  &   $12\,279\,076$& {${11\,949\,642}$}\\
\hspace{0.4cm}  $\piepi>3$                     &  $2\,645\,014$  &  $3\,518\,388$& {${3\,369\,456}$}\\  
\hspace{0.4cm} photometry                    & $12\,618\,364$&   ${13\,706\,954}$ & {${11\,436\,625}$ }\\  
\hline   
2. ACV       &  $24\,578\,296$   & {${25\,835\,286}$} & {${21\,835\,927}$}\\  
\hspace{0.4cm}  $\varpi < 0.1$\,mas      & $4\,974\,104$   &  {${4\,879\,087}$} & {${4\,509\,263}$}\\  
\hspace{0.4cm}  $\varpi < 0$\,mas      &  $3\,945\,985$   &  {${3\,781\,306}$}& {${3\,496\,645}$} \\  
\hline   
3. {Clusters} &  & {} & {} \\
\hspace{0.4cm} Berkeley 29     & 365   & {370} & {334} \\  
\hspace{0.4cm} Saurer 1     &   283  & {284} & {263} \\  
\hline
\hline                  
\end{tabular}
\end{table}

In this work we explore the Galactic anticentre region using different data selections obtained from  \Gaia EDR3 \citep{Brown2020} that can be accessed through the \Gaia Archive (\url{https://gea.esac.esa.int/archive/}). More details on the data and validation are given also in \citet{Lindegren2020b}, \citet{Riello2020} and \citet{Fabricius2020}. The main datasets are shown in Fig.~\ref{fig_AC} and listed below. The number of stars for these samples {and a comparison with DR2} are in Table~\ref{tab_data}.

\begin{enumerate}[label=(\arabic*)]
	\item AC20: A square {on} the sky centred at $(\ell,b)=(180,0)\,\deg$ of {20$\,\deg$ on a side} (blue square in Fig.~\ref{fig_AC}). This sample is used to explore Galactic disc kinematics in Sect.~\ref{sect_disc}. It contains ${14\,120\,029}$  stars but most of the time we use only the selection with $\piepi>3$ (see Sect.~\ref{sect_distances}), which comprises  ${3\,518\,388}$ sources (\ACf). The data {were} retrieved from the archive using the query \ref{q1} in Appendix~\ref{queries}. Similar queries were used for other samples.
	\item ACV: A {rectangle on} the sky centred at $(\ell,b)=(180,0)\,\deg$ with a width of $20\deg$ in $\ell$ and height $100\deg$ in $b$ (green rectangle in Fig~\ref{fig_AC}). This sample is used to explore the {halo and the} structures in the outer disc such as Monoceros or the Sagittarius stream in {Sect.~\ref{sect_haloouter}}. For parts of our analysis, we performed a selection of $\varpi<0.1$ mas to favour distant stars. We note that with this selection there are {2\%} fewer stars  in EDR3 than in DR2 (Table~\ref{tab_data}). While the total number of stars in that region has increased 	with respect to DR2, many of the stars added are nearby faint dwarfs (see Sect.~\ref{sect_quality}) and the overall quality of the parallaxes has improved significantly as proven by the decrease in the number of sources with a negative parallax (and spurious sources). As a consequence, our parallax cut is now able to reject the nearby sources more efficiently, thus resulting in a slightly smaller sample.
	\item Two clusters in the anticentre. All sources brighter than $G=19$ within 4 arcmin of the centres of the extremely distant Galactic old open clusters Berkeley~29 and Saurer~1. These data are analysed in Sect.~\ref{sect_clusters}.
\end{enumerate}

In all our analysis, our fundamental observables are the astrometric quantities $\varpi$, $\mu_\alpha*$, $\mu_\delta$ (parallax and proper motions) and the photometric bands $G$, $BP$, $RP$. In order to use the best quality data, we applied several selections. First we applied the following astrometric quality selection on the Renormalised Unit Weight Error ({\tt RUWE}) as recommended in \citet{Lindegren2020}: 

\begin{equation}\label{eq_astromcuts}
\mathtt{RUWE}<1.4.
\end{equation}
On the other hand, whenever the photometry was used we selected {good photometry} sources with:
\begin{multline}\label{eq_photcuts}
0.001+{0.039}(BP-RP) < \log_{10}\,\mathtt{excess\_flux}  \\
< 0.12+{0.039}(BP-RP).
\end{multline}
{Sources out of these limits have inconsistent $G$, $G_{\rm BP}$ and $G_{\rm RP}$ fluxes due to blends (more than one source in the BP and RP windows), contamination by a nearby source (outside the window) or a sign of the extended nature of the source.}
{Additionally, we corrected the fluxes in $G$ for 6p sources following \citet{Riello2020} -their Table 5- using directly an ADQL query as suggested in \citet{Brown2020}. The last column in Table~\ref{tab_data} indicates the number of stars after these selections.}

\subsection{Complementary datasets}\label{sect_selection}

For validation and other purposes, we also used the following complementary data:

\begin{enumerate}[label=(\roman*)]
	\item 6Dsample: a full sky sample with stars that have DR2 {line-of-sight} velocity in EDR3 \citep{Seabroke2020}, thus with full 6D phase space information. {After filtering, }this sample contains ${6\,156\,684}$ stars and is used  mainly in Sect.~\ref{sect_disc}.
	\item DR2: the same selections as above (AC20, ACV) but for DR2. These are used for comparison with EDR3.
        \item GOG {\& GUMS}: the same selections as above but for GOG (Luri et al. 2014; GEDR3 documentation Chapter 2) {which is a mock \Gaia catalogue based on the Besan\c{c}on model \citep{Robin2003}, and for GUMS } \citep{Robin2012} {which} contains the intrinsic properties of the  {mock} sources before applying the \Gaia instrument modelling. 
        Here we used the GOG version 20.0.3. with uncertainties that have been scaled to 34 months of data (but see Fig.~\ref{fig_astrometricerrors}). These samples are used for the evaluation of completeness and extinction effects and they do not contain any kinematic substructure or asymmetries. Furthermore, GOG and GUMS were used in Appendix~\ref{app_distances} for testing how robust each of the distance estimation methods is. These simulated data were retrieved through the Gaia Archive querying from the corresponding tables ({\tt gaiaedr3.gaia\_source\_simulation} and {\tt gaiaedr3.gaia\_universe\_model}).
	\item 2MASS: {We used the official crossmatch of EDR3 with the \twomass{} point source catalogue \citep{twomass2006}, provided in {\tt gaiaedr3.tmass\_psc\_xsc\_best\_neighbour.} For the AC20 sample,} this {yields} about {$55\%$} anticentre objects with \twomass{} photometry. 
		These data were used to select red clump (RC) stars and compute their photometric distances (Sect.~\ref{sect_populations} and \ref{sect_distances}, Appendix~\ref{app_RCsel} and \ref{app_RCdist}). 
\end{enumerate}

\subsection{EDR3 data quality and completeness}\label{sect_quality}

\noindent\textcolor{blue}{}

\begin{figure*}
\centering
\includegraphics[width=\hsize]{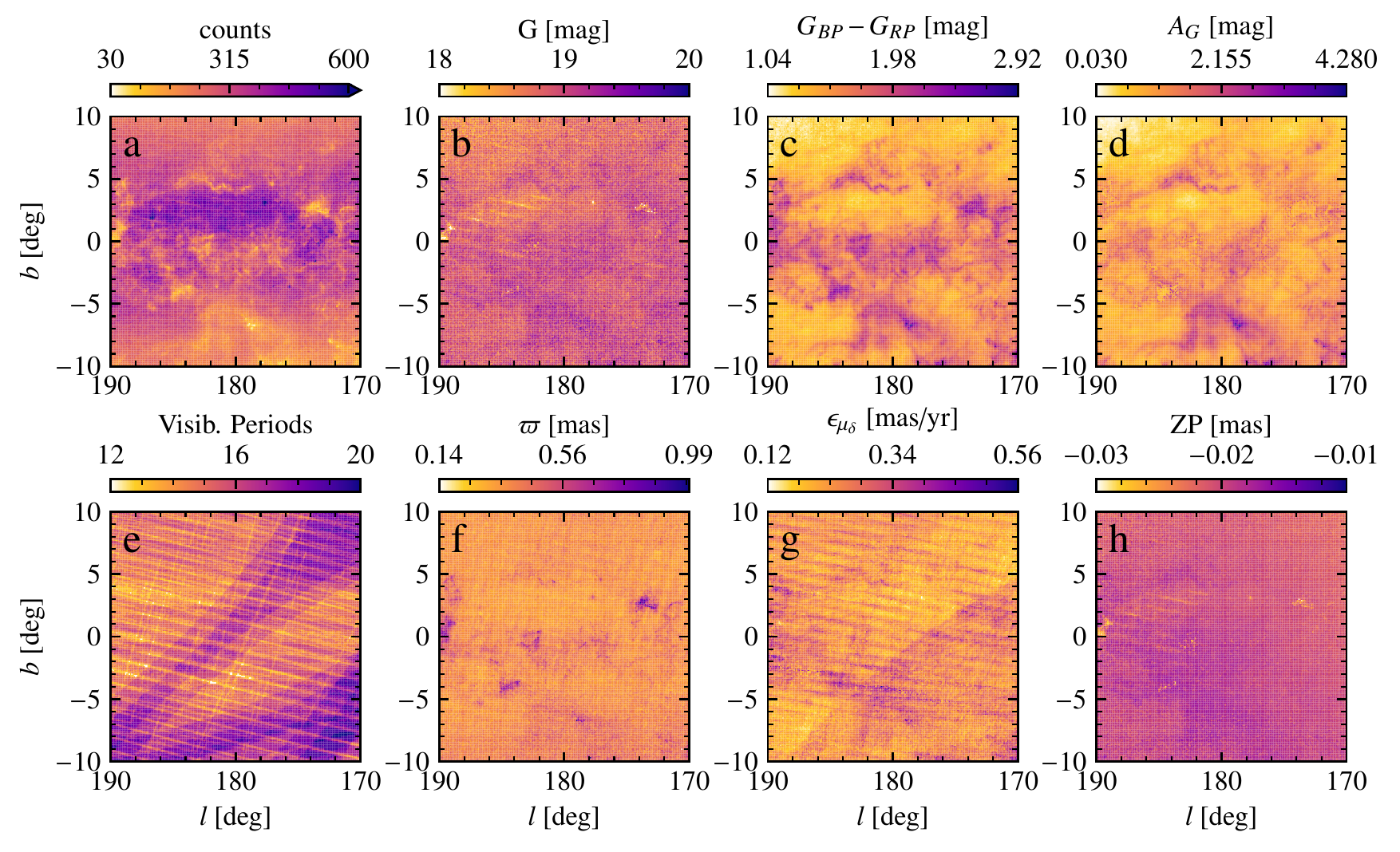}
	\caption{Characteristics of the anticentre AC20 sky. Panel (a) shows the number of sources and the rest of panels show median quantities for bins of $0.1\deg$ of the magnitude (b), colour (c), extinction in the $G$ band (d, {only for stars with $\varpi/\sigma_{\varpi}>3$}, see Sect.~\ref{sect_populations}), {\tt visibility\_periods\_used} (e), parallax (f), uncertainty in the proper motion in the $\delta$ direction (g), and zero point correction to the parallax $ZP_{56}$ (h, see Sect.~\ref{sect_zpt})}
\label{fig_sky_medians}
\end{figure*}

\begin{figure}
\centering
\includegraphics[width=0.8\hsize]{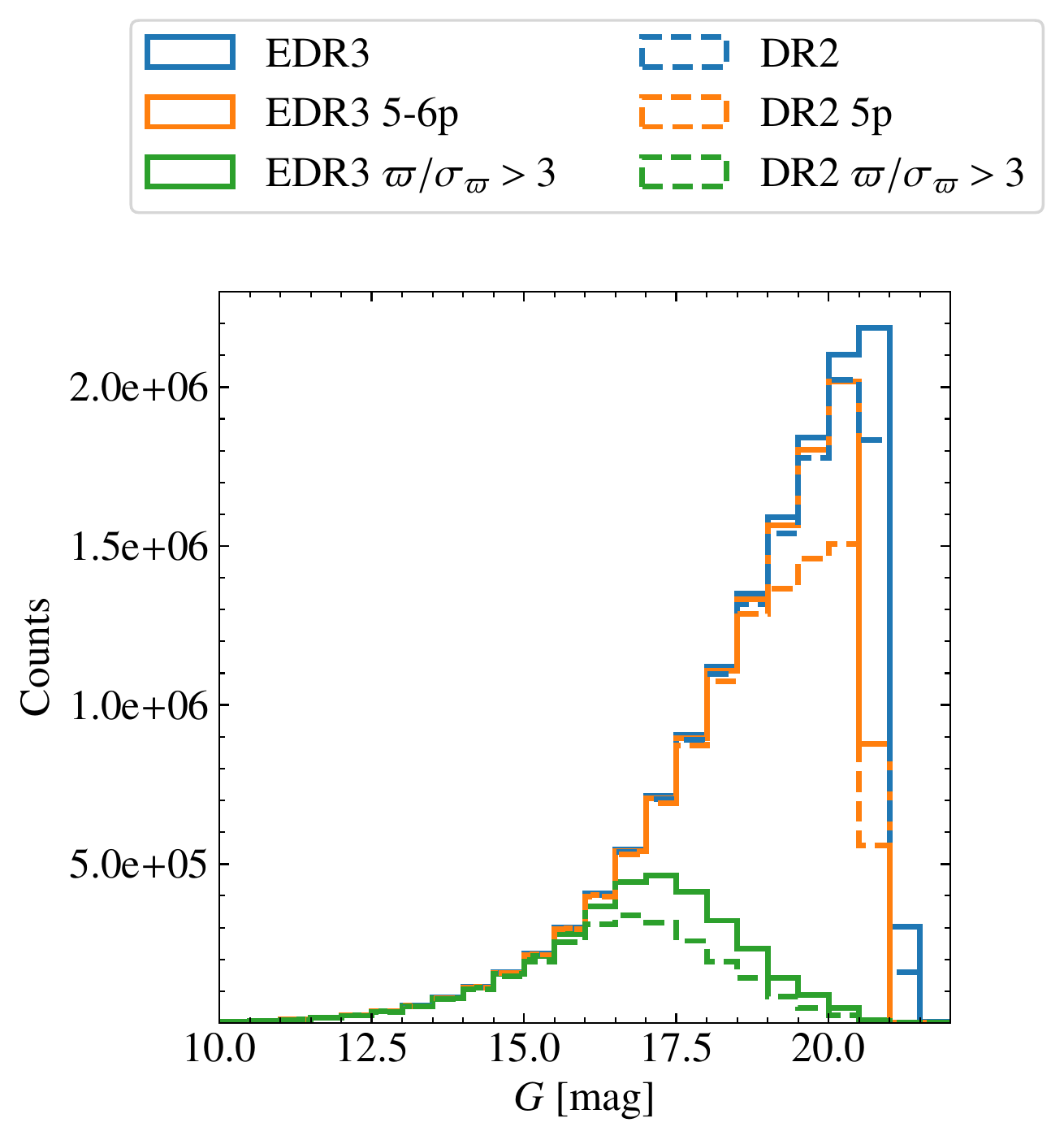}
\caption{Distribution of G magnitude in the anticentre. We show the number of stars in bins 0.5 mag for the AC20 case. The gain from DR2 to EDR3 is mostly at the faintest magnitudes where some sources did not have enough observations {to appear in} the past release.}
\label{fig_Gbins_1d}
\end{figure}

In this section we examine the quality of the EDR3 data and compare it to DR2. The most relevant improvements in EDR3 for our study include a larger number of sources at the faintest magnitudes {and} a significant decrease of the astrometric uncertainties and thus a significantly larger number of stars with a certain parallax precision. Below we show these aspects in more detail focusing {mostly} on the AC20 region as an example.

\paragraph{1. General description.} Figure~\ref{fig_sky_medians} shows the AC20 region in Galactic coordinates coloured according to different quantities in bins of $0.1\deg$. In panel a we show the number of stars per bin while the rest of the panels show median quantities. {As expected, the counts anti-correlate with the patterns seen in the extinction map {from \citet{Green2019}} (d, {see details in}  Sect.~\ref{sect_populations}) combined with the decrease with Galactic latitude $|b|$. The median magnitude (b) and median colour (c) also correlate with extinction (d): } higher extinction regions have, on average, more reddened sources that have fainter (more extinguished) apparent magnitudes. Panel d shows that there is higher extinction for $b<0$. Additionally, there is a horizontal elongated window at $b\sim2.5\deg$ of low extinction with far more counts and brighter magnitudes, which is seen also in other panels where brighter magnitudes essentially translate into smaller astrometric errors (e.g. panel g) and also smaller parallaxes (panel f, stars reaching farther distances). {Whether this feature with larger counts reflects more than simply lower extinction (e.g. a flexing of the disc) requires a deeper analysis of the extinction and the selection function. We also note that the thin nearly-horizontal lines in panels a and b are a consequence of the {\tt RUWE} selection.}  

\begin{figure}
\centering
\includegraphics[width=\hsize]{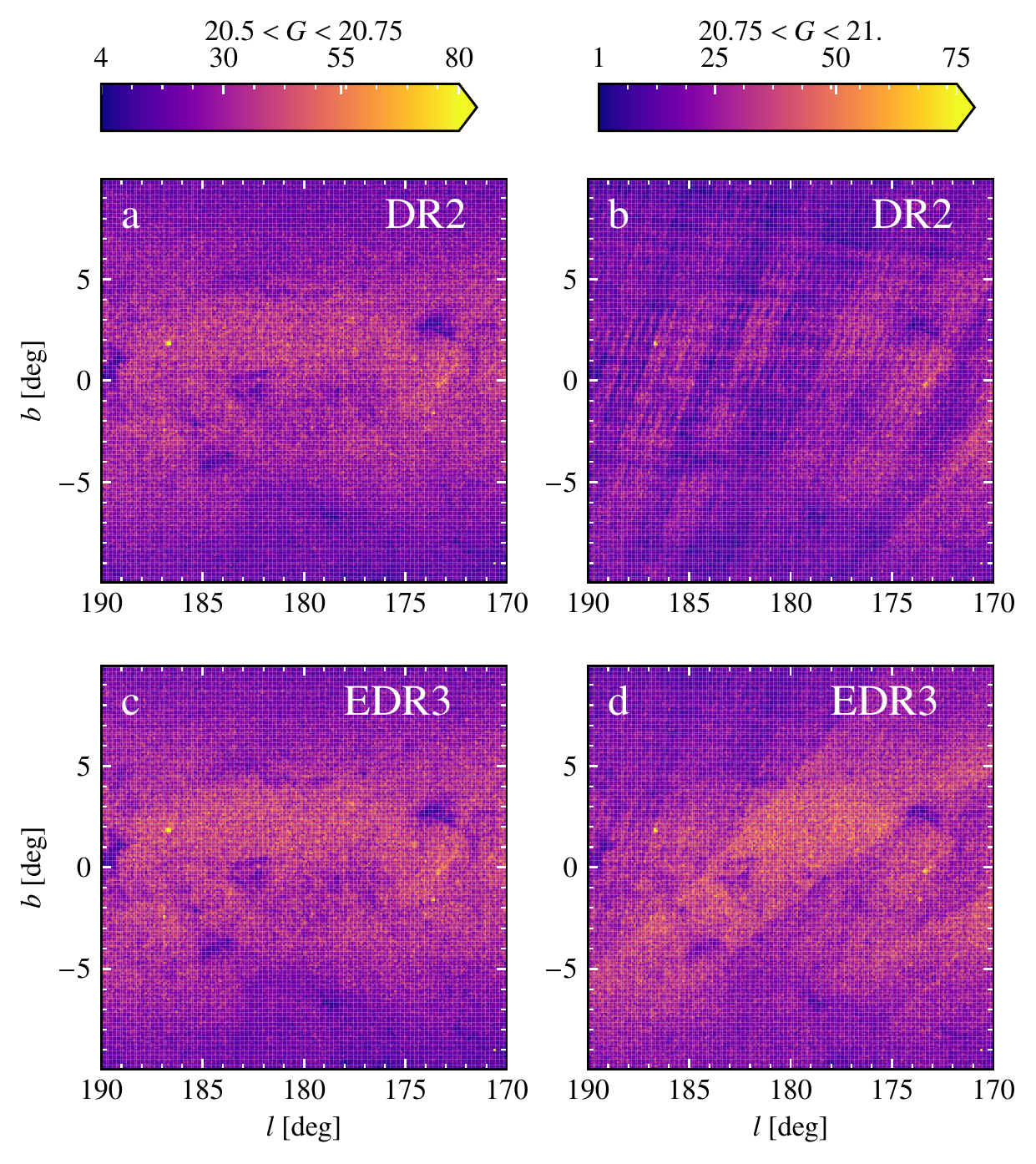}
\caption{DR2 and EDR3 counts for different magnitude ranges. The panels show the number of sources in bins of $0.1\deg$ in two different ranges of magnitude: $20.5<G<20.75$ (left) and $20.75<G<21.$ (right). To facilitate the comparison, the same colour bars has been used for each vertical pair of panels and the upper limit of the colour scale does not correspond to the maximum number of counts to avoid dominance of bins with clusters. An increase in the counts in EDR3 is observed, together with the decrease of some of the small scale patterns, although some bands remain in the faintest magnitude range.}
\label{fig_Gbins_2d}
\end{figure}

\paragraph{2. Completeness.} The evaluation of the completeness of the \Gaia data is a difficult task given that there is 
no deeper survey with a comparable resolution. Distinct methodologies to assess the data completeness can be found in \citet{Boubert2020b}, \citet{Lindegren2020b} and \citet{Fabricius2020}. Here we examine it in a simpler way. 
First we note that the AC20 sample (without any cuts) has about one million more stars in EDR3 compared to DR2 (see Table~\ref{tab_data}). Figure~\ref{fig_Gbins_1d} shows histograms of the $G$ magnitude for stars in the AC20 sample in DR2 and EDR3 (blue {solid and dashed} histograms) showing a great increase in the number of sources at the faint magnitudes with respect to DR2. This was expected given that the detection on board prioritises bright magnitudes and the effect of more months of observations produces new detections mostly at the faintest bins. 

Figure~\ref{fig_sky_medians}e shows the map of median {\tt visibility\_periods\_used}. This panel shows bands at different spatial scales that correspond to regions with higher and lower number of observations and thus higher and lower completeness, {respectively}. The thin, nearly-horizontal, yellowish pattern, 
separated by roughly 0.7$\deg$, similar to the width (across scan) of \Gaiaf's
FOV, corresponds to consecutive scans that did not overlap
in across scan.
The wider {purple} bands, indicating areas
where the coverage is better, are close to some `nodes'  in the scanning law, that is, the
positions in the sky that get repeated coverage during
some consecutive scans. 
 Figure~\ref{fig_Gbins_2d} shows the star counts for different ranges in $G$ for DR2 (top) and EDR3 (bottom) using the same colour scale. The bands of the scanning law appear clearly and correlate with Fig.~\ref{fig_sky_medians}e. Comparing to DR2, we can clearly notice the larger number of stars in EDR3 in these two magnitude ranges as well as the reduction of some of the bands (at scales of $\sim3\deg$) imprinted in DR2. 
 In the range of $20.75<G<21$ some scanning bands are still present. 
 
\begin{table*}
\caption{Indicative completeness of the kinematic samples. Absolute number of stars and fractions for all magnitudes and for distinct magnitude ranges are given for the cases with 2p and 5-6p solutions and for the selection of $\piepi>3$. To compute the percentages for DR2, the total number of sources in EDR3 for each case {has} been used. {These numbers are for samples without the {\tt excess\_flux} and {\tt RUWE} selections since these are not equally defined in the different releases and the selection in {\tt RUWE} eliminates the 2p solutions.}}             
\label{tab_data2}      
\centering          
\begin{tabular}{l |r r r| r r r}\hline\hline       
                                    &  \multicolumn{3}{c}{DR2}& \multicolumn{3}{|c}{EDR3}\\  \hline                    
\hspace{0.8cm}                          &  2p & 5p&    5p /ALL-EDR3      &     2p      &    (5p$\cup$6p) &  (5p$\cup$6p) /ALL-EDR3\\ 

\hspace{0.8cm}  $\forall G$      &   2\,5564\,48   & 10\,750\,864& 76\%&         1\,840\,953      &    12\,279\,076 &  87\%\\ 
\hspace{0.8cm}  $G<19$           &  111\,638   & 5\,860\,281& 96\%&        {67\,217}     &{6\,010\,199}     &   99\% \\ 
\hspace{0.8cm}  $19<G<20$     &   492\,849   & 2\,825\,129&82\% &       {60\,360}      & {3\,369\,371}    &  98\%\\ 
\hspace{0.8cm}   $20<G<20.7$ &  940\,298     &1944\,641 & 64\%&      {312\,553}      & {2\,705\,704}   &  90\%\\ \hline   
\hspace{0.8cm}                           &  $\piepi < 3$  &    $\piepi>3$   &  $\piepi>3$ /ALL-EDR3&    $\piepi<3$ &  $\piepi>3$ & $\piepi>3$ /ALL-EDR3 \\ 
\hspace{0.8cm}  $\forall G$     &  10\,662\,298      &2\,645\,014 & 19\%&   10\,601\,641      & 3\,518\,388   &  25\%\\  
\hspace{0.8cm}  $G<15$         &   30\,930    & 478\,565& 93\% &         {13\,098}       &  {500\,913}  & {97\%} \\  
\hspace{0.8cm} $15<G<17$  & 360\,149 & 1\,096\,109&74\% &        {170\,528}      &  {1\,301\,329}  &{88\%} \\  
\hspace{0.8cm} $17<G<19$  &  6\,284\,445  & 1\,039\,701&14\%  &     {5\,860\,892}     & {1\,660\,388}   & 22\% \\  
\hspace{0.8cm}  $19<G<21$   & 3\,826\,133     &  30\,637& 0.7\% &         {4\,232\,117}  & {55\,142}   & 1\% \\  

\hline   
\hline                  
\end{tabular}
\end{table*}

\paragraph{3. Completeness of the kinematic samples.} Some of the \Gaia sources have only partial astrometric solutions, from which only sky coordinates are derived (2p solutions) while others have full astrometric solution (positions, parallax and proper motions available) and are dubbed 
 5p and 6p solutions \citep{Lindegren2020b}{, where the 6th parameter is the colour}\footnote{\label{56p} 2p partial solutions (only positions) are indicated as {\tt astrometric\_params\_solved}=3 in the \Gaia Archive. 5p solutions are those for which the \Gaia colour is used in the astrometric solution, while in the 6p cases, this quantity, more precisely, the pseudocolour, is derived simultaneously in the solution \citep{Lindegren2020b}. The 5p and 6p solutions correspond to {\tt astrometric\_params\_solved}=31 and 95, respectively. In DR2 all full astrometric solutions were included under the  {\tt astrometric\_params\_solved}=31 case, {even if} in some cases a chromaticity different from the {photometric} colour was used.}.
 In the first rows of Table~\ref{tab_data2} we {give} the number of stars with partial (2p) and full (5p, 6p) solutions comparing DR2 and EDR3 for the whole AC20 sample and for different ranges of magnitude.  {In} EDR3, there are two million more stars with full astrometric solution {than in} DR2. The table also shows the percentage of full solutions relative to all sources in EDR3, which gives an indication of the internal completeness of the kinematic data. Most notably, in the range of  $19<G<20$ there is now a 98\% internal completeness {compared to} the 82\% in DR2, and in the range  $20<G<20.7$ the percentage is now 90\% versus the old 64\%, verifying that, as shown also with the orange {solid and dashed} histograms of Fig.~\ref{fig_Gbins_1d}, there is an outstanding gain at the faintest magnitudes. These stars have never been used before in kinematic studies with \Gaia data. 

We note that 6p solutions tend to be associated to fainter sources and their astrometric solutions are worse than for 5p ones. They have on average fewer {\tt visibility\_periods\_used} (i.e. less observations), and larger {\tt ipd\_frac\_multi\_peak} (i.e. relatively large probability of being a double source, either visual or real binary), having larger astrometric errors. While for the AC20 sample the fraction of 6p solutions is comparable to the 5p (42 and 45\%, respectively, the remaining 13\% being 2p), for the \AC case they represent only a 14\% (86\% being 5p) since we require good relative parallax errors.

After selecting stars with $\piepi>3$ (\AC sample) we find approximately one milion more stars in EDR3 than in DR2 (bottom rows of Table~\ref{tab_data2}), which represents an increase of 33\%. Figure~\ref{fig_Gbins_1d} (green {solid and dashed} histograms) shows an improvement of the completeness of the parallax quality selection at magnitudes fainter than $G=16$, which means better sampling {at} all distances and probing larger ones. Table~\ref{tab_data2} also shows that at the relatively bright magnitudes $15<G<17$, there were 74\% of stars in DR2 satisfying this condition but we have now 88\%. It is nevertheless important to remark that the completeness of the sample with good parallaxes is low even at intermediate magnitude ranges both for the DR2 and EDR3 (as low as 14\% and 22\% in the range $17<G<19$, respectively), although we see an overall improvement for the new release. 
 
\paragraph{4. Astrometric quality, systematics and parallax zero point.} The improvement in astrometric quality of EDR3 with respect to DR2 is discussed in \citet{Lindegren2020b} and {is reflected in smaller uncertainties and} a reduction of the number of negative parallaxes (e.g. for the ACV sample where there are 164679 less sources with negative parallaxes, Table~\ref{tab_data}). 
Figure~\ref{fig_astrometricerrors} in Appendix~\ref{app_material} illustrates the improvement in the uncertainties for the anticentre (similar to figure A.1 by \cite{Lindegren2020b} 
 for all EDR3). {There is} a reduction by a factor of 0.79 and 0.5 in parallax and proper motion uncertainties, respectively, as expected for the increase in the number of months of observations, {and} even a larger reduction for sources with $G<14$. This improvement in the astrometric quality allows us to have now a much larger sample of stars with very good relative parallax errors, and reach farther distances from the Sun. 
 We also note that the uncertainty in $\mu_\delta$ is smaller than for $\mu_\alpha*$. This is due to a geometrical scaling factor on the uncertainties resulting from the 
  scanning law which in the direction of the anticentre favours $\mu_\delta$. 
    
 \begin{figure}
\centering
\includegraphics[width=\hsize]{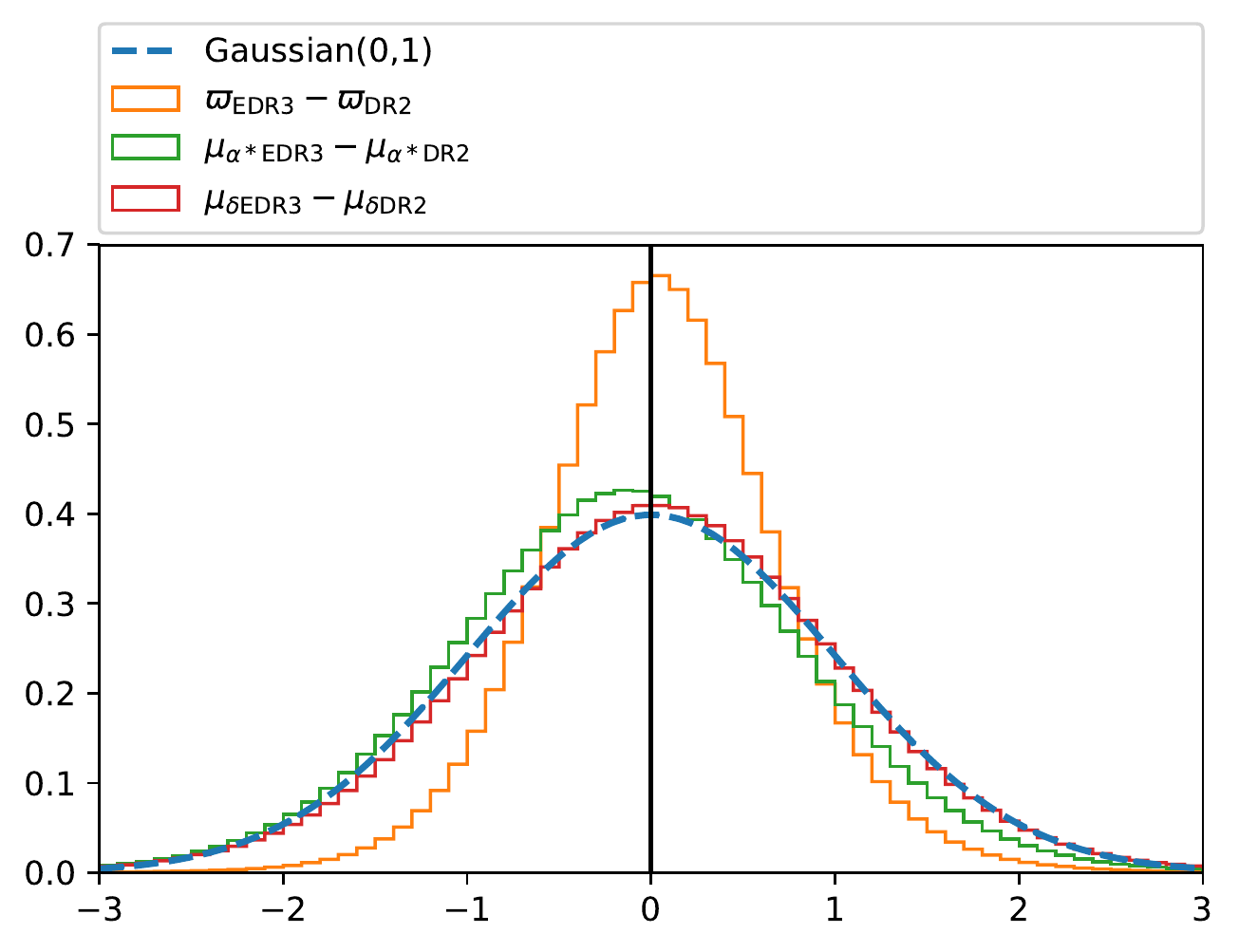}
\caption{Consistency between astrometric values in DR2 and EDR3. The histograms show the differences in parallax and proper motion normalised to the errors ${(x_{\rm{EDR3}}-x_{\rm{DR2}})/\sqrt{\sigma_{x,\rm{EDR3}}^2+\sigma_{x,\rm{DR2}}^2}}$, {where $x$ is $\varpi$, $\mu_\alpha*$ or $\mu_\delta$}, and compared to a Gaussian distribution with 0 mean and variance of 1. The differences in $\mu_\alpha*$ are due to a systematic in DR2 that has been now corrected. We have corrected the parallaxes (Eq.~\ref{eq_zpt}) using the median offset for DR2 (-27 $\mu$as) and for EDR3 (-17$\mu$as).}
\label{fig_systematics}
\end{figure}

As in previous releases, the astrometric \Gaia data suffer from some systematics. The median astrometric quantities and their uncertainties show checkered patterns that somehow correlate with the scanning law, as illustrated for the median parallax (Fig.~\ref{fig_sky_medians}f) and median uncertainty in $\mu_\delta$ (Fig.~\ref{fig_sky_medians}g). The later shows additionally some of the large scale bands mentioned above. The amplitude of these systematics has, however, been reduced in EDR3 (see \citealt{Lindegren2020b}). Another known systematic is a zero point in parallax \citet{Lindegren2020} that has also been reduced and which we examine in detail in Sect.~\ref{sect_zpt}.

Figure~\ref{fig_systematics} shows the differences in all astrometric quantities between DR2 and EDR3 normalised to the errors\footnote{{We have used the {\tt gaiaedr3.dr2\_neighbourhood} 
 table for the correspondence between sources.} 
 }. The median absolute differences between EDR3 and DR2 are {15}$\muasyr$ in $\varpi$, $-48\muasyr$ in $\mu_{\alpha*}$ and ${7}\muasyr$ in $\mu_\delta$. For comparison, we show a Gaussian distribution with 0 mean and 1 as variance with a blue curve, although the quantities from DR2 and EDR3 are not independent and thus these distributions are not expected  necessarily to follow this curve. 
{The large} systematic differences in $\mu_\alpha*$ (green histogram) is explained by a correction of the reference frame (spin) 
 for EDR3 that has largely reduced the medium-scale (1-20\,$\deg$) inhomogeneities in the median parallax and proper motion of the quasars, which actually were quite large precisely in the direction of the anticentre for $\mu_\alpha*$ (about $0.1\muas$, figure 13 of \citealt{Lindegren2020b}). 
   The histogram of parallax differences is narrower than the Gaussian case  
and is slightly positively biased. In  Fig.~\ref{fig_systematics} the zero point has been corrected using the median estimated values for quasars respectively in DR2 and EDR3 (more details are given in Sect.~\ref{sect_zpt}). We note that the bias was even larger if we neglected the corrections (giving a median of $25\,\mu$as). The persistent bias after the correction could be attributed to underestimation of the zero point in DR2 (for which there is some evidence, see \citealt{Lindegren2020}) or effects of considering a fixed value of the offset (Sect.~\ref{sect_zpt}). %

\paragraph{4. Photometric quality.} 
The improvement of the photometry of EDR3 with respect to DR2 is described in \cite{Riello2020}. 
In summary, the increase in the number of observations and the improvement of  several
steps of the pipelines (image parameter determination, LSF and PSF calibrations, cross-match
and photometry) have led to a significant decrease of the systematics at the bright end ($G<15$).
The effects of blends and contamination by nearby stars are mainly filtered out using Eq.~\ref{eq_photcuts}.

\subsection{Extinction and selection of tracer populations}\label{sect_populations}

We used two different approaches to select the tracer populations. The first {one} uses only \Gaia data and the populations were  selected in the de-reddened HR diagram. In the second approach, external photometric data were used  to define a sample of RC sources. 

\paragraph{Method I: Using the \Gaia HR diagram.} 
We obtained the de-reddened HR diagram for the \AC sample 
using  the 3D {dust-reddening maps from \bayestar{} \citep{Green2019}.}
We used the $d_{PM}$ distances (Sect.~\ref{sect_distances} and Appendix~\ref{sect_bayesiandistances}) and the galactic $(\ell,b)$ coordinates to infer the line-of-sight visual extinction $A_V$ for each source. Then we transformed $A_V$ to $A_G$ and $E(BP-RP)$ 
using {the \Gaia extinction factor from  \citet{2018MNRAS.479L.102C}. We compared our results with the ones using the extinction model by \citet{Lallement2019} and the conversion to $A_G$ and $E(BP-RP)$ following Appendix A in \cite{Ramos2020}, finding only mild differences in the classification of stars without consequences on the conclusions from the subsequent sections.}

 \begin{figure}
   \centering
   \includegraphics[width=0.95\hsize]{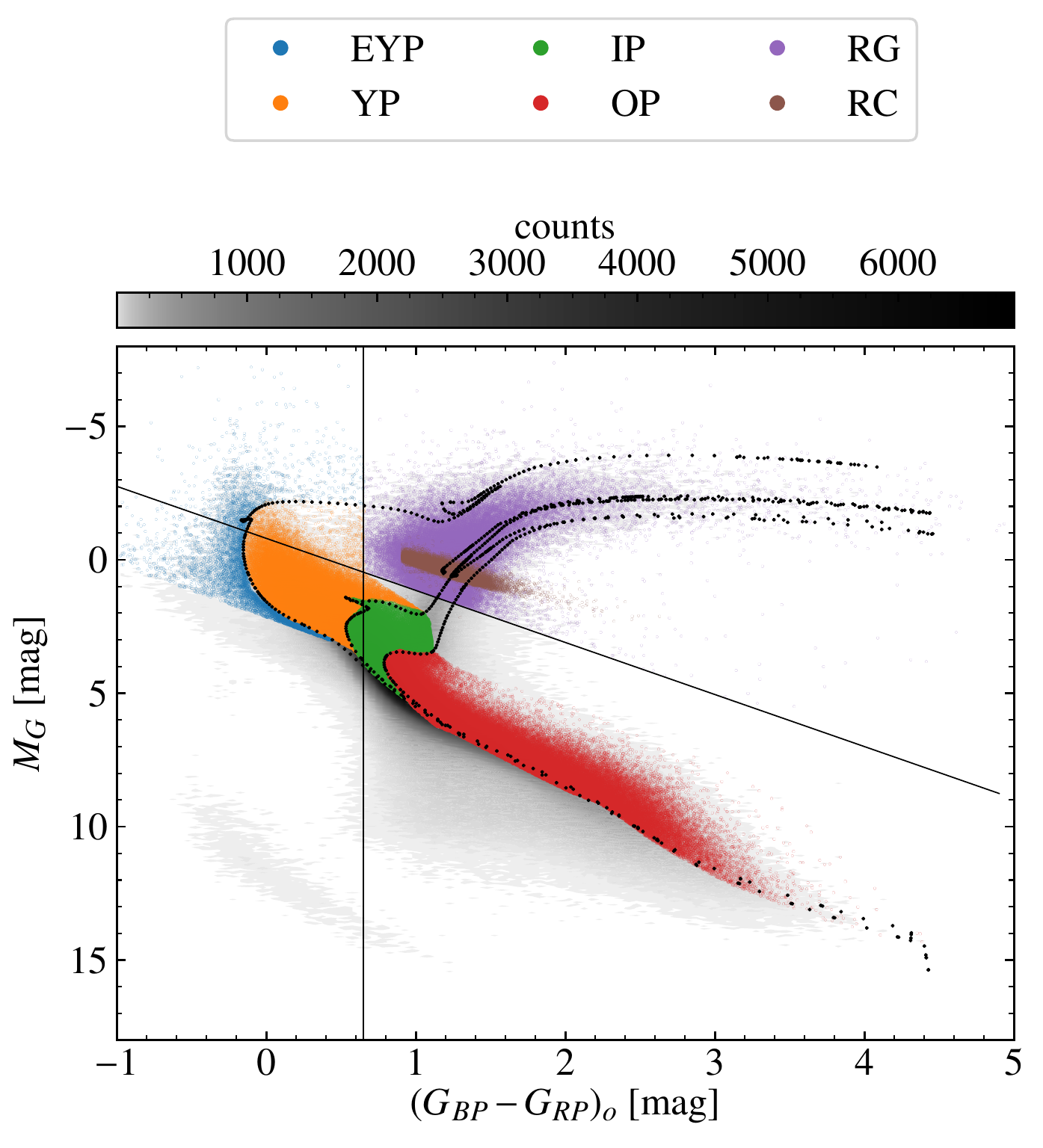}
         \caption{De-reddened HR diagram of the anticentre region and different selected populations. The diagram is shown for the {3\,369\,456} sources of the \AC sample with available photometry ($G$, $BP$, $RP$) and extinction data {and absolute magnitudes} derived considering the $d_{PM}$ distances. We over-plot three \texttt{PARSEC} stellar isochrones with $[M/H] = 0$  for the ages of 0.2, 2 and 8 Gyr, a line at $BP-RP={0.65}$ and a diagonal line following the extinction slope used for the selection of populations (Method I) {which} appear in different colours. The RC have been selected using a different method  (Method II).}
         \label{fig:populations}
   \end{figure}

\begin{table}
\caption{Populations in the AC20 sample. The selections are obtained following the  Method I, except for the RC of the last row (Method II).}             
\label{tab_pop}      
\centering        
\setlength{\tabcolsep}{1.2pt}  
\begin{tabular}{l l r}\hline
   & Population & Sources \\
   \hline 
        & All & {3\,369\,456} \\
   EYP & Extremely young massive  ($\tau \lessapprox 0.2\, Gyr$) &  {12\,792}  \\
   YP  & Young main sequence with  $\,  0.2 \lessapprox \tau \lessapprox 2\, Gyr$ & {276\,344}  \\
   IP & Intermediate main sequence with $\,  2 \lessapprox \tau \lessapprox 8\, Gyr$&  {807\,910} \\  
   OP & Old main sequence &  {1\,032\,916} \\  
    RG & Red giants & {176\,193} \\
    RC & Red clump & {121\,857}  \\
\hline                  
\end{tabular}
\end{table}

Figure~\ref{fig:populations} shows the Gaia de-reddened HR diagram for the \AC sample. We used the direction of the extinction line in this diagram ($M_G = 1.95(BP-RP)_o - 0.8$, black diagonal line) combined with the vertical cut $(BP-RP)_o > {0.65}$ to select (conservatively) giants (purple dots). 
Then we used the \texttt{PARSEC} isochrones \citep{Bressan2012,Marigo2017} and respective updates\footnote{\url{http://stev.oapd.inaf.it/cgi-bin/cmd}} to 
perform a statistical partition by ages of the main sequence sources into extremely young (EYP, 0.2 Gyr), young  (YP, 0.2 - 2 Gyr), intermediate (IP, 2 - 8 Gyr) and old populations (OP), as specified in Table~\ref{tab_pop} and illustrated in Fig.~\ref{fig:populations}. The massive sources of the EYP   were constrained to have $(BP-RP)_o < {0.65}$, while the YP, IP and OP were selected 
between the lower (ZAMS) and the upper (TAMS) luminosity boundary of the main sequence band defined by the \texttt{PARSEC} stellar evolutionary tracks at $[M/H]$ = 0. We note that while the OP has contribution from young stars, we can claim it is on average older that the IP: we expect an average age of 4-5 Gyr (e.g. figure 13 in \citealt{Bland-Hawthorn2019}) but with an important contribution of the oldest stars in the disc. In general, these selections are contaminated by stars of different ages due to several aspects (stars with different metallicities to the ones used in the isochrones, inaccuracies of the extinction model used, confluence of isochrones around the ZAMS, binarity, etc). Nevertheless, we expect our samples to be dominated by the age ranges desired, which is enough for our basic purposes here.

\paragraph{Method II: Using Gaia \& 2MASS.}
\label{sec_RCsel0} 
We combined EDR3 parallaxes and $G$-band photometry with that from \twomass{} $K$-band for the AC20  sample. 
	The passbands in \twomass{} are narrow and in the infrared, and are thus weakly affected by errors in the extinction estimation. For \twomass{}, the flag (qfl) = `AAA' indicates the highest photometric quality. However, this would significantly reduce our sample (to only 15\% of the entire AC20 sample). Instead, we chose to enforce a quality cut only at the distance estimation stage using the photometric errors, $e\_jmag$ \& $e\_kmag$. We first computed the extinction of each source using the 3D dust-reddening maps from \bayestar{} \citep{Green2019} with the inverse of the parallax as a prior for distance. The RC sources were selected in a Bayesian manner around the literature values for the absolute magnitude of the RC simultaneously for the $G$ \Gaia band and the \twomass{} $K$ band. {We compared these distances with the ones using Bayesian distances for the prior ($d_{PM}$, see Sect~\ref{sect_distances}) and our results do not change significantly.}  More details of the procedure and a validation with an external sample are given in Appendix~\ref{app_RCsel} and \ref{app_RCdist}.

\section{Distances and phase space coordinates}\label{sect_phasespace}

In this section we describe how the distances and phase coordinates are computed in our study. We start by discussing the zero point in the \Gaia parallaxes (Sect.~\ref{sect_zpt}), which needs to be corrected in order to estimate first distances (Sect.~\ref{sect_distances}), and subsequently Galactic cylindrical positions and velocities (Sect.~\ref{sect_velocities}).

\subsection{Parallax zero-point correction}\label{sect_zpt}

As for previous releases, the \Gaia parallaxes have a zero point\footnote{We use a different notation compared to \cite{Lindegren2020} to distinguish with the vertical cylindrical coordinate $Z$.} ($ZP$) that needs to be {considered. In EDR3} the median parallax of the quasars is ${-17}\muas$ {\citep{Lindegren2020}}. This negative correction needs to be subtracted from the EDR3 parallaxes ({parallaxes need to be increased by $\sim17\muas$)}:
\begin{equation}\label{eq_zpt}
\varpi^{\rm corrected}=\varpi-ZP,
\end{equation}
\noindent or equivalently, reducing the inferred distance.
Here we corrected all parallaxes by subtracting   $ZP={-17}\muas$. Additionally, when relevant, we also compared our results with the more sophisticated approach presented in \cite{Lindegren2020}.
In that work, they estimate the parallax zero-point $ZP_{56}$ as a function of magnitude, colour (more precisely, {\tt nu\_eff\_used\_in\_astrometry} for the 5p solutions and {\tt pseudocolour} for the 6p solutions, hence the names $ZP_5$ and $ZP_6$), and ecliptic latitude, by looking at the parallaxes of quasars, binary stars and sources in the Large Magellanic Cloud for EDR3.
Here we computed $ZP_{56}$ using the Python implementation that {is} available online {as part of the \Gaia EDR3 access facilities\footnote{\href{https://gitlab.com/icc-ub/public/gaiadr3_zeropoint}{https://gitlab.com/icc-ub/public/gaiadr3\_zeropoint}}}. Panel h in Fig.~\ref{fig_sky_medians} shows the median zero-point $ZP_{56}$ in the AC20 region. We observe a mild dependency of its value on the sky position. The median value for all stars in AC20 region is $ZP_{56}={-20}\muas$, thus {similar to that of} the quasars, with the 10 and 90 percentiles being ${-32}$ and $-14\,\muas$, respectively, and  $ZP_{56}>0$ only for 0.02\% of the stars. For the \AC sample, which has a significantly different magnitude distribution compared to the case without the parallax quality cut (Fig.~\ref{fig_Gbins_1d}), we find a median $ZP_{56}$ of $-30 \muas$ and percentiles of $-38$ and $-20\muas$, respectively. 
The $ZP_{56}$ case{, thus, yields} the largest differences {between uncorrected and corrected distances} (Fig~\ref{zpt} in Appendix~\ref{app_zpt}). The velocities, which depend linearly on the distances, are consequently scaled as well. All these will be important in order to determine, for example, the exact distances to some kinematic features that we detect but we do not observe any qualitative difference in our results. More details are given in Appendix~\ref{app_zpt} and throughout the paper.


\subsection{Distances}\label{sect_distances}

 To convert the astrometric measurements by \Gaia into phase space coordinates, we require an estimate of the distance to a given star. The complications of estimating distances to stars given their measured parallaxes have been discussed by a number of authors over a long period of time \citep[e.g.][]{Stromberg27,LutzKelker73,BailerJones15,Luri18}. The transformation between parallax and distance is non-linear, which leads to a number of issues, including the extreme case of negative measured parallaxes. Simply taking the inverse of the measured parallax gives a biassed estimate of the distance of a star, and this bias grows more serious as the relative uncertainty grows larger. 
It has therefore become extremely common to apply a Bayesian approach to the problem of providing distance estimates from parallaxes, and/or to use photometric information to produce a better estimate of the distance.

We work primarily with distance estimates from a Bayesian approach ($d_{PM}$), similar to that applied by \citet{Schonrich2017}, with a prior that is derived iteratively to be consistent with the data.
 These distances use a {prior $P(d) \propto d^2 P_\rho({r}(d)) S(d)$, where} ${r}(d)$ is the position at distance $d$ along a given line-of-sight, {so $P_\rho({r}(d))$ is proportional} to the density of a model Galaxy. The term $S(d)$ is the selection function -- i.e. the probability that a randomly chosen star at a distance $d$ enters the catalogue. The distance estimate, $\tilde{d}$ is then found as the expectation value of $d$ given this prior and the measured parallax ({and its} uncertainty). As explained in the previous section, default distances are computed considering a fixed {parallax zero point of $-17 \muas$.} The distances and the code can be found at \url{https://zenodo.org/record/4415706} and \url{https://zenodo.org/record/4415669}, respectively. More details can be found in Appendix~\ref{app_distances}. 
 
To check that our results are robust, we compared to results when we estimate the distance as simply the inverse of the parallax, and also with a different Bayesian approach based on that from \citet{BailerJones2018}. We tested each of these approaches on GOG data, and further details are given in Appendix~\ref{app_distances}.
From these tests, we conclude that using a parallax quality cut of $\piepi>3$ is a good compromise between the performance of the estimate and the number of stars of our samples. However, we emphasise that all the estimators tested here return somehow imperfect distances, which in the Bayesian case depend also on how close the assumed prior on the Galaxy distribution is to the Galaxy model {used in GOG (i.e. the Besan\c{c}on Galaxy Model)}. We find that the median relative difference between the simulated true distances and the estimated ones can be as large as 20\% at 4 kpc and larger than 50\% for 25\% of the stars even with the $\piepi>3$  selection.


On the other hand, for the stars classified as RC (see Sect.~\ref{sect_populations}), we inferred their distance ($d_{RC}$) in a Bayesian manner using complementary photometric data from \twomass\, 
 (details are given in Appendix~\ref{app_RCdist}). {Typical uncertainties are of 1 kpc at a distance of 6 kpc (see Fig.~\ref{fig:absmag_hist_gmag}).}

\subsection{Positions and velocities}\label{sect_velocities}

\begin{figure}
\begin{center}
\includegraphics[width=\columnwidth]{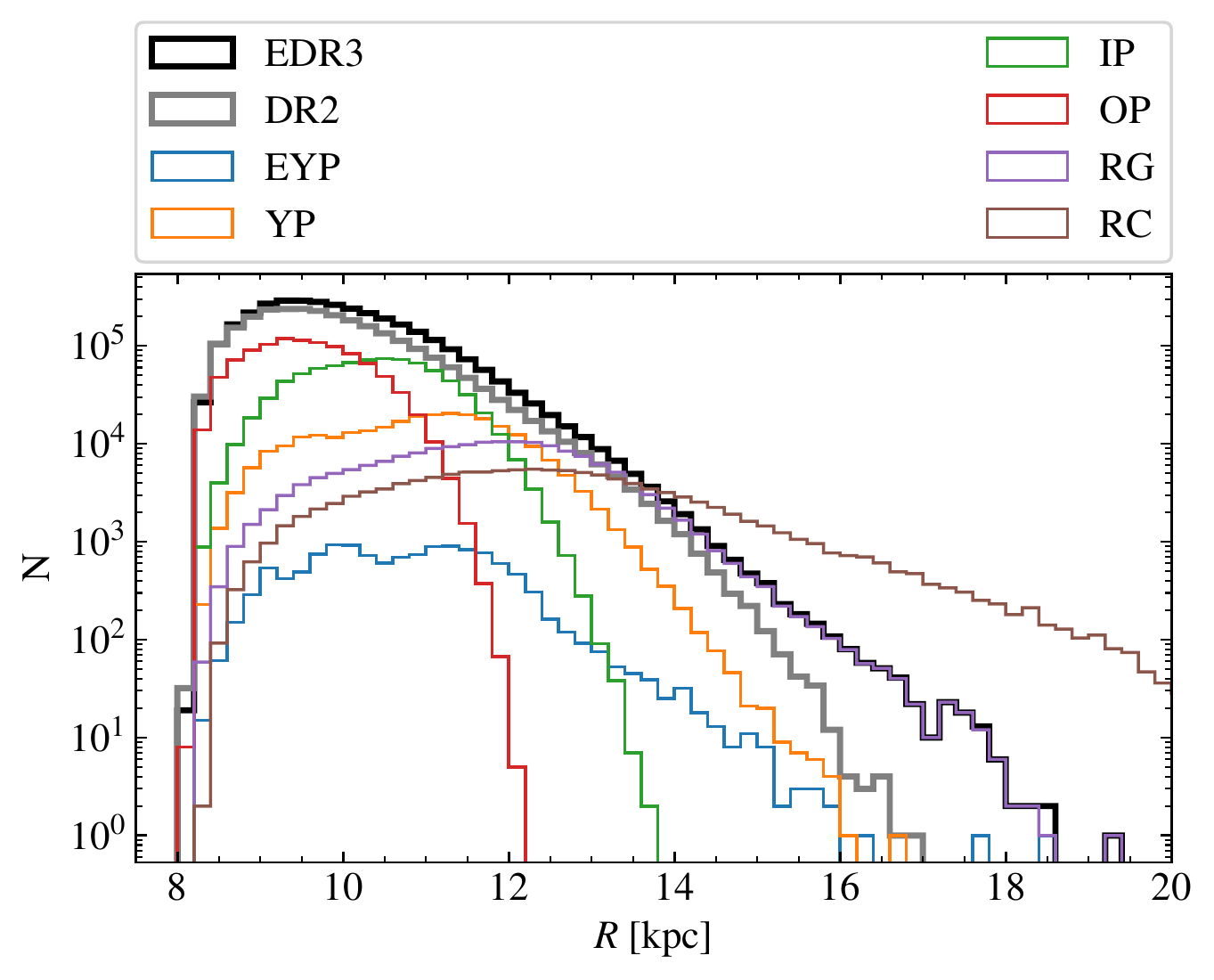}
\caption{Distribution of stars in Galactocentric radius. Number of stars per radial bin of 200 pc for the whole \AC sample for EDR3 (black line) and DR2 (grey line), and for each stellar population (colour lines as indicated in the legend). The RC do not have the constraint $\piepi>3$ and appears more {extended in $R$} than the black distribution.}
\label{fig:nbstarsperradialbin}
\end{center}
\end{figure}

From the distances obtained in Sect.~\ref{sect_distances} and the sky positions, we computed the Galactic Cartesian ($X,\,Y,\,Z$) and cylindrical ($R,\,\phi,\,Z$) positions, assuming that the Sun is located at  $d_{\odot-GC}=8.178\kpc$ from the Galactic centre \citep{Gravity2019} and a height above the Galactic plane of $Z_\odot=0.0208\kpc$ \citep{Bennett2019}. Figure~\ref{XYZ} in Appendix~\ref{app_material} shows the spatial distribution of the AC20 sample. By construction, the vertical $Z$ and azimuthal $Y$ distributions are wider for larger distances from the Sun, with some stars at $R\sim X=14\kpc$  reaching heights of 1 kpc above and below the plane. Figure~\ref{fig:nbstarsperradialbin} shows the number of stars as a function of Galactocentric radius. The gain in EDR3 for sources with $\piepi>3$ (black line) compared to DR2 (grey line) at large radii is very significant with an increase of one order of magnitude already at 16 kpc and notably more stars at almost all radii.

For the different populations detailed in Sect.~\ref{fig:populations} (colour lines in Fig.~\ref{fig:nbstarsperradialbin}), the samples with younger ages have distributions that, as expected, extend to larger radii compared to older populations. The distribution for the whole sample with $\piepi>3$ is dominated by dwarfs for $R<12\kpc$ while giant stars take over beyond that. We see some hints of an over-density at around 12 kpc for the EYP and YP that could be the Perseus spiral arm but a good assessment of this requires more investigations of the selection function and the extinction.  For the RC whose distances are computed photometrically without the $\piepi>3$ constraint (brown line), there is a larger number of stars at large distances compared to the whole sample with  $\piepi>3$ (black line).

For the velocities, we computed $\vl$ and $\vb$ and corrected them for the reflex of the solar motion using the following equations:
\begin{equation}
\label{eq:vl}
\vl  = k d \mu_\ell -U_\odot\sin(\ell)+(v_{c,\odot}+V_\odot)\cos(\ell)\\	
\end{equation}
\vspace{-0.5cm}
\begin{equation}
\label{eq:vb}
\vb =k d \mu_b+[-U_\odot\cos(\ell)-(v_{c,\odot}+V_\odot)\sin(\ell)]\sin(b)+W_\odot\cos(b),
\end{equation}
\noindent where $k=4.7404705$ is the usual factor for units conversion, and we assume $U_\odot=11.1$, $v_{c,\odot}+V_\odot=248.5$,  $W_\odot=7.25\kms$ for the solar motion \citep{Schonrich2010,Reid2020}, where $v_{c,\odot}\equiv v_c(R=R_0)$ is the value of the rotation curve at the Sun's position. In the anticentre direction, $\vl$ and $\vb$ are approximately aligned with the usual cylindrical velocities $V_\phi$ and $V_Z$ and, thus, we use:

\begin{align}
V_\phi^* &\equiv - \vl\label{eq:fx}\\
V_Z^* &\equiv\vb. \label{eq:dfx}
\end{align}

We note that $V_\phi^*$ {is positive for most of the disc stars with this definition and that $V_\phi^*$} is not exactly equivalent to $V_\phi$, nor is $V_Z^*$ to $V_Z$, due to a geometric difference in the vector orientation and the contribution of the line-of-sight velocity, but the differences are small in the anticentre. In the Appendix~\ref{app_material} we used GOG to quantify this and we find that $80\%$ of the sources with $\piepi>3$ have absolute differences smaller than 2.9 and 3.3 $\kms$ for $V_\phi^*$ and $V_Z^*$, respectively (Figs.~\ref{5d6d_1}, \ref{5d6d_2} and \ref{5d6d_3}). 
We see that $\Vp$ is mainly {smaller} than $V_\phi$ with a median of {-0.4}$\kms$. When examining how these differences are distributed in the $\ell$-$b$ projection, we see, as expected, larger differences in $\Vp$ the farther from the exact anticentre line ($\ell=180\deg$). The differences in $\VZ$ show a quadrupole symmetry, indicating that any kinematic signature following this same shape in the sky would be clearly suspicious but that for most of the cases, since we average over the whole area, the global effect of these differences is null. 
For stars in the Gaia 6D phase space sample (thus a more realistic case), the differences are similar though slightly larger ($80\%$ of the stars with $\piepi>3$ have absolute differences smaller than {3.2} and 4.0 $\kms$ for $V_\phi^*$ and $V_Z^*$, respectively).

\begin{table}
\caption{Uncertainties in phase space coordinates for the different \Gaia releases. In the first rows we show the median uncertainties (first three numerical columns) and upper limit uncertainty for 80\% of stars (three columns from the right) for stars in the \AC sample for DR2 (first row), for the stars from  EDR3 in common with DR2 (using the {\tt gaiaedr3.dr2\_neighbourhood} 
 table for the correspondence between sources), and for EDR3. The last rows compare the heliocentric velocity uncertainties in DR2 and EDR3 for the sample with 6D velocities (6dsample) when the error in $v_{\rm{los}}$ is not (left) and is considered (right). }             
\label{tab_velerrors}      
\centering          
\setlength{\tabcolsep}{4.pt}
\begin{tabular}{l r r r r r r}\hline\hline   
\multicolumn{7}{c}{\AC}\\
&\multicolumn{3}{c}{median} &\multicolumn{3}{c}{80\% of sources} \\
                                        &$\epsilon_R$    &$\epsilon_{\Vp}$ &$\epsilon_{\VZ}$ &$\epsilon_R$    &$\epsilon_{\Vp}$ &$\epsilon_{\VZ}$ \\
DR2                                     &0.30 &         3.8&       2.2&$<$0.57 &    $<$8.4&    $<$4.6\\
EDR3 ($\cap$ DR2)           &  0.18 &  2.3&1.2&$<$ {0.42} &    $<$ {5.6}&   $<$  {3.0}\\
EDR3                                 &{0.3} &     3.1&       1.7&$<$0.58 &    $<$ {7.4}&  $<$  3.9\\

\hline 
\multicolumn{7}{c}{6dsample}\\
&\multicolumn{3}{c}{median $\epsilon_{v_{\rm{los}}}=0$} &\multicolumn{3}{c}{median} \\
                                        &$\epsilon_U$    &$\epsilon_V$ &$\epsilon_W$ &$\epsilon_U$    &$\epsilon_V$ &$\epsilon_W$ \\

DR2                                    &0.09 &    0.09&    0.07& 0.43  &   0.44&    0.38\\
EDR3                                 &  0.04&  0.04   &   0.04  &0.38 &0.39&  0.34\\

\hline  \hline                  
\end{tabular}
\end{table}

 Another reference system for the velocities that we used in Sect.~\ref{sect_halo} is the tangential velocity $V_t$ defined as:
\begin{equation}\label{eq_vt}
V_t \equiv k d \sqrt{{\mu_\alpha*}^2+{\mu_\delta}^2}.
\end{equation}
In particular for that section we used as distances the inverse of the parallax with a more strict selection of $\piepi>5$.

\begin{figure}
   \centering
   \includegraphics[width=0.8\hsize]{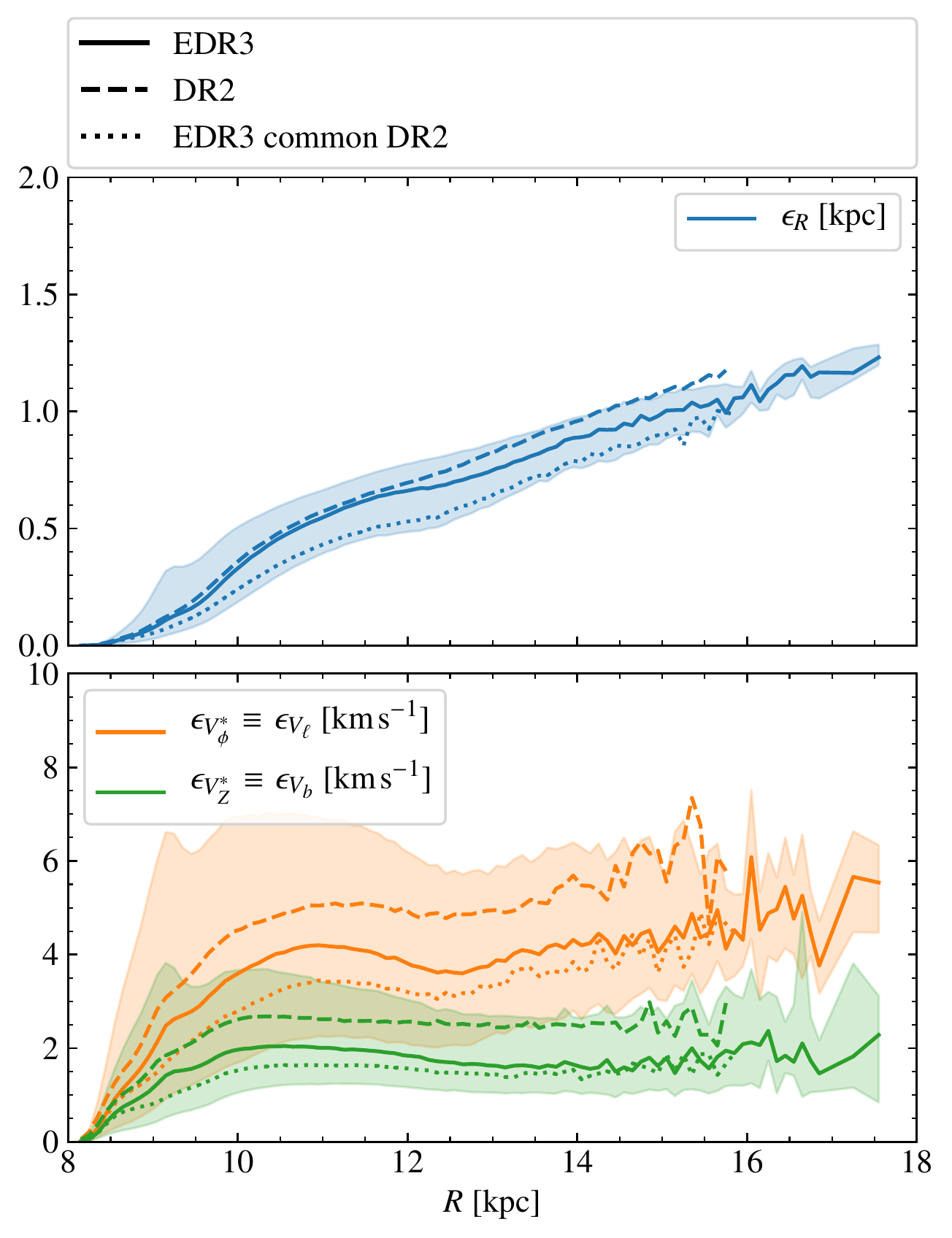}
      \caption{Errors in phase space coordinates in the anticentre. The curves are for the \AC sample and show the median errors for $R$ (blue) in the top panel, and $\Vp$ (orange) and $\VZ$ (green) in the bottom panel, while the shaded regions show areas enclosing 50\% of the stars (that is, limited by the 25 and 75 percentiles). We show the values for EDR3 (solid), DR2 (dashed) and sources in common (dotted).}
         \label{errors_RVlVb}
 \end{figure}

We used the Jacobian matrix to compute the errors in the positions and velocities from the errors (and correlations) of the astrometric quantities. We neglected the errors in the angular positions since they are extremely small. In the case of the Bayesian and photometric distances, no correlation between distance $d$ and proper motions $\mu$ was considered (but see discussion Appendix~\ref{app_distances}). Figure~\ref{errors_RVlVb} shows the median uncertainty in the radius $R$ (top) and velocities (bottom) as a function of $R$ for EDR3 (solid lines) for the \AC sample, and the area delimited by the 25 and 75 percentiles (shaded regions). The median errors in $R$ (solid blue line) remain lower than 1 kpc for $R<{14} \kpc$ and the velocity uncertainties (solid orange and green lines) are smaller than 5 and 2$\kms$ for $\Vp$ and $\VZ$, respectively, for most of the radii probed. The slight change of trend in the solid curves at around $12\kpc$ is due to the contributions of different stellar types, in particular giants stars that are intrinsically brighter at a given $R$ and have, thus, smaller astrometric uncertainties. Table~\ref{tab_velerrors} gives a summary of these position and velocity errors: 80\% of stars have errors $<0.6\kpc$ in Galactocentric radius, and  $<7\kms$ and $<4\kms$, respectively for $\Vp$ and $\VZ$. The errors for $\VZ$ are smaller than for $\Vp$ due to the better alignment of $\mu_b$ with $\mu_\delta$ which in turn has smaller errors than $\mu_\alpha$  {in this sky direction} as seen in Sect.~\ref{sect_quality} (Fig.~\ref{fig_astrometricerrors}). 
 
In Fig.~\ref{errors_RVlVb} we also show the equivalent errors in DR2 (dashed lines). However,  a fair comparison 
requires that we compare {only} the common sources (otherwise the new  sources of fainter magnitudes {in EDR3} at each bin in $R$ contribute in a negative way to the overall values). The dotted lines obtained for the sources of EDR3 in common with DR2 show a quite significant improvement. For the velocities, the uncertainties are now smaller by about $\lesssim2\kms$ at a Galactocentric distance of $R=12\kpc$, which represents an improvement  of 30\%. 
  
\begin{figure}
   \centering
      \includegraphics[width=\hsize]{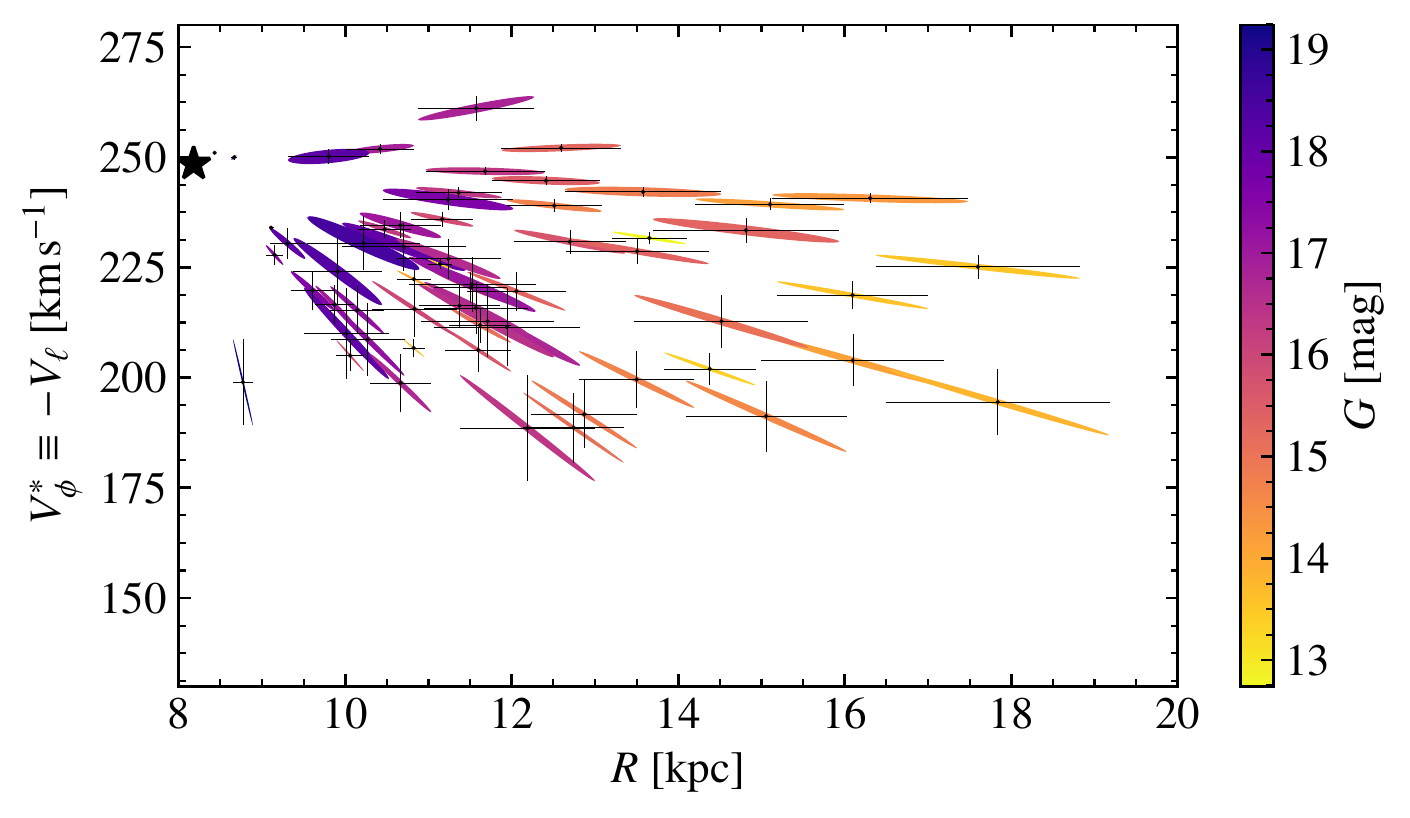}

      \caption{Error ellipses in the $R$-$\Vp$ plane for stars in the anticentre. The ellipses have been drawn for 60 stars from the \AC sample chosen randomly but with weight of ${\exp{R}}$ in order to sample {better the} different $R$. The ellipses are coloured by magnitude $G$ and the error bars are included as black lines. The error ellipses are oriented pointing towards the $R_0$-${(v_{c,\odot}+V_\odot)}$ point (black star).}
         \label{ellipses}
   \end{figure}

Figure~\ref{ellipses} shows the full error ellipses for a few stars chosen to sample different values of R in the $R$-$\Vp$ projection that we explore later. While the black error bars show the errors on the individual quantities, the error ellipses show large correlations between these two variables. This correlation is  induced by the coordinate transformations, which in both {axes} have an approximately linear dependency with the distance error. As expected, the ellipses are all oriented pointing towards the position {and velocity} of the Sun assumed (indicated with a black star).

Finally, another good illustration of the improvement in the astrometry is the comparison of the uncertainties in the heliocentric velocities $U$, $V$ and $W$ for the 6dsample in DR2 \citep{Katz2019} and EDR3 \citep{Seabroke2020}, which is shown in the last three rows of Table~\ref{tab_velerrors}. Assuming that there are no {line-of-sight} velocity uncertainties, the median uncertainties (left columns) are reduced by around $50\%$ in EDR3.  Including the {line-of-sight} velocity uncertainties (rightmost columns) does not show such a reduction, highlighting that the {line-of-sight} velocity uncertainties dominate.  
This will change in DR3 where these uncertainties are expected to decrease substantially and millions of additional sources will have {line-of-sight} velocity measurements for the first time.

\section{Disc kinematics}\label{sect_disc}

\begin{figure*}
\begin{center}
\includegraphics[width=0.45\hsize]{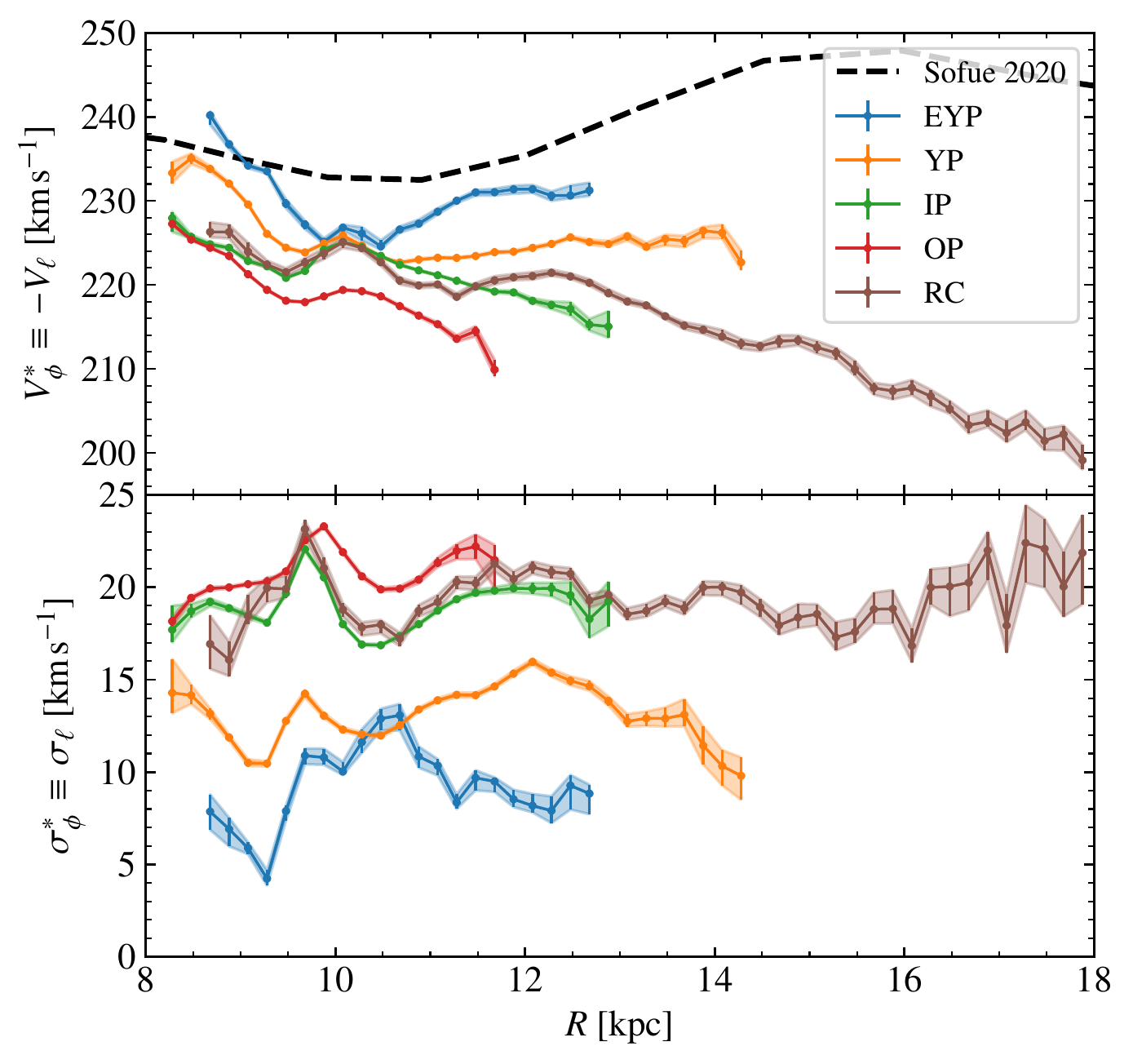}
\includegraphics[width=0.45\hsize]{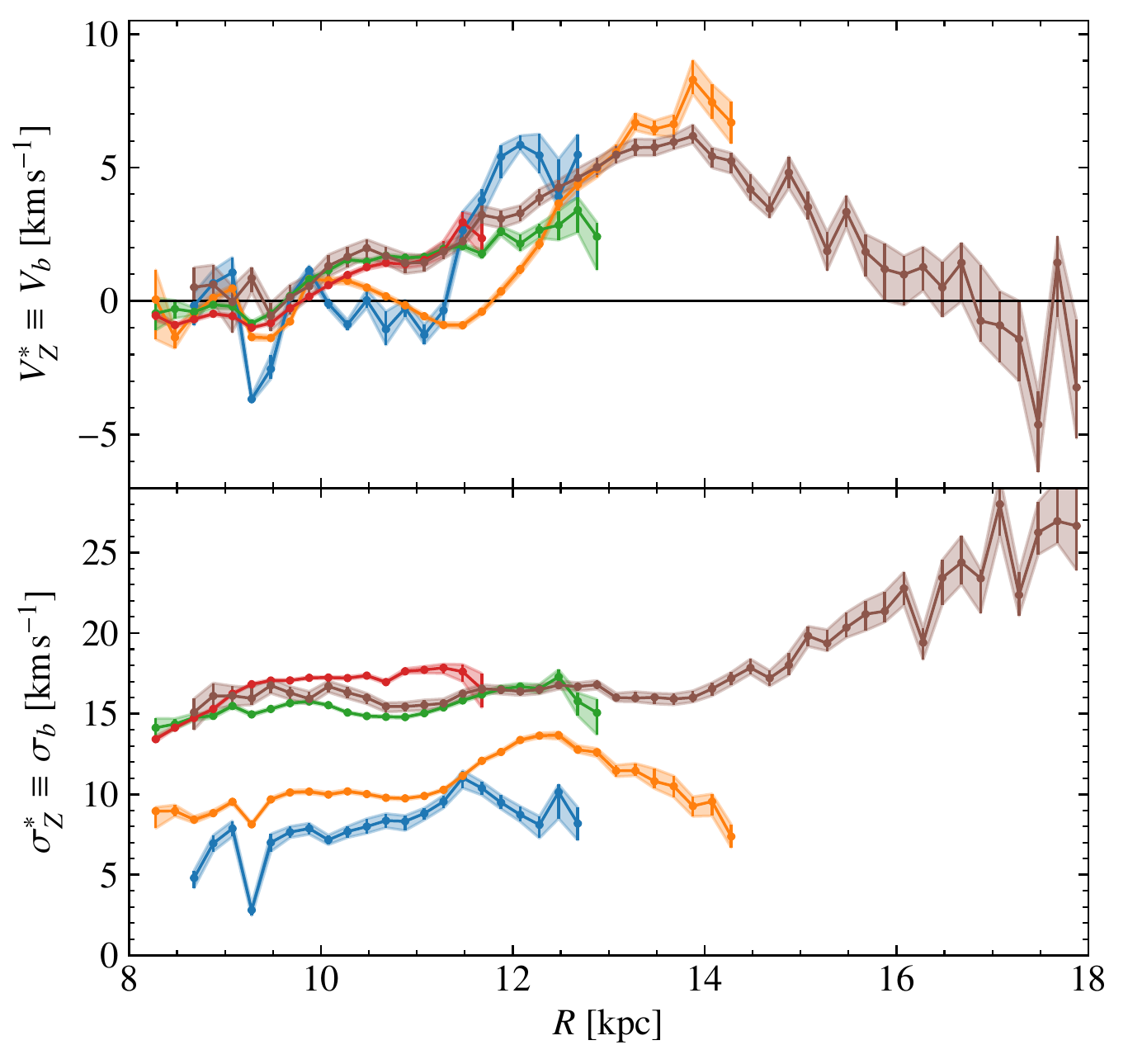}
\caption{Rotation  and vertical velocity profiles in the anticentre. Top: Median azimuthal and vertical velocities of the populations EYP, YP, IP, OP and RC as indicated in the legends (same as in Fig.~\ref{fig:nbstarsperradialbin}). Shaded areas represent the uncertainties (see text) {but they are very small and barely visible in most of the cases}. The rotation curve by \citet{Sofue2020} is over-plotted in the top left panel. Bottom: Same as top but for the velocity dispersions {(computed as $1.5MAD$ values)}. Apart from the expected differences due to the different ages of the populations and the asymmetric drift, we see significant oscillations in all curves.}
\label{fig:vphivzprof}
\end{center}
\end{figure*}

In this section we explore the dynamics of the  {MW} disc, analysing the velocities as a function of positions. As seen in Sects.~\ref{sect_quality} and \ref{sect_velocities}, the improvement in the EDR3 astrometry allows us to probe the disc's outer regions. We start by examining the median velocities and velocity dispersions  (Sect.~\ref{sect_vphivprof}) as a function of Galactocentric radius. We then look at large scale velocity asymmetries and phase space correlations in Sect.~\ref{sect_assym}, to end with the analysis of small scale velocity substructures (Sect.~\ref{sect_substructure}) that are now resolved for the first time.

\subsection{Azimuthal and vertical velocities and dispersions} 
\label{sect_vphivprof}

We measured the median velocity profiles and dispersions of $\Vp$ and $\VZ$ for each stellar population. {We used $\sigma^*\equiv1.5MAD$ (where MAD is the median absolute deviation) as a robust estimate of the standard deviation, to which we subtracted the median error in each bin in quadrature. Using a robust estimator, rather than the standard deviation, prevents outliers from producing a noisy dispersion profile. Although the 1.5 factor is strictly valid only for normally distributed data, this approximation puts our values on the same scale as the standard deviation for a more easy comparison.}
{We used bins of 200 pc and discard those with less than 100 stars.}
 The uncertainties were then obtained by performing 1000 bootstrap resamplings of these distributions at each radius, choosing the 16th and 84th percentiles, {and are indicated as shaded colour bands and error bars in the following the panels}. 
  
The rotation velocity curves for the different populations are shown in the top left panel of Fig.~\ref{fig:vphivzprof}. A difference in the median $\Vp$  is  observed for the different stellar populations with the older stars rotating slower as a result of the asymmetric drift.
 On average, the EYP stars rotate {between 10 and} ${\sim}{{20}} \kms$  faster than the OP or the RC. 
{The curve of the EY stars presents the best agreement with the rotation curve (black dashed lines) derived in \citet{Sofue2020} from a compilation of kinematic data from molecular gas and stars in the infrared.}  This is consistent with the expectation 
that younger stars  
rotate as fast as the cold interstellar gas, thus at velocities closer to the true circular velocity of the  {MW}.  Globally,  all the rotation curves decline for $R \lesssim 9.5$ kpc and show a bump at {around} 10 kpc. 
Beyond {this}, the curve of YP stars is flat, while those of older stars decrease again.

\begin{figure}
\begin{center}
\includegraphics[width=0.9\hsize]{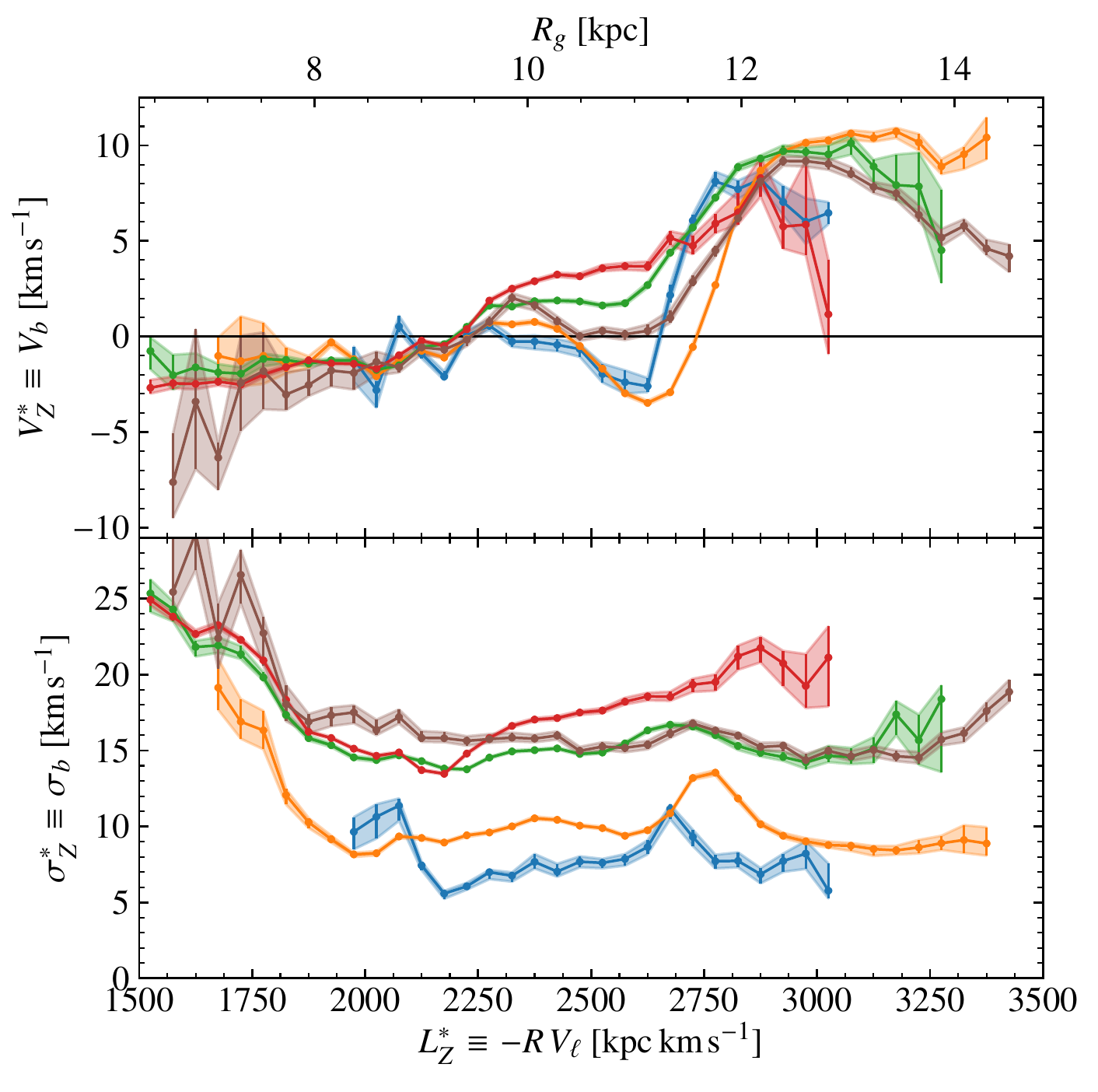}
\caption{{Vertical velocity profiles as a function of angular momentum. The plot and legend is equivalent to the left panels of Fig.~\ref{fig:nbstarsperradialbin} but as a function of $\Lz$. To guide the eye, we also show an approximate guiding radius $R_g =\Lz/(236\kms)$ on the top axis.}}
\label{vzprofL}
\end{center}
\end{figure} 

The effects of the parallax zero point are examined in Appendix~\ref{app_zpt} where we show, as an example, the effects of the different adopted values of this offset on the rotation curve of our \AC sample (Fig.~\ref{figprofiles_zpts}). 
As expected and discussed in Sect.~\ref{sect_zpt},  we see slight differences in the curves due to a decrease of the distance and scaling of the velocities when the correction is used. However, the general features of the curves remain the same.

\begin{figure*}
\begin{center}

\includegraphics[width=0.45\hsize]{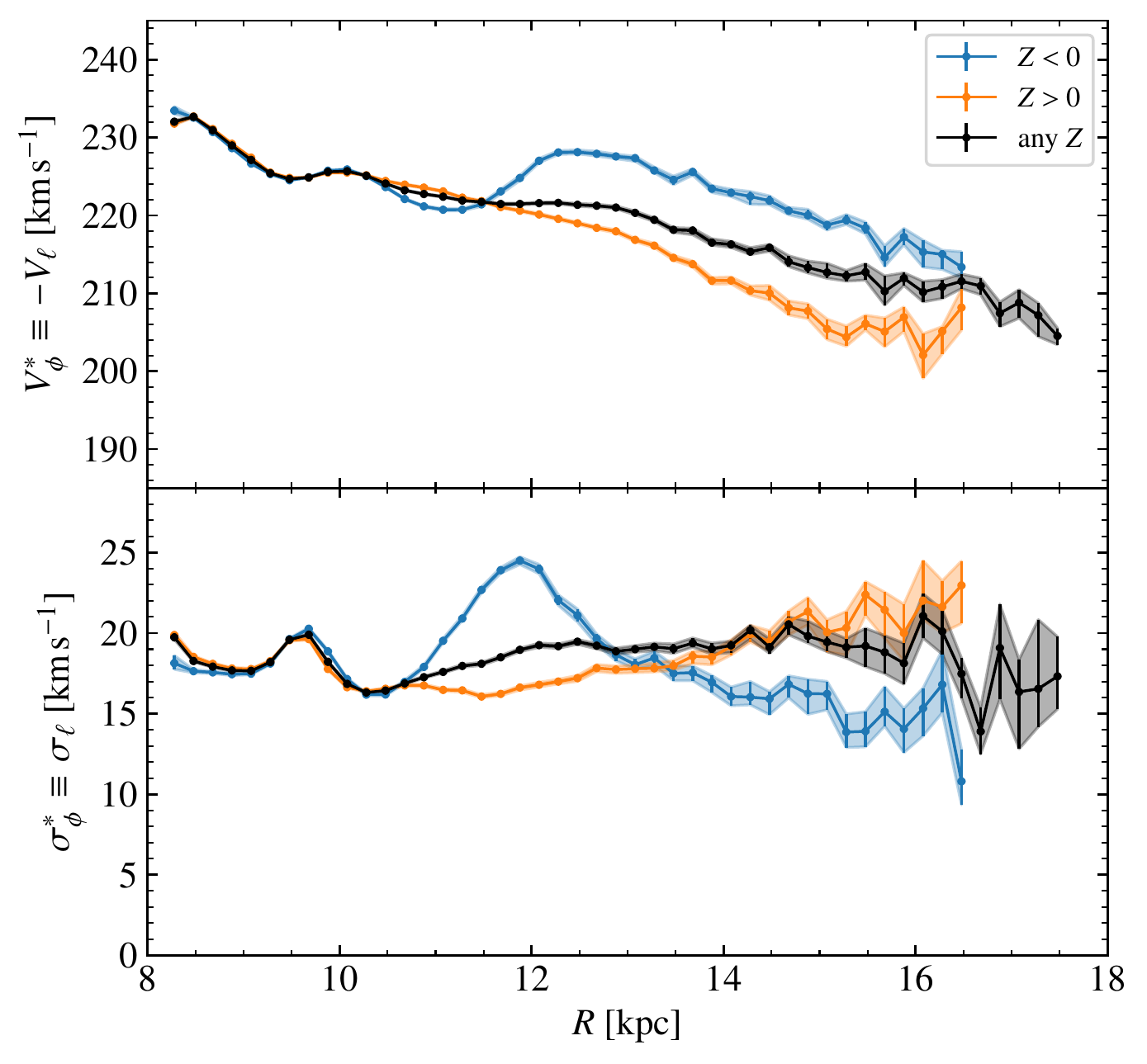}
\includegraphics[width=0.45\hsize]{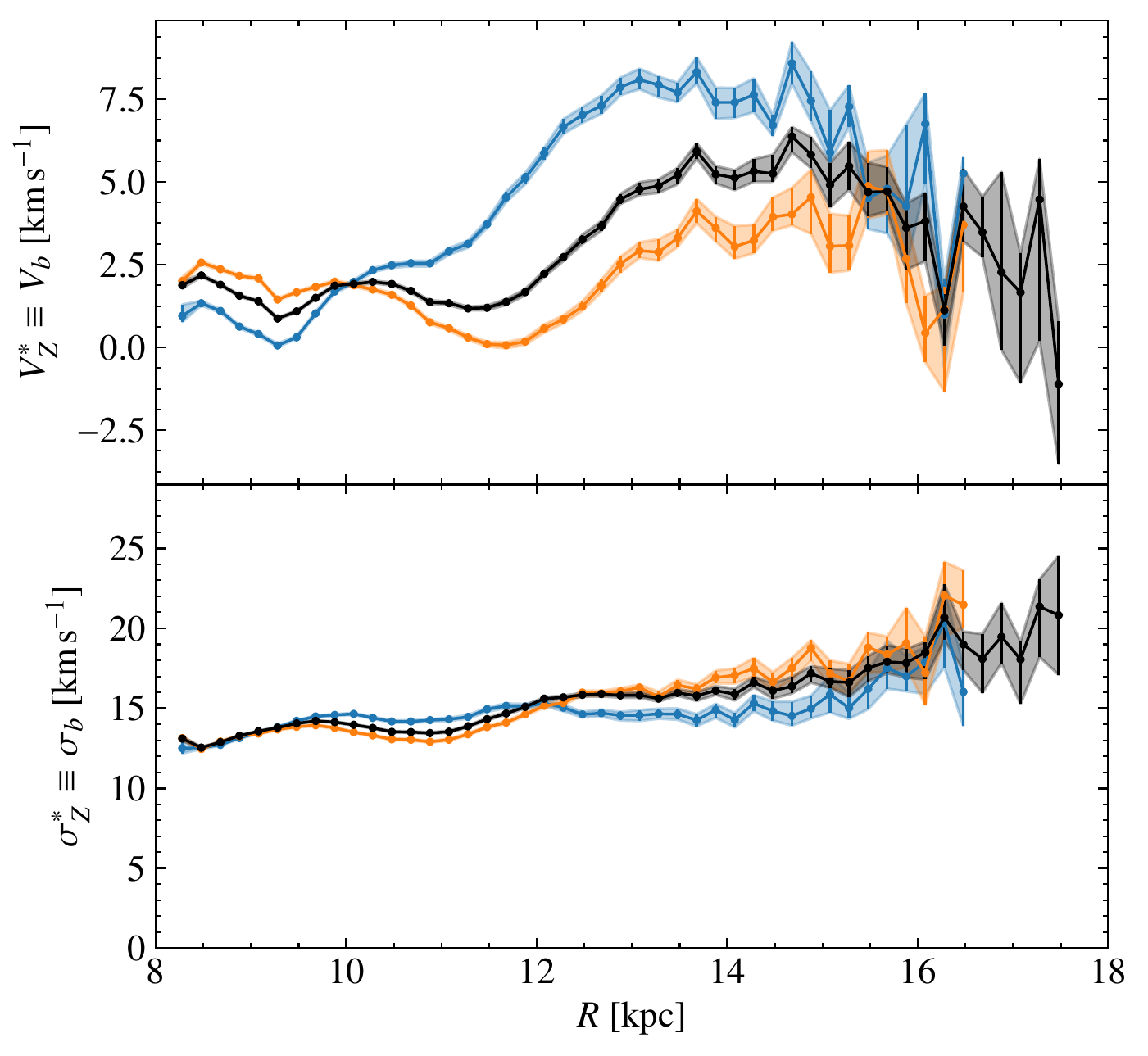}

\caption{Comparisons of the velocities above and below the Galactic plane.  We plot the median azimuthal and vertical velocities (top) and velocity dispersions ({computed as $1.5MAD$ values,} bottom) for the whole sample \AC (black lines, {not including the sample of RC with photometrically derived distances}), and for stars with $Z\ge 0$ (orange lines) and for $Z<0$ stars (blue lines). Shaded areas represent the uncertainties. We observe notable asymmetries beyond 10-11 kpc.}
\label{fig:zasym}
\end{center}
\end{figure*}

Interestingly, we observe stars from the YP rotating as far as 14.5 kpc from the Galactic centre (see also Figs.~\ref{fig:nbstarsperradialbin},  \ref{fig:zasym} and \ref{figridges}). 
In total, we find as many as {1186} stars with $16<R<18\kpc$ and $\Vp>{200}\kms$ for the $ZP={-17}\,\mu$as (with median uncertainties of $\epsilon_R\sim1.\kpc$), and {275} for the case of $ZP_{56}$. 
This establishes a lower limit to the disc size although a detailed analysis is required, in particular in the context of the biases of the distance estimators, which can be large at these distances (see Fig.~\ref{disterrdistrueL}). 

The top right panel of Fig.~\ref{fig:vphivzprof} shows the median vertical velocities. These  velocities appear to have small oscillations of the order of $2\kms$ inside $R\sim 12$ kpc. 
There are clear dips at $R\sim{9.5}$ kpc, coinciding with the dip in $\Vp$, and at $R\sim 11$ {more prominent} for the young stars.  {Figure~\ref{vzprofL} (top) shows the same velocities but as a function of $\Lz$ where the oscillations appear clearer with dips especially at around $2200$ and $2600\kmskpc$. In the later one, we can see a strong age dependence with younger populations showing a deeper valley. We note that at the extremes of $\Lz$ in this plot, the populations are biased towards high eccentricity orbits that are those that manage to reach the observed volume.}
 Beyond the location of the dips, $\VZ$ increases and stars move in median upwards ($\VZ > 0$). The profile of the RC stars {with $R$ (and of all populations in the $\Lz$ plot)} draw a clear wiggle (with a subsequent decrease), with maxima of $\sim {{5}}\kms$ at $R\sim 14$ kpc {($\sim10\kms$ at $\Lz\sim 3000\kmskpc$)}. In Fig.~14 of \citet{Katz2018} only the first part of this positive vertical velocity wiggle was observed and seemed to have certain dependencies on the Galactic azimuth $\phi$  and vertical position $Z$ of the stars, as we confirm in Sect.~\ref{sect_assym}. 
 {The oscillations and the outer increase {in} the vertical velocities as a function of angular momentum were also observed in \citet{Schonrich2018}, \citet{Huang2018} and \citet{Cheng2020}.}
 

The bottom panels of Fig.~\ref{fig:vphivzprof} show the diversity in the velocity dispersions $\sigma^*_{\phi}$ and $\sigma^*_{Z}$  in the Galactic anticentre direction. 
 Although we expect decreasing dispersions with $R$ \citep[][and references therein]{vanderKruit1986} supported by observations in external galaxies \citep{Martinsson2013} {including} the Large Magellanic Clouds in the \Gaia data \citep{Luri2020}, the general behaviour here shows bumpy dispersions in all the populations that correlate with the oscillations in the median velocities. 
 
 Apart from the oscillations, overall we observe dispersions that are quite flat as a function of $R$, and even increasing at larger radii for RC stars. {For the vertical velocity dispersion as a function of angular momentum (bottom panel of Fig.~\ref{vzprofL}) the oscillations are even clearer. In the inner parts the dispersions decrease with $Lz$ but this could be due to the selection effects explained above, while the profiles are overall flat in the outer parts.} We note that the geometry of our \AC samples have larger ranges of $Z$ for increasing $R$ (Fig.~\ref{XYZ}).   This together with a complex selection function in the more distant regions and the approximation in the velocities of Eqs.~\ref{eq:fx} and \ref{eq:dfx} could {produce artificial trends in} the velocity dispersion. 
A similar flattening of the vertical velocity dispersion outside the solar radius was observed in \citet{Sanders2018} where the authors also discuss different biases that could explain this behaviour but also the possibility of being related to the flare (see also \citealt{Mackereth2019,Sharma2020}).  
  
As for the amplitude of the dispersions, younger stars unsurprisingly present lower velocity dispersions
than more evolved stars, {most likely} because these populations
have not had the time to be heated by various internal and external processes, unlike older populations. 
On average, the {azimuthal}  and vertical velocity dispersions of the EYP stars are {around 8 and 5} $\kms$ lower than those of {old} stars, respectively {for $\Vp$ and $\VZ$}.

The flattening of the velocity ellipsoid $\sigma^*_{Z}/\sigma^*_{\phi}$ inside $R \sim 14.5$ kpc is {quite} homogeneous
 among the various populations and within the whole sample, all of them showing {an azimuthal}  dispersion larger than the vertical component ($\sigma^*_{Z}/\sigma^*_{\phi} = 0.7-0.8$, on average). 
The vertical random motion only exceeds the {azimuthal}  component for RC stars beyond 15 kpc, and for EYP stars at $R=9$ kpc {and after 11 kpc}. Interestingly, the random motions of the EYP stars with dispersions of values of {9} and ${8} \kms$ for $\sigma^*_{\phi}$ and $\sigma^*_{Z}$ on average, respectively, are comparable to the typical velocity dispersions seen in the gas
($\sim 9$ and $4.5 \kms$ respectively for neutral atomic and molecular gas for $R < 8$ kpc,  \citealt{mar17}), for a gas velocity ellipsoid assumed isotropic.
 {Thus, in 200 Myr (the maximum age of the EYP), the youngest stars present already a slight velocity anisotropy as expected since they have oscillated of the order of one-two vertical periods.}

 \subsection{Velocity correlations and asymmetries}\label{sect_assym}

We study here  kinematic differences as a function of the location with respect to the Galactic mid-plane, and other phase space correlations. First, we compare the 
 kinematics of $Z <0$ stars  with those at $Z \ge 0$ for the whole \AC sample (Fig.~\ref{fig:zasym}). 
 There is a notable asymmetry in the median velocities  and the velocity dispersions (Fig.~\ref{fig:zasym}), starting approximately at 10-11 kpc, 
 thus coinciding with the starting position of the large vertical velocities of Fig.~\ref{fig:vphivzprof}.
The rotation of 
   $Z< 0$ stars (blue curves)    clearly leads that of stars at $Z\ge0$ (orange) beyond $R \sim 11$ kpc typically by up to $10 \kms$.  
   A significant asymmetry is also seen for $R> 10$ kpc in the vertical motion where stars at $Z < 0$  move at larger velocities than $Z \ge 0$ stars, 
   with a difference of up to $\sim 6 \kms$ (already noticed in \citealt{Katz2018} and \citealt{Wang2020b} for example). 
   The asymmetries in $\VZ$ start close to the Sun, though with opposite trend compared to $R>10\kpc$. 
   The {azimuthal dispersions} are comparable at lower {radii} but asymmetric beyond $R \sim 10.5$ kpc (larger values for $Z < 0$ stars, by up to ${5}\kms$) {and reversing beyond 13 kpc}.
   There is also a vertical velocity dispersion asymmetry but it is weaker ($\lesssim 1 \kms$).
   In any case, the dispersions observed correspond to the typical thin disc velocity dispersions \citep[e.g.][]{Robin2003,Aumer2009}.
   
   \begin{figure*}
   \centering
   \includegraphics[width=\hsize]{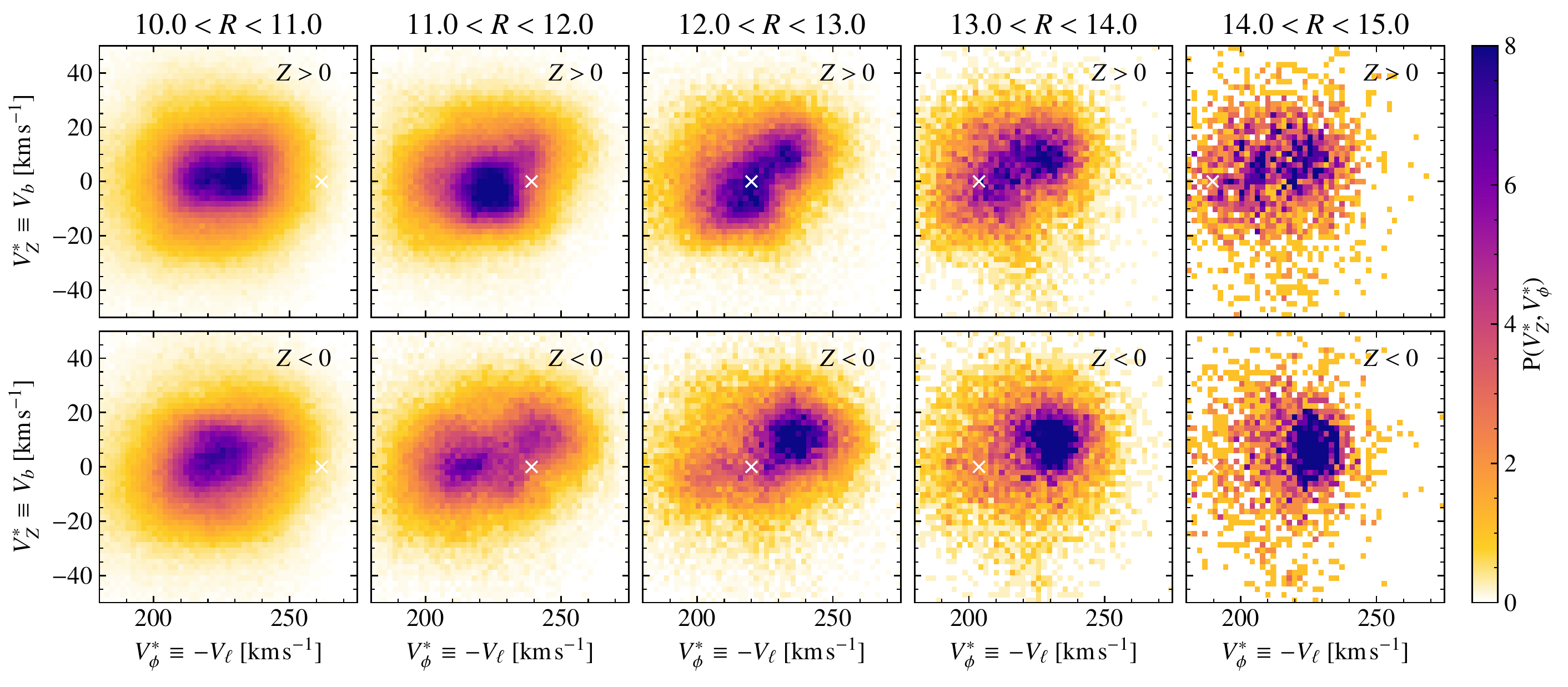}
   \caption{{Density} in velocity space at different distances above and below the plane. Stellar density in the $\Vp-\VZ$ plane, for bins in $R$ from $10$ to $15 \kpc$ for the \AC sample for $Z>0$ (top) and $Z<0$ (bottom). We see division into two components in the outer radial bins. To guide the eye a cross has been placed at $\Vp$ corresponding to $\Lz=2750\kms\kpc$ for a point in the centre of the radial bin (see also Fig.~\ref{fig:LzVz}). }
\label{fig:VphiVzclumps}
   \end{figure*}
   
We now follow up these asymmetries by looking with more detail at the density of stars in the $\Vp$-$\VZ$ plane. We show the counts in this projection in $1\kpc$-wide radial bins for Galactocentric distances ranging from 10$\kpc$ to 15$\kpc$, and for the north ($Z>0$, top) and south ($Z<0,$ bottom) Galactic plane (Fig.~\ref{fig:VphiVzclumps}).  
One of the clearest features in Fig.~\ref{fig:VphiVzclumps} is the lack of symmetry for stars above and below the plane. Secondly, for the bins {at} $R>12\kpc$ we observe a bimodality where stars are sitting mainly in two clumps, one with negative $\VZ$ at lower $\Vp$, which is more prominent in the north, and one with positive $\VZ$ at higher $\Vp$, more conspicuous in the south.
The different proportions of the clumps of the bimodality at different $Z$ seems to be the cause of   the vertical asymmetries seen at the top panels of Fig.~\ref{fig:zasym}, moving the median velocities to higher or lower $\Vp$ and higher or lower $\VZ$.
 However, we emphasise that the bimodality appears on both sides of the disc, just in different ratios.

   \begin{figure*}
   \centering
   \includegraphics[width=0.99\hsize]{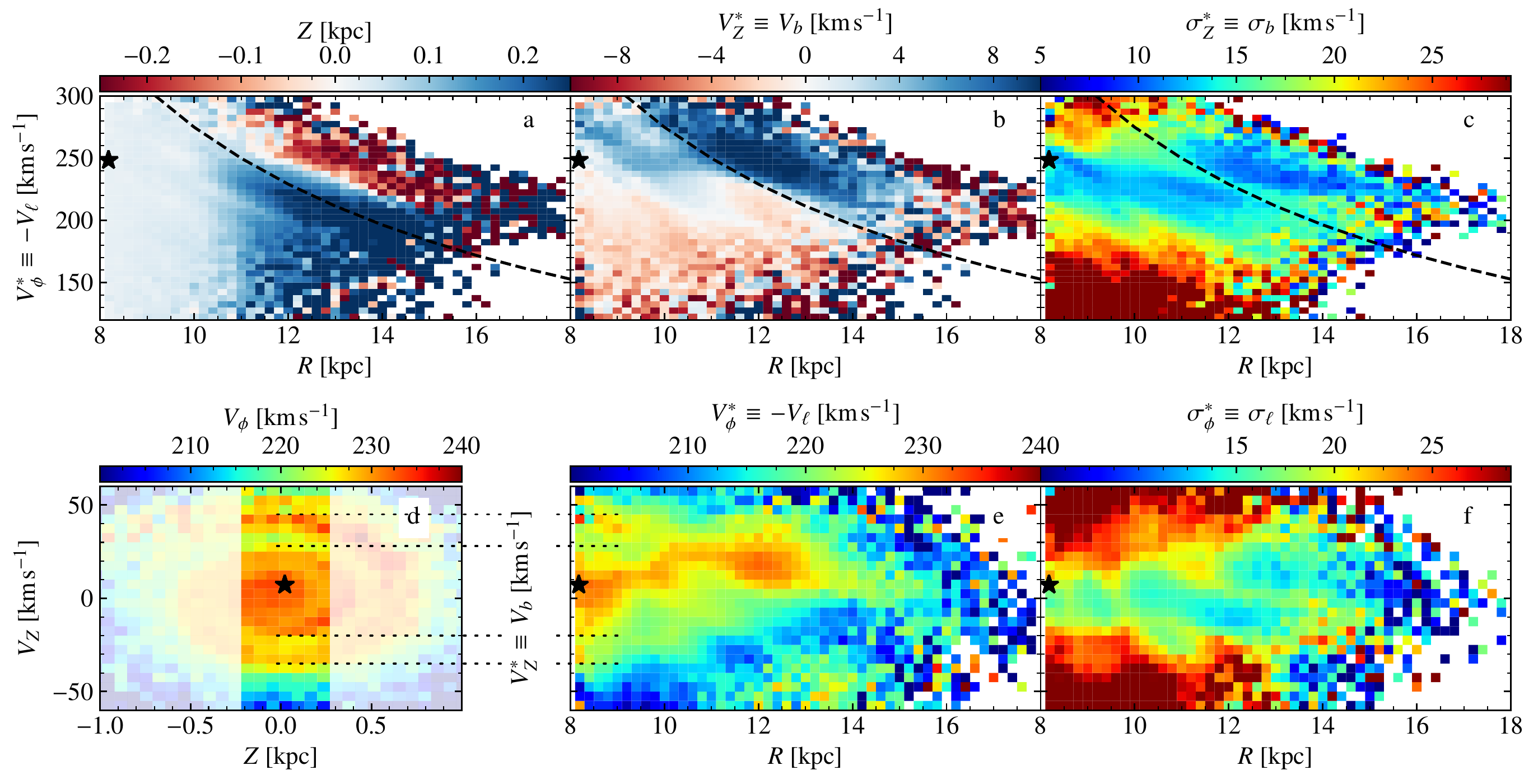} 
         \caption{Phase space projections of the Galaxy disc. The plots show for the \AC sample: a) median $Z$ coordinate in the $R$-$\Vp$ plane; b) median vertical velocity $\VZ$ in the same projection; c) dispersion in the $\VZ$ velocity in the same projection {(computed as the $1.5MAD$)}; d) phase spiral in the 6dsample in EDR3 for stars in the Galactic radial range $|R- {R_0}|<0.2\kpc$;  e) median azimuthal velocity $\Vp$ in the $R$-$\VZ$ plane; f) dispersion in $\Vp$ in the same plane. The bimodality appears  in the outer parts of the disc in panels a, b and c, with the separation marked with a line of constant angular momentum $\Lz=2750\kms\kpc$. In panel e, the evolution of a slice of the phase spiral (marked in brighter colours in panel d) is seen for {smaller radii}, while a signature related to the above bimodality is seen beyond $\sim 12\kpc$ in panels e and f.}
         \label{figRVb}
   \end{figure*}
   
Figure~\ref{figRVb} shows other phase space projections, allowing us to study this {phenomenon} in a more continuous way: 
 the top panels show  $\Vp$ as a function of $R$, colour-coded by either the $Z$ position (a), the median $\VZ$ (b) and the {dispersion} $\sigma_{\VZ}$ (c). {At R>11 kpc,} the population having large $\Vp$ ($\sim30\kms$ larger than the other group) and positive $\VZ$ ($\sim10\kms$, blue colours in panel b) is predominantly at negative $Z$ (red colours in panel a), and vice-versa for the population having smaller $\Vp$ and negative $\VZ$ (of about $-2$ to $-5$ $\kms$), as seen before. {We note that the bimodal kinematic behaviour is also present at smaller $R$ but with smaller amplitude in the vertical velocities.  In fact some hints of this bimodality were seen in the maps of, for example, \citet{Khanna2019} and \citet{Wang2020} but those reach only 12 kpc from the Galactic centre and the bimodality appears marginally at the borders of their distributions.} Additionally, we note now a clear spatial evolution, with the region at large $\Vp$ and positive $\VZ$ smoothly diminishing its $\Vp$ when $R$ increases. 
A line of constant angular momentum $\Lz=R\Vp=2750\kms\kpc$ has been plotted that roughly marks the transition in the sign of $\VZ$ in panel b. There is not an exact match between the transition zone in panels a and b, indicating that the dominance of one clump over the other does not occur exactly at $Z=0$.  We note that the velocity dispersion of both groups of stars is typical of the thin disc ($\sigma_{\VZ} \sim 15\kms$), as already inferred from the bottom-right plot of Fig.~\ref{fig:zasym}. {We also see that the dispersion profiles of panels c and f are vey different from the ones for the GOG mock \Gaia data in Fig.~\ref{figRVlVb_GOGUM}.}

  \begin{figure}
   \centering
    \includegraphics[width=0.9\hsize]{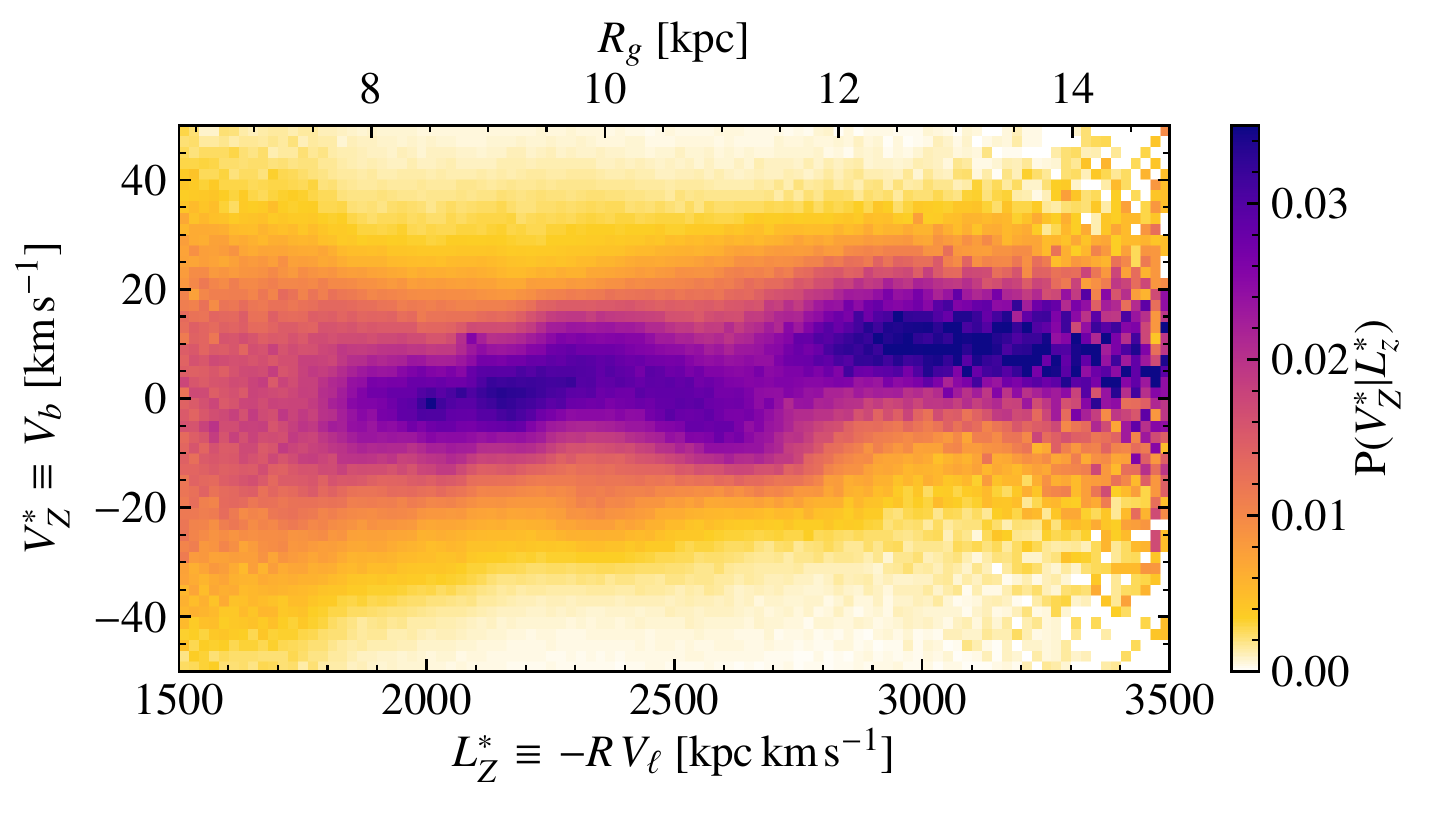}   
    
    \includegraphics[width=0.9\hsize]{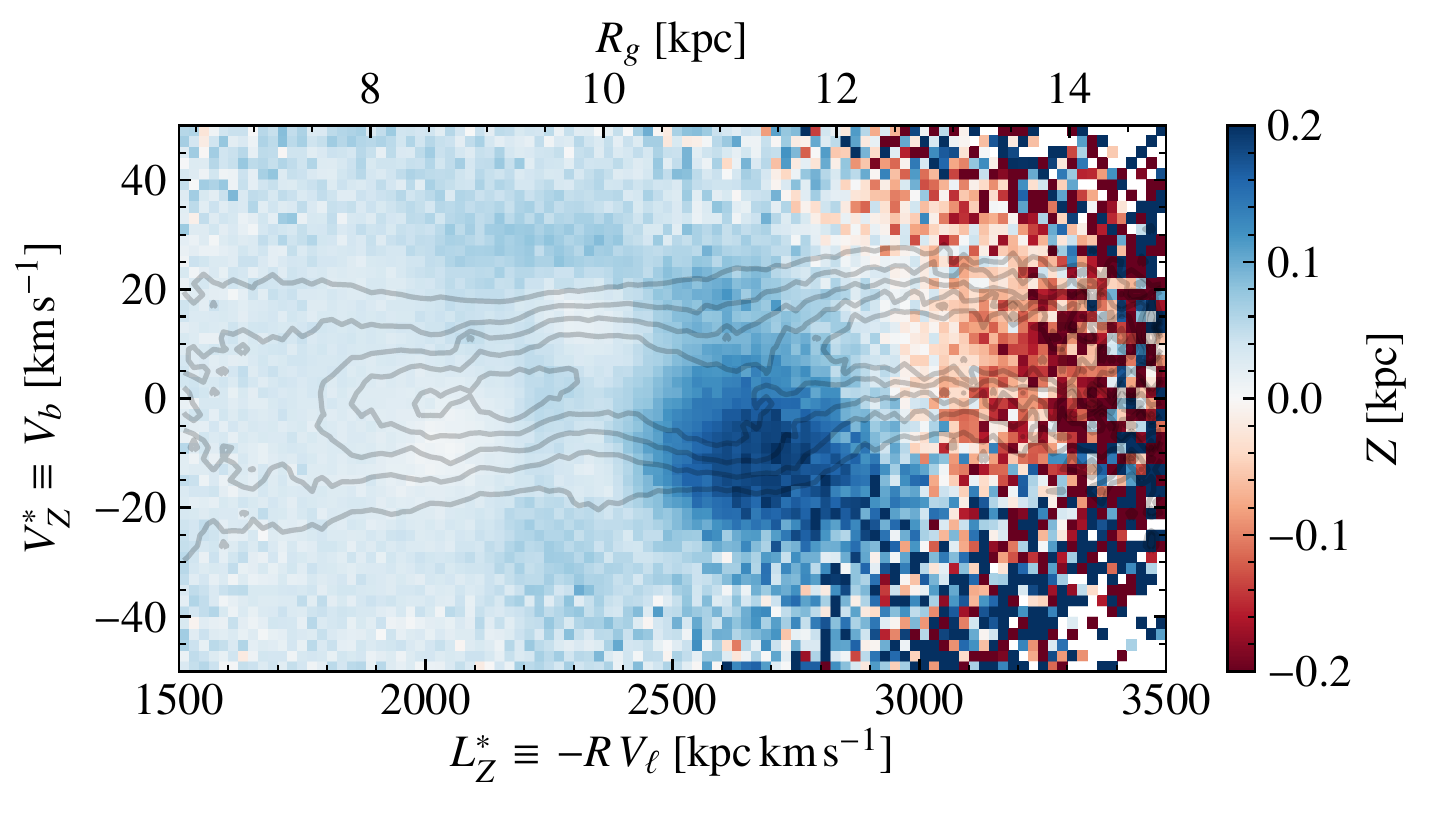}
   \caption{Structures in vertical velocity and angular momentum. Top: Column normalised histogram of star numbers in the $\Lz$-$\VZ$ plane for the \AC  sample (the colour represents the fraction of stars in a given $\Lz$ bin that have a certain $\VZ$). Bottom: Average $Z$ of stars in each bin in $\Lz$-$\VZ$ in our \AC sample. Contours are the same as the colour plot in the top panel. {To guide the eye, we also show an approximate guiding radius  $R_g = \Lz/(236\kms)$. }}
         \label{fig:LzVz}
   \end{figure}
   
   The phase spiral identified in  \citet{Antoja2018} with DR2 data for stars in the immediate Solar vicinity (within ${R_0}\pm200$ pc), is illustrated in panel d  of Fig.~\ref{figRVb} now with astrometry from EDR3. The morphological change of this phase spiral (or more precisely a slice of it, centred around $Z\sim0\kpc$, highlighted with brighter colours in panel d) is traced {for radii between 8 and 10} in panel e, by plotting $\VZ$ as a function of $R$ colour-coded by $\Vp$. Up to $R\sim{10}\kpc$, one can still see the different arms of the phase spiral at positive and negative $\VZ$ {(dotted lines linking panels d and e)}, with a diminishing envelope as one moves outwards, due to smaller restoring forces. While this has been observed already in \citet{Laporte2019} for discrete ranges of $R$, we see it here in a continuous way.
However, farther out than $R\sim{11}\kpc$, we see a clump (red colours) of large $\Vp$ and positive $\VZ$ dominating, which corresponds to one of the modes of the bimodality discussed above. Whether this is a manifestation of the same {spiral (but blurred since we are now considering a wide range of $Z$ given the cone geometry of the sample and because of the errors),} another phase spiral at larger radius or a different phenomenon {such as the warp } --perhaps with the same origin-- is not clear at this point.

In Appendix~\ref{app_material} we repeat some of the plots presented thus far for the GOG and UM samples {(Fig.~\ref{figRVlVb_GOGUM})}. From those we conclude that selection effects due to extinction can induce some features in projections such as $R$-$\Vp$ coloured as a function of $Z$. This is because a different extinction below and above the plane favours distinctly the different types of stars (different ages) that have different asymmetric drift (thus different $\Vp$) creating correlations between these variables. However,  we do not observe any special vertical kinematics for these features in the mock data. 
We also checked that the effects of the zero point in parallax does not induce or remove the features observed but merely change the distance scale with the pattern arriving farther or closer, independently whether a constant $ZP$ or $ZP_{56}$ is used (Fig.~\ref{figprofiles_zpts} in Appendix \ref{app_zpt}). Moreover, these features preferentially occupy positive or negative Galactic latitudes but do not correlate with the smaller scale checkered patterns seen in the astrometry. 
We note also that the stars participating in this phenomenon are relatively bright stars (Fig.~\ref{figRVlVb_phot} in the Appendix~\ref{app_material}), thus with good astrometry in general. Also the difference of 10$\kms$ seen in the velocities of the two distinct features mentioned above which are at a 
typical distance of 4 kpc, correspond to a proper motion difference of around 0.5 \masyr, which is much bigger than any known systematics.

 Finally, Fig.~\ref{fig:LzVz} shows the angular momentum-vertical velocity ($\Lz,\VZ$) space, {coloured by density in $\VZ$ at each $\Lz$ (top) and average $Z$ coordinate (bottom).}  
  In this plot, we see oscillations in $\VZ$ for the smaller $\Lz$ {(better seen in the top panel of  Fig.~\ref{vzprofL}, and seen also in \citealt{Huang2018} and \citealt{Cheng2020})} that  likely correspond to the vertical oscillations also seen in the {top right} panel of Fig.~\ref{fig:vphivzprof} at nearby Galactocentric radii. Most notably, these plots show that the clumpy features seen for $R>11.5\kpc$ in Fig.~\ref{fig:VphiVzclumps} correspond to
a clear break in the ($\Lz,\VZ$) {density} at $\sim2750\kms\kpc$ rather than a smooth transition. We note that when we separate our sample into young population (YP+EYP), main sequence (IP+OP) or Giants (Fig.~\ref{fig:LzVzPop} in Appendix~\ref{app_material}), this trend is seen for all the populations (albeit most clearly in the young one, as it has the lowest velocity dispersion)  implying that this break is most likely of dynamical origin.  In particular, the change in proportions between the two populations that we see in Fig.~\ref{fig:VphiVzclumps}  as we move outwards is related to the fact that the population with $\Lz\lesssim2750\kms\kpc$ (and $\VZ<0$) does not reach as large radii as the population with $\Lz\gtrsim2750\kms\kpc$ (and $\VZ>0\kms$). According to the bottom panel of  Fig.~\ref{fig:LzVz}, and as seen above, the part of the disc at lower angular momentum $\Lz$ corresponds to stars predominantly at positive $Z$ while the one with higher $\Lz$ mostly has  negative $Z$, though without perfect one-to-one correlation. 

 
 \subsection{Small scale velocity structures}\label{sect_substructure}
  
    \begin{figure}
   \centering
      \includegraphics[width=0.9\hsize]{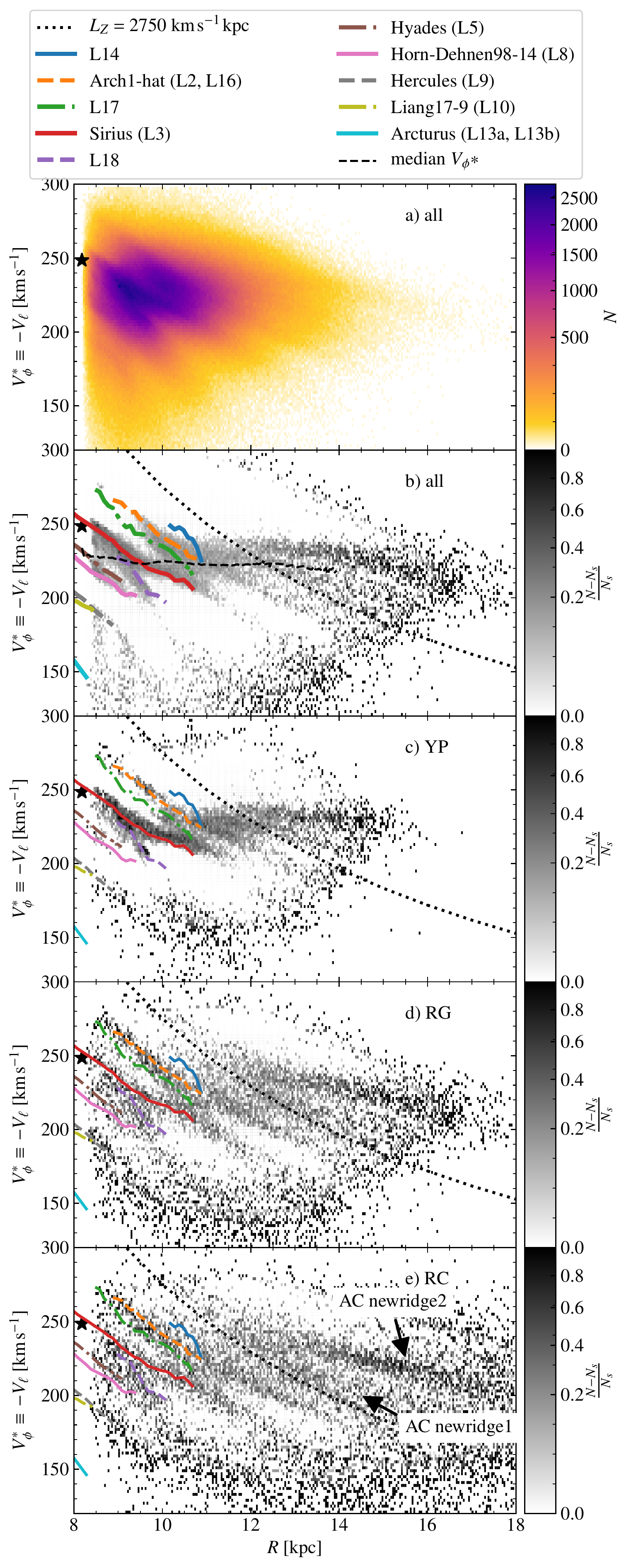}

         \caption{Substructures in the $R$-$\Vp$ projection in the anticentre direction. a) Number counts in the $R$-$\Vp$ plane in bins of size $\Delta R=0.02\kpc$ and $\Delta \Vp =1\kms$ for all stars in the \AC sample. b) Same but applying a substructure mask to highlight the ridges (see text). c-e) Same as b but for different stellar types. We also plot: some ridges from \citet{Ramos2018} with coloured lines, the separation of the bimodality (black dotted line), and the median velocity (black dashed line in panel b). We see the ridges extending beyond their limits in DR2 and new ridges resolved here for the first time.}
         \label{figridges}
   \end{figure}

  Apart from the two clumps discussed in Sect.~\ref{sect_assym}, finer substructures in the phase space of the disc can already be seen in the top panels of Fig.~\ref{figRVb} for nearby radii. 
  These structures are better visualised in Fig.~\ref{figridges} showing the 2-dimensional histogram of the $\Vp$-$R$ projection (panel a). Diagonal ridges (i.e. substructures with decreasing $\Vp$ as a function of $R$) can be seen, as already discovered in the \Gaia DR2 \citep[][]{Antoja2018,Kawata2018}. 
To enhance the contrast of these substructures, in panel b we show the density relative to a smoothed density obtained from a Gaussian filter $\frac{N-N_s}{N_s}$, where $N$ are the counts and $N_s$ are the smoothed counts with a $\sigma=5$ times the bin size {(}similar to what is done in \citealt{Laporte2019}). 
Panels c to e show this relative density for different stellar types. We do not note any difference between using $ZP={-17}\,\mu$as and $Z_{56}$ except for the already mentioned distance scaling.

The location of the main ridges obtained in \citet{Ramos2018}  with the DR2 \Gaia~RVS sample are over-plotted with colour lines in Fig.~\ref{figridges}b-e. {Following their notation,} we can identify the ridges associated to Hercules, Hyades, L18 (with a different slope compared to the rest) and one that could be linked to L16 or the so called hat \citep[e.g.][$V\sim40 \kms$ on their figure 22]{Katz2018} - also related to L14 and L17. 

Interestingly, for the YP the Sirius ridge appears to have slightly higher $\Vp$ velocities than the marked ridge (red line), as if following the asymmetric drift relations, and the ridges look thinner  {than in} the RG or RC {plots}. 
{We estimate the fraction of stars forming the ridges by calculating $\frac{\sum (N-N_s) >0}{\sum N_s}$. These fractions are 30\%, 13\%, 8\%, 8\% for the EY, YP, IP and OP, respectively. The fractions are 11\%, 14\% and 8\% for the RG, RC and all \AC stars, respectively. This fraction depends on the $\sigma$ used to smooth the distribution but the relative trends are the same, from which we see that the younger the population, the higher the fraction of stars in substructures.} 
On the other hand, we do not have enough stars in the lower $\Vp$ region in any of the populations to notice the low angular momentum ridges suggested in \citet{Laporte2020b}.

More importantly, in Fig.~\ref{figridges} the ridges are now seen at much larger distances than before. The Sirius ridge is detected  up to $R\sim12.5\kpc$, while in  \Gaia DR2 a sophisticated method to detect very low contrasts was needed to reach even 
$R\sim11\kpc$ \citep{Ramos2018}. We can also spot three ridges that reach outer regions of the disc, up to $16\kpc$ and beyond in the case of the RC. The one at lowest $\Vp$ could be the extension of L16. The other two were previously unknown and have been marked with arrows in the bottom panel (new anticentre ridges 1 and 2).
The new structures do not point towards $\Vp\sim {v_{c,\odot}+V_\odot}$  and $R={R_0}$ (black star in the panels) as expected for structures stretched by errors in distance (see Fig.~\ref{ellipses}). 
 In addition, we do not see any similar ridge induced by selection effects, uncertainties, or extinction, in the GOG equivalent sample.

In the panels b to e of Fig.~\ref{figridges} we {also} plot the line of angular momentum $\Lz=2750\kms\kpc$ (dotted black line) which marks the approximate separation of the bimodality described in Sect.~\ref{sect_assym}. While this line seems to coincide with the new anticentre ridge 1 (especially in panel e), {no} dynamical connection is clear at this stage. 
The median rotation velocities from Fig.~\ref{fig:vphivzprof} are over-plotted as a black {dashed} line in panel b and we see that the bump at around 10 kpc seems linked to the appearance of the L16 ridge that, with higher $\Vp$, moves the median curve slightly upwards. The connection between ridges and bumps in the rotation curve was already suggested by \citet{MartinezMedina2019,MartinezMedina2020}. The bump at 13 kpc could also be connected to the new anticentre ridge 2.

{In Fig.~\ref{figRVb}a, we see some correspondence between the median $Z$ and the density ridges seen in Fig.~\ref{figridges} (e.g. the white ridge in panel a with lower median $Z$ overlaps with the Sirius ridge). Similarly, in Fig.~\ref{figRVb}b the ridges exhibit a complex pattern of positive and negative vertical velocities, {thus indicating } 
 coupling between in-plane and off-plane kinematics. These effects were also noticed in \Gaia DR2 with the RVS sample \citep[][]{Khanna2019b,Laporte2019}, where the ridges were stronger at lower $Z$ and had some amplitude in $\VZ$ though typically lower than $5\kms$. }


\section{Halo, thick disc, and distant structures}\label{sect_haloouter}

In this section, we investigate several constituents of the Galaxy through the powerful combination of \Gaia astrometry and photometry. In Sect.~\ref{sect_halo} we look at the {stars} of high tangential velocity {which are contributed by} the halo and the hot thick disc and secondly, in Sect.~\ref{sect_outer} we explore the structures in the outer parts of the Galaxy disc.
  
\subsection{Halo and thick disc}\label{sect_halo}

Our goal in this section is to establish the extent and properties of the
{stellar} halo populations beyond the solar vicinity and
towards the galactic anticentre. To enhance the contribution of halo stars and partially mitigate the effects of
high-extinction near the disc plane, we used the {ACV} sample, defined in
Sect.~\ref{sect_selection}, with an additional selection of $|b| < 40^{\circ}$.
We selected on $\piepi > 5$ and compute distances as the inverse parallax. Since
we are interested in precise intrinsic colours and magnitudes, we chose only
stars that have $G$-band extiction $A_{G} < 1.0$. Here the extinction is
computed using the \citet{Schlegel1998} maps (with the correction of
\citealt{Schlafly11}) and a \citet{Cardelli89} extinction curve with $R_{V} =
3.1$. Although this extinction correction does not yield  intrinsic magnitudes
as accurate as in Appendix ~\ref{app_RCsel}, the main goal here is simply to
remove high-extinction regions from our analysis, while producing accurate
enough colours at large distances.

\begin{figure}
\centering
\includegraphics[width=\hsize]{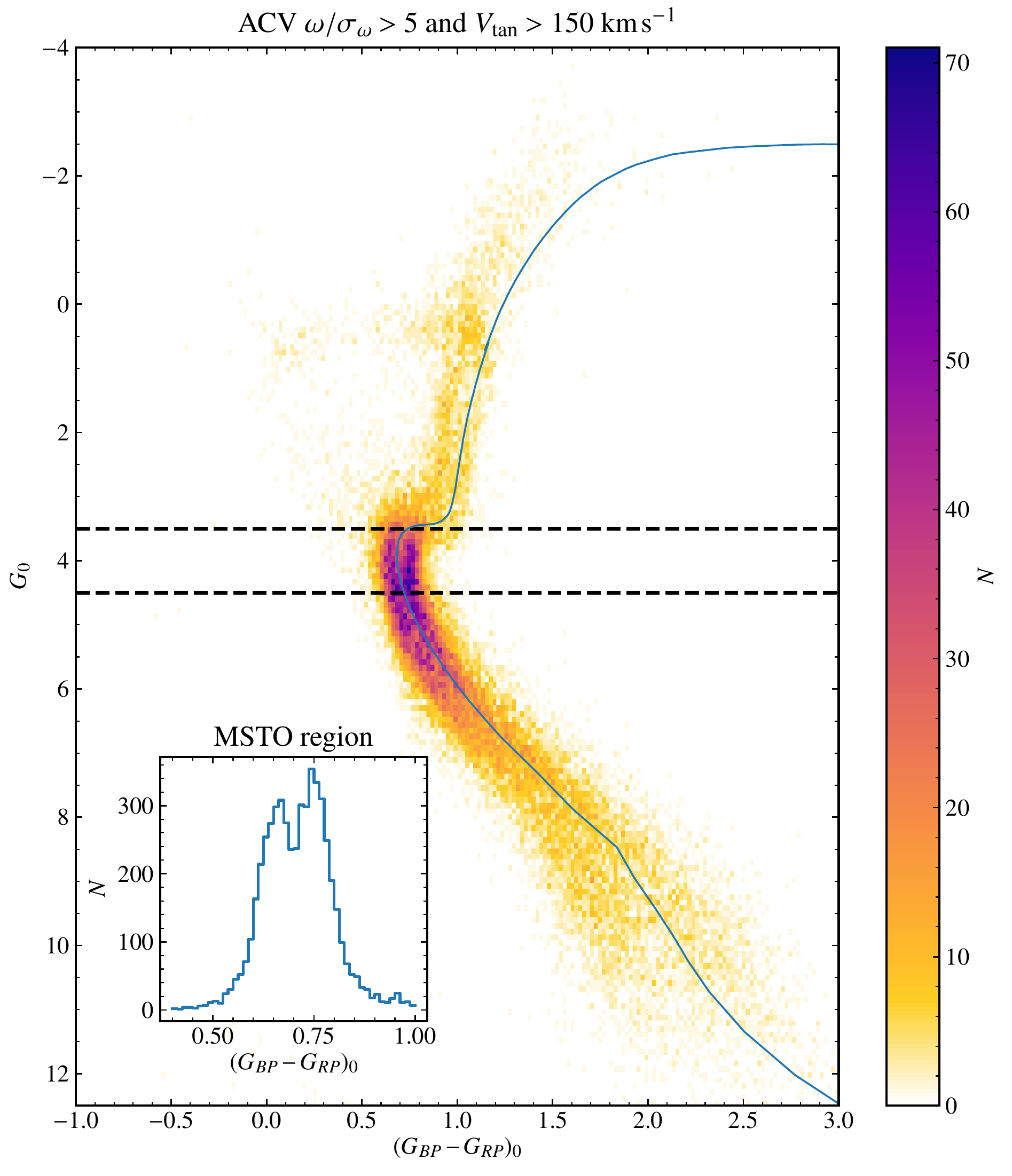}
\caption{Red and blue sequences for high tangential velocity stars. $\grp - \gbp$ vs $G$ Hess diagram for the ACV sample {with $|b| < 40^{\circ}$ and} with a $\piepi
    > 5$ and $\Vt > 150 \kms$. A \texttt{PARSEC} isochrone with ${\rm [M/H]} =
    -0.5$ and age of 11 $\Gyr$ is shown in blue (but shifted by
    0.04 in colour and 0.2 in magnitude in order to match the gap between blue
    and red sequences). The inset histogram shows the colour distribution in the
    magnitude range of the MSTO (shown as dashed lines in the main figure). A clear separation in two
    sequences is clearly seen as originally noted in \citet{Babusiaux2018} with
    DR2 data. An
    animated version of this figure for varying
    $\Vt$ limits is available at {\url{https://www.astro.rug.nl/~balbinot/files/ms.mp4}}. }
      \label{fig:double_seq_CMD}
\end{figure}

Following the approach of \citet{Babusiaux2018}, we focus on the HR diagram for
stars in the ACV sample that pass the cuts described above. We find
that when selecting only stars with high {heliocentric} tangential velocity $\Vt$
(Eq.~\ref{eq_vt}), two sequences arise as shown in
Fig.~\ref{fig:double_seq_CMD}, and that at the value of $\sim 150 \kms$ both
sequences seem to be found in equal numbers around the main-sequence turn-off
point. For completeness, see also our animation of how the HR diagram varies as
$\Vt$ is {increased} in 5 $\kms$ slices that is available online (see caption).
When $\Vt$ is low, there is a significant contribution from the thin
and the canonical thick discs, whereas at $\Vt \gtrsim 250$~km/s
mainly the blue sequence (that {locally is dominated by stars from the
accreted Gaia-Enceladus-Sausage}) is apparent.

In Fig.~\ref{fig:double_seq_CMD} the double sequence extends beyond
the turn-off point, but with fewer {luminous} stars in the
red sequence {than in} the blue one, suggesting that the distance
distribution of the two populations is different {(since at the largest
  distances only the brightest stars are apparent, and there are fewer
  of these on the red sequence)}. In order to select stars in either sequence we use a
\texttt{PARSEC} isochrone \citep{Bressan2012,Marigo2017} with
${\rm [M/H]} = -0.5$ and age of 11 $\Gyr$ (blue line). The isochrone
was shifted by 0.04 in colour and 0.2 in magnitude in order to match
the gap between blue and red sequences. Both the isochrone
and extinction coefficients use \gdrtwo~ transmission curves.

\begin{figure}
\centering
\includegraphics[width=\hsize]{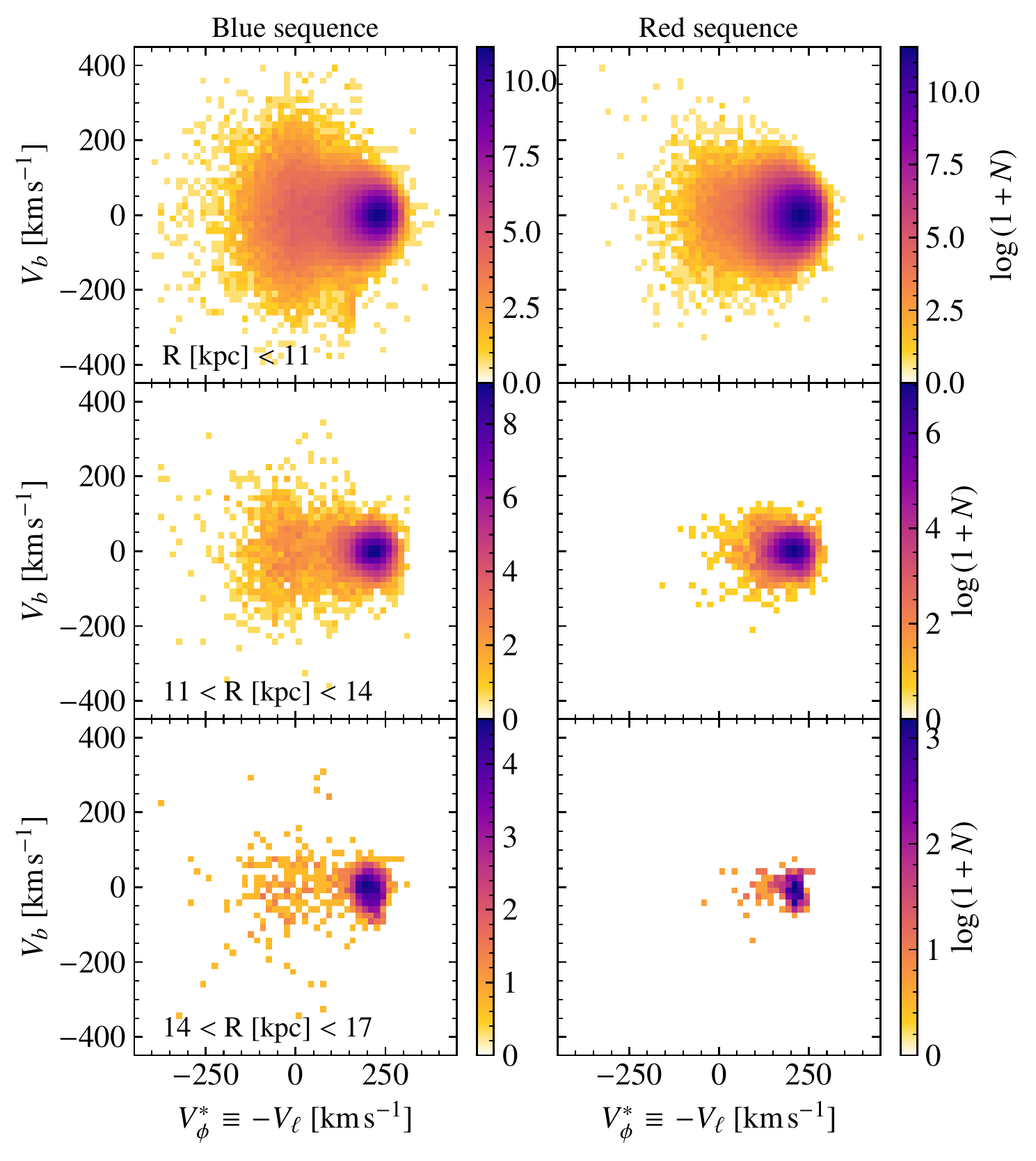}
\caption{Velocity distribution of the blue and red sequences. ${-\vl\sim\Vp}$ vs ${\vb}$
    distribution showing in the left (right) column the stars in the blue (red)
    sequence. Each row shows the distributions for a given distance slice,
    indicated in the left panels. The stars with low rotation (even
    the retrogrades ones) are far more prominent in the blue sequence and
    extend to larger Galactocentric radii.}
\label{fig:VlVbdist}
\end{figure}

We now explore the dynamical distributions of the stars belonging to
these two sequences in more detail. {To this end we explore the
    velocity distribution in $\vl$ and $\vb$ (Fig.~\ref{fig:VlVbdist}),
  for three cylindrical galactocentric distance bins for the stars in
  the blue (left) and red (right) sequences. {We note that at the higher latitudes within this sample $\VZ$ becomes a poorer approximation to $V_z$, but that is still a reasonable approximation to the non-radial, non-$V_\phi$ velocity of stars, and to low latitude stars. On the other hand, $\vl$ is good proxy for $V_\phi$ given the small range in $\ell$.} 
    The densest structure at
  $\Vp \sim 220 \kms$ is comprised mainly of disc stars, while the
  more extended and sparser structures belong to the halo and thick
  disc. Firstly, we note the presence of the \citet{Helmi1999} streams
  in the top-left panel at $(-\vl, \vb) \sim (150, -250) \kms$
  (indicating that these streams are a relatively local feature, in agreement with the results and
  predictions of \citealt{Koppelman2019}). We see, however, some hints
  of structures at similar velocities (and mirrored ones) in the {second left}
  panel of more distant stars that could
  potentially be related to these known streams. For the local sample
  (top panels) we observe a higher $\vb$ velocity spread for the
  halo (i.e. at $\vl \sim 0$) blue sequence stars compared to the red
  sequence. In the intermediate distance bin (middle panels) the
  velocity distribution of red sequence stars barely extends to
  $\vl\sim 0$ and for the most distant stars (bottom panels) only the
  blue sequence is apparent in the halo population, with the red
  sequence mostly appearing as a low-dispersion disc-like component.}

\begin{figure}
\centering
\includegraphics[width=\hsize]{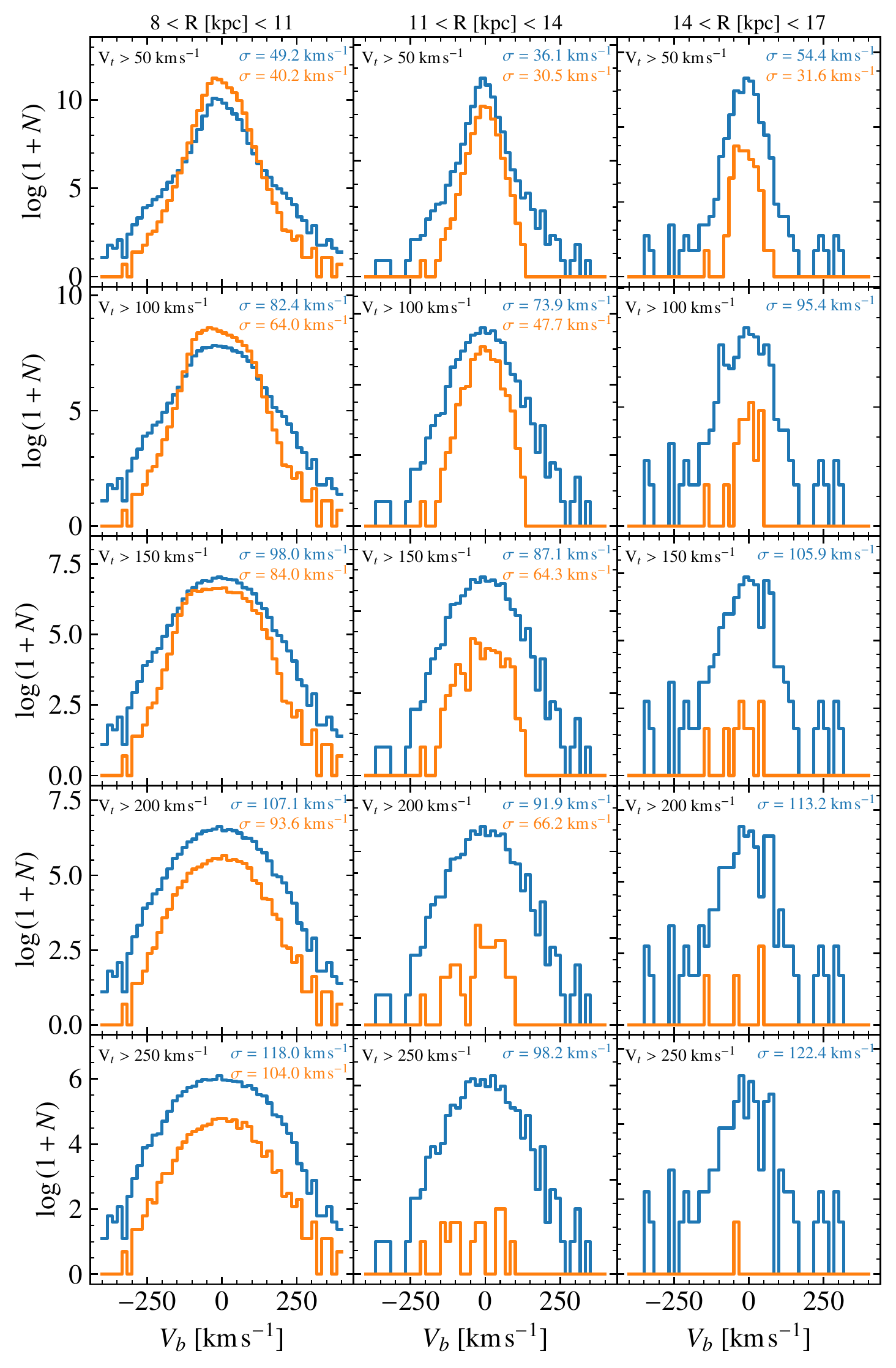}
\caption{Vertical {tangential} velocities ${\vb}$ for the blue and red sequences. The plots are 
    for different distance slices, indicated in the top of each column. Each
    row shows a different $\Vt$ selection, indicated in each panel. The blue
    (orange) curve shows the distribution for the blue (red) sequence. The
    different relative contribution of the sequences in the different panels is
    indicative of the spatial distribution of the accreted component and the
    ancient heated disc, and in particular of a shorter extent of the later
    one.}
\label{fig:Vbhist}
\end{figure}

{Similar conclusions can be drawn from Fig.~\ref{fig:Vbhist} which
  shows the distribution of $\vb$ velocities for the blue and red
  sequence for the same distance bins (columns) as in the previous
  figure, and for five $\Vt$ selections (rows).  We note again that
  the red-sequence distribution generally has a lower velocity
  dispersion than the blue sequence stars {(indicated with numbers for cases with at least 50 stars)}. We   also see that for
  $\Vt \gtrsim 100\kms$ the dispersion increases significantly,
  indicating the transition from the canonical thick disc to a hotter
  component. For the more distant bins, the contribution of
  {this} hot thick disc becomes smaller (bottom right panels), and it is
  basically absent beyond 14-17 kpc (whereas the canonical thick disc
  still is apparent in the top panels at these radii). On the other
  hand, the blue sequence is apparent at all radii, and has a relatively
  large $\vb$ velocity dispersion.} 

Therefore, the analyses presented in this section show that the hot
thick-disc component, {which locally has been associated with the
  heated disc at the time of the merger with Gaia-Enceladus-Sausage},
has a smaller extent presently than the canonical thick disc. This
{suggests} that the disc present at that time was smaller in size,
as indeed expected from cosmological models. A more quantitative
estimate of its size would require a careful assessment of the density
distribution of the older stars in the red sequence, which is beyond
the scope of this work. On the other hand, we see that the component
{locally} associated with Gaia-Enceladus-Sausage {extends} 
out to large distances from the Sun, as we detect the presence of a
retrograde component out to $\sim 17\kpc$ from the Galactic centre.

\subsection{Distant structures}\label{sect_outer}

 \begin{figure}
   \centering
   \includegraphics[width=1.\hsize]{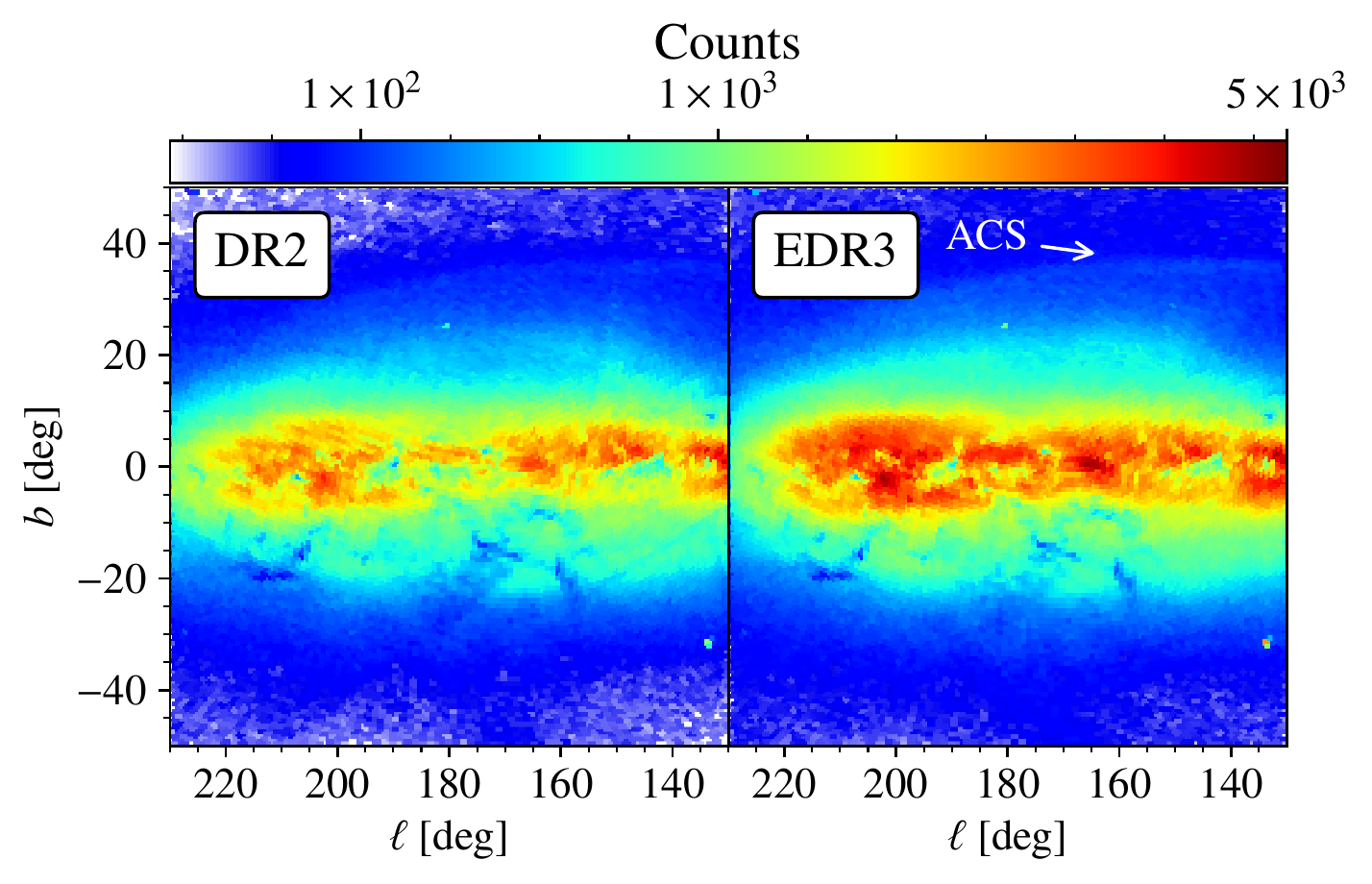}
            \caption{Counts in the sky for a selection of stars that favours the outer disc structures. The selection is for stars with $\varpi<0.1\mas$, $-1<\mu_\alpha*<1\mas\yr^{-1}$ and $-2<\mu_\delta<0\mas\yr^{-1}$. Left: DR2 (we observe marks of the scanning law). Right: EDR3 (with more stars and better homogeneity) {without filters nor parallax zero-point correction}. The ACS can be seen more clearly in the right panel.} 
         \label{fig:distant_counts}
   \end{figure}

 \begin{figure*}
   \centering
   \includegraphics[width=0.99\linewidth]{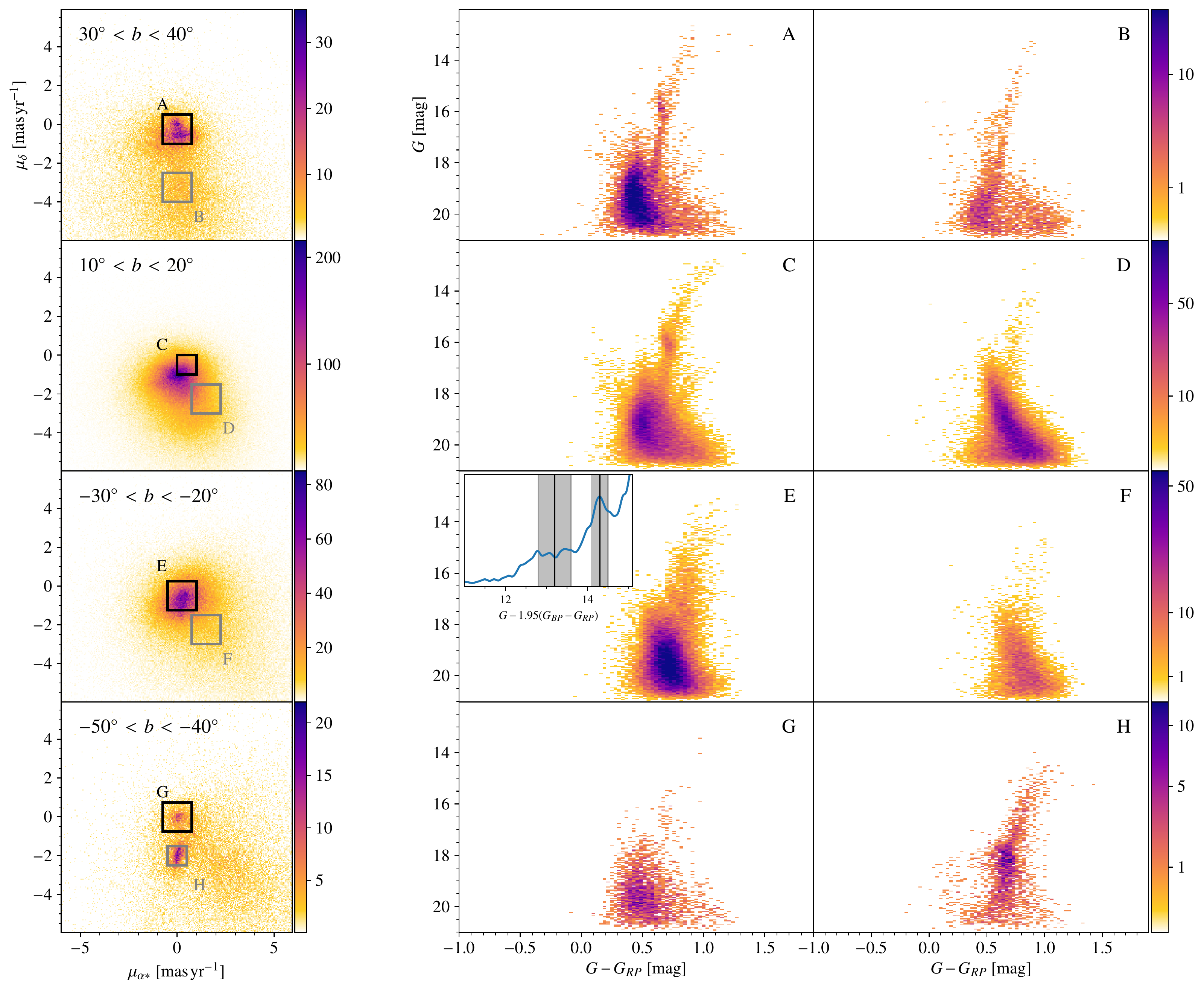}
         \caption{Colour-magnitude diagrams of different features in the anticentre. The diagrams are for the kinematic groups selected in proper motion in the range 170$^\circ$\,<\,$\ell$\,<\,190$^\circ$. Each row corresponds to a different latitude. The first column shows their distribution in proper motion and the definition of two regions of interested. The stellar population in {these selections are} shown in the second {and} third columns. We see structures such as ACS (A), Monoceros (C), the Sagittarius stream (H, C) and other outer disc structures (E). }
         \label{fig:distant_cmds}
  \end{figure*}

Studying the outskirts of the disc is a difficult task since the anticentre is mostly {outshone} by the nearby stars which are more numerous due to  both the density gradient of the Galaxy and the magnitude limitations inherent to any survey. The majority of the studies of the outer disc detected unexpected overdensities in counts {such as Monoceros and ACS} and focused on 
a specific stellar type, generally main sequence turn-off stars or M giants. An alternative way is now possible with \Gaiaf, which allows us to detect them by applying the right astrometric selection. First, we can  significantly reduce the amount of foreground contamination with a cut in parallax selecting only stars with $\varpi<0.1\mas$. By doing so, we guarantee that most of the stars closer than 10\,kpc are not selected, although the probability of failure is related to the parallax error of the source (the fainter sources being more likely to pass the filter regardless of their true distance). Then, we applied a kinematic selection since the proper motion signatures of these structures, given that they are relatively far from the Sun, are significantly different from the nearby disc and halo stars. The latter tend to have large proper motions due to the large relative velocity with respect to the Local Standard of Rest, while the former also tend to have large proper motions, but in this case due to the small heliocentric distance. Figure~\ref{fig:distant_counts} is an example of such parallax and kinematic selection (-1\,<\,$\mu_\alpha*$\,<\,1\,mas\,yr$^{-1}$ and -2\,<\,$\mu_\delta$\,<\,0\,mas\,yr$^{-1}$) where, in contrast to Fig.~\ref{fig_AC}, we can observe a perfectly defined and thin ACS{, as indicated by the arrow.} 
The difference between DR2 (left) and EDR3 (right) is clear: we now have more stars (7\,624\,697 compared to 5\,951\,302), mostly due to the higher completeness of stars with proper motions in EDR3, and the sample is less affected by the scanning law and other artefacts. 

In the first column of Fig.~\ref{fig:distant_cmds} we show the proper motion {2d} histograms for different slices in latitude around the anticentre (170$^\circ$\,<\,$\ell$\,<\,190$^\circ$) using the sample ACV{, now with the astrometric and photometric filters, as well as the parallax zero-point correction}. As we move from the north to the south Galactic hemisphere (top to bottom), different structures can be observed. 
 We examine them by selecting stars in the rectangles A to H and plotting their Colour-Magnitude diagrams (CMDs) in the second and third columns using $G-G_{RP}$ instead of $G_{BP}-G_{RP}$ since, as exposed in \citet{Riello2020} --see their Fig.\,26--, the flux in the $BP$ band can be overestimated for faint sources.
 
First, we note that the large concentration of sources close to the proper motion origin in the boxes A and G are mostly quasars for several reasons. Firstly, they are faint and too blue, with $G-G_{RP}\,<\,0.5$\,mag, which is equivalent, incidentally, to the cut used in \cite{Newberg2002} ($g-r\,<\,0.3$\,mag) to remove the SDSS quasars\footnote{We used the values in Table~5 from \cite{Jordi2010} to convert the SDSS colours to \Gaia colours.}. Secondly, the fraction of primary sources (\begin{tt}astrometric\_primary\_flag\end{tt}), a significant fraction of which are quasars (\citealt{Lindegren2018}), 
is abnormally high in both A and G. Finally,  {$\sim$32\%} of the sources in A and  {$\sim$75\%} in G are found in the {\begin{tt}agn\_cross\_id\end{tt} table}.

More interestingly, box A contains other kinematic structures apart from the aforementioned quasars. There is a more extended giant branch formed by the ACS \citep[c.f. Fig.~2 from][]{Laporte2020} and, tentatively, two fainter tips of a giant branch that could be related to the Sagittarius stream (similarly to box C, as explained below). The other box (B) at the same latitude corresponds to the distribution of halo stars, their proper motions larger due to the Sun's {reflex motion} and their CMD compatible with an old isochrone at $\sim$10\,kpc or farther, where stars accumulate due to our parallax cut. In the second row, panel C contains parts of both Monoceros, which provides the giant branch, a well defined RC and a very blue turn-off consistent with previous observations \citep[e.g.][]{Newberg2002,Yanny2003}, and the leading tidal tail of Sagittarius, which is only {evident} by its AGBs at magnitudes between 17 and 18. Boxes D and  F are dominated by the disc which, after the selection in parallax, is expected to have a thick main sequence created by faint dwarfs with large parallax uncertainties, and a few Red Giants\footnote{A RC star fainter than $G\sim$15\,mag at latitudes $b$>10$^\circ$ is bound to be higher than 2\,kpc from the disc, which is unlikely, but stars brighter than that tend to have a reliable parallax and are therefore more likely to be removed with our parallax cut.} above magnitude $G\sim$17\,mag.

In the south, at latitudes  $-30\deg<b<-20\deg$, we observe that the CMD of the small proper motion population is dominated by two RCs (panel E), the densest at magnitude $\sim$17\,mag and the other at $\sim$15.5\,mag. To confirm their existence, we have obtained the Gaussian kernel of $G^* = G-1.95(G_{BP}-G_{RP})$, therefore marginalising the apparent magnitudes along the extinction line (see Sect. \ref{sect_populations}). This kernel ({shown within panel E}) presents two {overdensities} corresponding to each of the mentioned RCs. By approximately selecting stars in these clumps and computing their distances assuming an absolute magnitude of the RC of $M_G = 0.495$ \citep{RuizDern2018} and the extinction by \citet{Schlegel1998} -and thus, upper limits-, we find that they are located at an average heliocentric distance of 9 and 14 kpc with variance of 3 and 2 kpc, for the bright and faint clumps, respectively. These corresponds to Galactocentric cylindrical radii of around 16 and 21 kpc, and heights below the plane of -4 and -6 kpc, respectively.
With this it is very likely that the bright RC corresponds to a nearby south extension of Monoceros \citep{Ibata2003,Li2017}, alternatively called south middle structure \citep{Xu2015}, at around $\sim12$ to 16 kpc from the Galactic centre. On the other hand, the faint RC could be related to TriAnd \citep{Rocha-Pinto2004, Xu2015,Li2017,Bergemann2018}, at a Galactocentric radius between 18 and 25. We note however that previous TriAnd detections were located around the range 100-$160\deg$ in $\ell$, thus not exactly in the anticentre direction, and our detection would then be a confirmation of the broadness of this structure and their extension up to $\ell=180$, albeit predicted by models such as that from \citet{Sheffield2018}.

\section{Clusters in the outer disc}\label{sect_clusters}

\begin{figure}
\begin{center} \resizebox{0.5\textwidth}{!}{\includegraphics[scale=0.8]{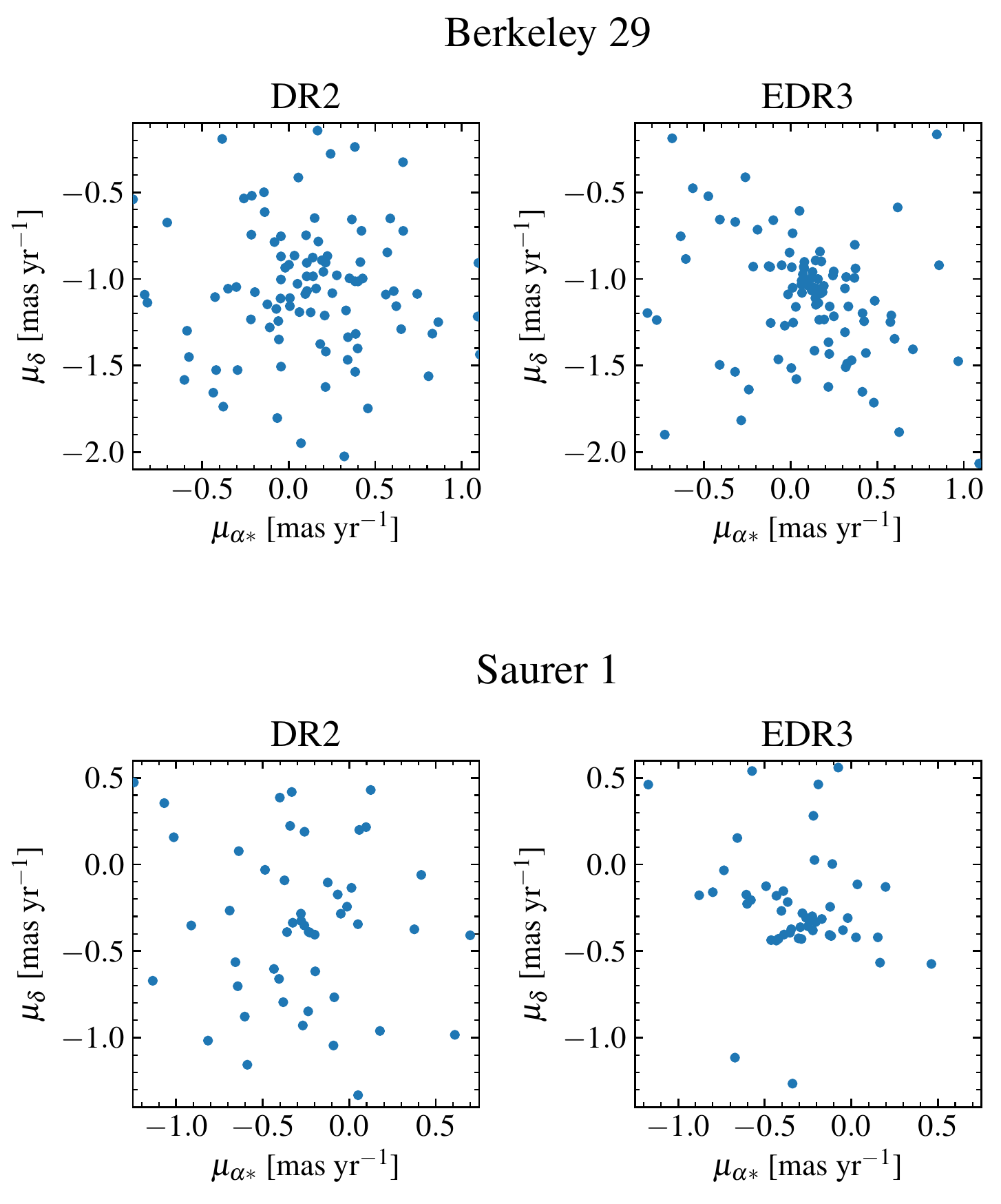}} 
\caption{Proper motions of the stars in Berkeley~29 and Saurer~1. The proper motions are  for Berkeley~29 (top) and Saurer~1 (bottom) for \textit{Gaia}~DR2 (left)
and \textit{Gaia}~EDR3 (right), for sources brighter than $G$=19 in the investigated field of view. The reduced uncertainties in EDR3 make the stars appear much more clumped than in DR2, allowing for a better selection of members and a better determination of the proper motion of the clusters.}\label{fig:pm_dr2_dr3} 
 \end{center}
\end{figure}

\begin{figure}[ht!]
\begin{center} \resizebox{0.5\textwidth}{!}{\includegraphics[scale=0.7]{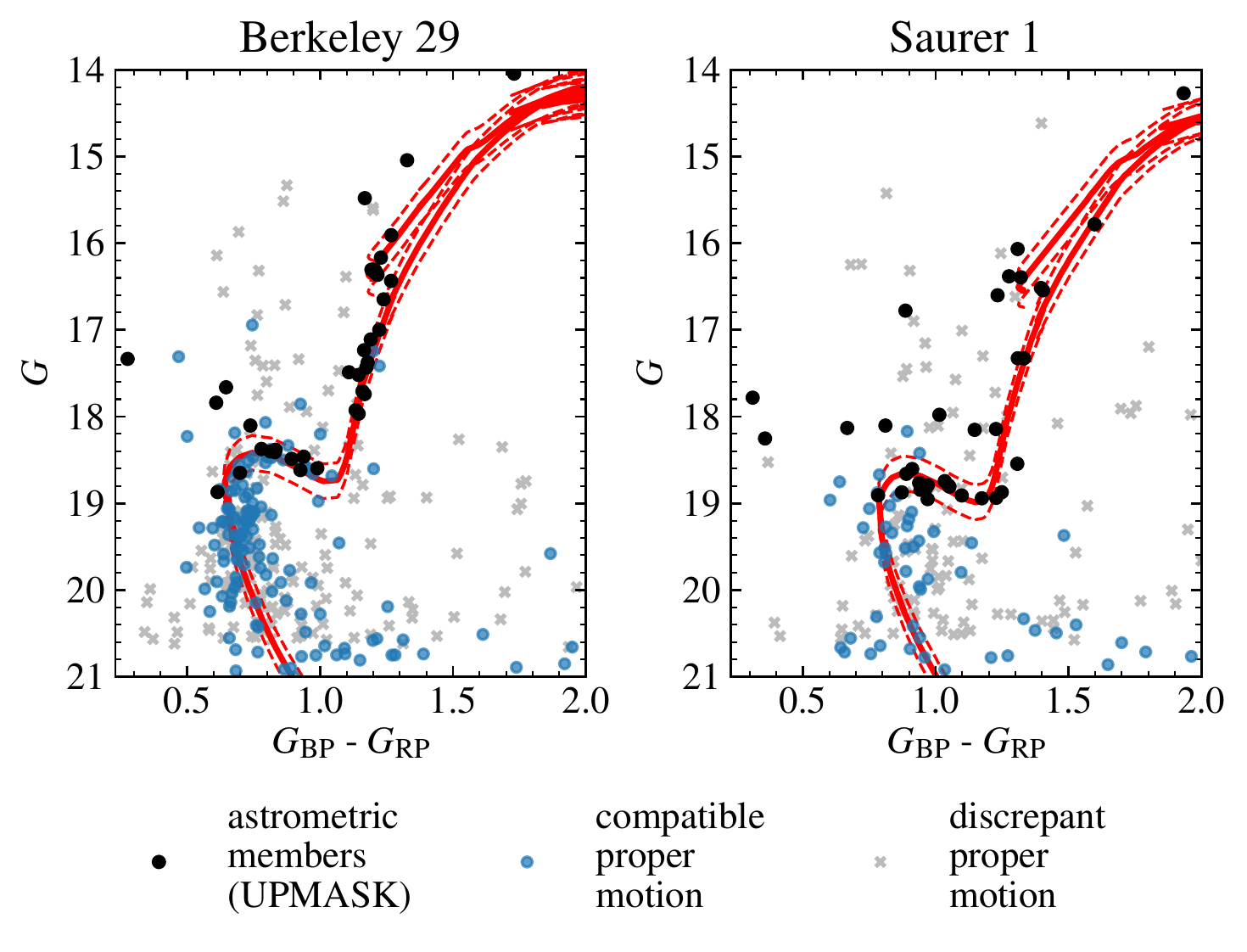}} \caption{\label{fig:be29sau1_CMDs} CMDs for Berkeley~29 and Saurer~1. The secure members identified with UPMASK are indicated and used to compute the mean proper motions. The blue points are sources with similar proper motions but large uncertainties, or magnitudes $G$>19. The red lines are PARSEC isochrones. The dashed lines correspond to offsets of $\pm$0.2\,mag in distance modulus.} \end{center}
\end{figure}

In this section, we investigate the peculiar clusters Berkeley~29 and Saurer~1. The \textit{Gaia}~EDR3 astrometric data allow us for the first time to perform a reliable member selection of these clusters and to constrain their proper motions in order to determine their orbits.
We retained all sources brighter than $G$=$19$ within 4\,arcmin of the cluster centres. The members were identified from their \textit{Gaia} proper motions and parallaxes with the unsupervised clustering procedure UPMASK \citep{KroneMartins14}.
\cite{CantatGaudin18}, {also} using UPMASK, analysed the stars brighter than $G=18$~mag of \textit{Gaia}~DR2 and 
detected Berkeley~29 but not Saurer~1. The improvement of \textit{Gaia}~EDR3 with respect to \textit{Gaia}~DR2
allows us to gain one magnitude and reliably detect both clusters. Figure~\ref{fig:pm_dr2_dr3} impressively shows how the stars in these clusters appear much more concentrated in proper motion space compared to DR2.
The CMDs of the clusters are shown in Fig.~\ref{fig:be29sau1_CMDs}, highlighting the sources that we consider the most secure members (with membership scores over 50\%).
We manually fitted PARSEC isochrones \citep{Bressan2012} to the observed CMDs. For Berkeley~29 we employed an isochrone with a metallicity $\feh=-0.5$ \citep{Yong05,CantatGaudin16}, an age $\log t$=9.55, and a distance modulus of 15.5\,mag with an extinction $A_V$ of 0.2\,mag.
For Saurer~1 we used an isochrone of metallicity $\feh=-0.4$ \citep{Carraro03,CantatGaudin16}, an age $\log t$=9.6, and a distance modulus of 15.4\,mag with an extinction $A_V$ of 0.4\,mag.

\begin{figure*}
\begin{center} \resizebox{0.95\textwidth}{!}{\includegraphics[scale=0.8]{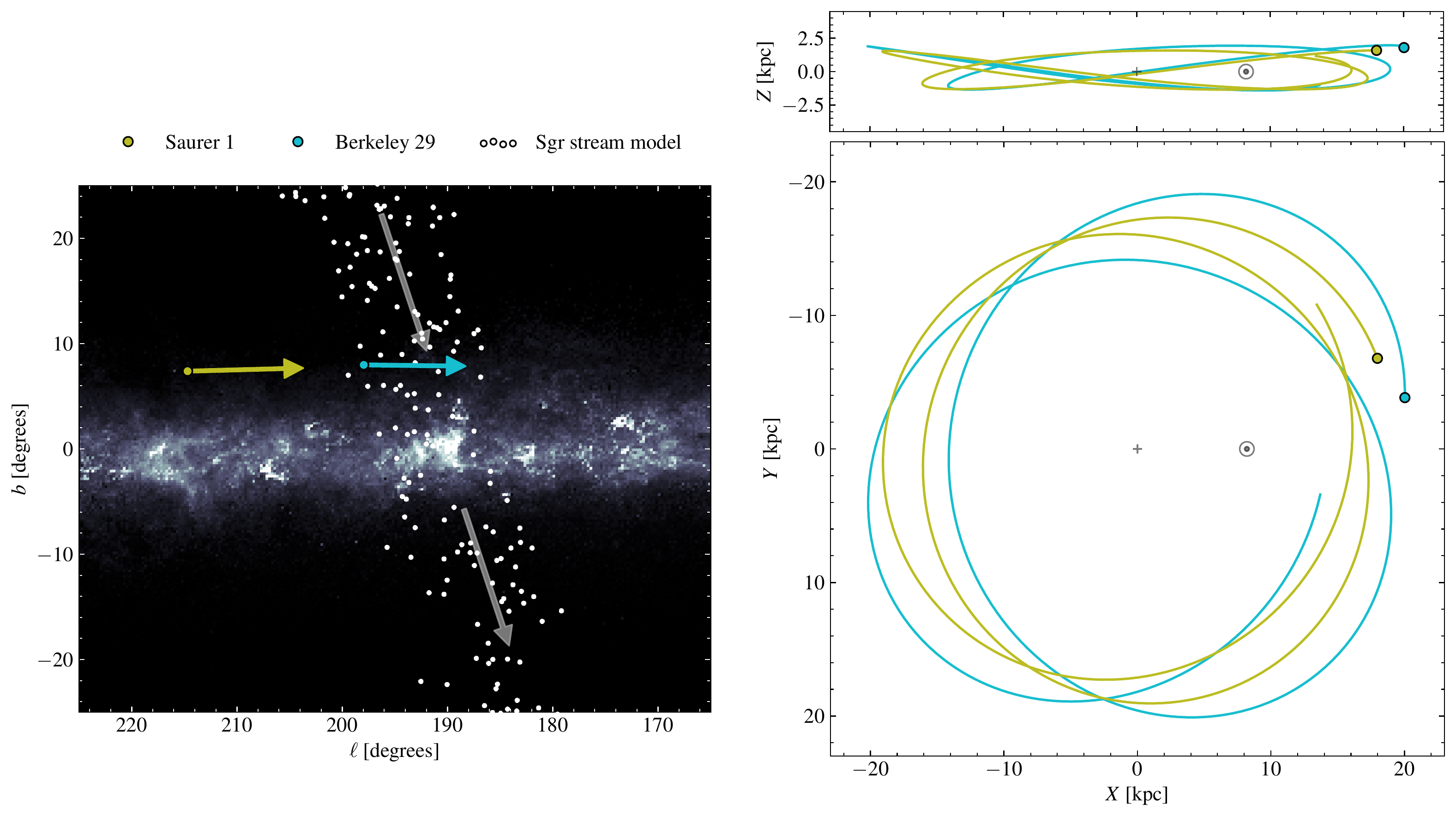}} 
\caption{Orbits if the Berkeley~29 and Saurer~1 clusters from EDR3 data. Left: location of Saurer~1 and Berkeley~29 in Galactic coordinates. The white symbols are the Sagittarius stream particles modelled by \citet{Law10}. The arrows indicate mean {tangential velocities} corrected for the solar motion {($\vl, \vb$)}. The background is the integrated extinction model of \citet{Green2015}, beyond 2\,kpc. Right: integrated orbits of Saurer~1 and Berkeley~29. The Sun's position and Galactic centre are indicated as the usual Sun's symbol and cross, respectively. We find  that the orbits of these clusters are very similar and typical of the disc.}\label{fig:be29sau1_3d} 
 \end{center}
\end{figure*}

The mean proper motions of the cluster members are $(\mu_{\alpha}*,\mu_{\delta})$=$(0.11,-1.05)$\,mas\,yr$^{-1}$ for Berkeley~29, and $(-0.26,-0.32)$\,mas\,yr$^{-1}$ for Saurer~1. {The mean tangential velocities} $\vl$ and $\vb$ {(Eq.~\ref{eq:vl}) are represented} in the first panel of Fig.~\ref{fig:be29sau1_3d}, along with {that} of the Sagittarius stream particles from the \citet{Law10} model.  In this panel, all proper motions were corrected 
from the effect of the Solar motion. 
The velocity vector of both clusters is mostly parallel to the Galactic plane, and differs significantly from that of the stream.

We used \texttt{galpy} MWPotential2014 model \citep{Bovy15} to integrate the orbits of these objects, shown in the left panels of Fig.~\ref{fig:be29sau1_3d}. For this, we supplemented the quantities derived from \textit{Gaia} data with {line-of-sight} velocities obtained from high-resolution spectra analysed by \citet{CantatGaudin16}. Their mean {line-of-sight} velocities are 24.8\,km\,s$^{-1}$ for Berkeley~29 (from eight stars), and 98.0\,km\,s$^{-1}$ for Saurer~1 (from two stars). All the stars they used to compute those mean velocities are part of the sample of secure members we obtained in the present study.  We estimated the uncertainty on the main orbital parameters by Monte-Carlo sampling of the uncertainties on the distance, {line-of-sight} velocity, and proper motion. We assumed an uncertainty of 0.2\,mag on the distance modulus, and 2\,km\,s$^{-1}$ on the {line-of-sight} velocities of both clusters. The precision on the mean cluster proper motion is limited by systematics, on the level of 11\,$\mu$as\,yr$^{-1}$ on each component of the mean cluster proper motion. All sampled orbits correspond to prograde, bound trajectories. The maximum altitude above the Galactic plane is $z_{max}$=1.80$^{+0.45}_{-0.09}$ for Berkeley~29, and $z_{max}$=1.59$^{+0.11}_{-0.09}$ for Saurer~1. They also exhibit small eccentricities $e$=0.03$^{+0.08}_{-0.01}$ and 0.05$^{+0.06}_{-0.05}$, respectively. Despite their large Galactocentric distance, the orbits of these clusters are typical of disc objects. We obtained similar results with the model by \citet{McMillan2017}.

\section{Discussion and conclusions}  \label{sect_conclusions}
\subsection{Summary of results}

With the combination of photometric and astrometric data from \Gaia EDR3, we explored the dynamics of different elements of the MW in the Galactic anticentre. The main results of this study are:

\begin{enumerate}
\item There are prominent oscillations in the median rotation and vertical velocities {and dispersions} of disc stars as a function of radius {and angular momentum} which depend on the evolutionary state of the stars (Sect.~\ref{sect_vphivprof}).
\item There are significant asymmetries in velocity when comparing stars above and below the standard Galactic plane for disc stars that can be as high as $5\kms$ for the vertical velocities and $10\kms$ for the rotation ones (Sect.~\ref{sect_assym}).
\item At the outer disc, stars are predominantly following a bimodal distribution, with a group of stars mostly below the plane moving upwards with velocities of $\sim10\kms$ and rotating faster by about $\sim30\kms$ than another group of stars predominantly above the plane moving downwards by 2-5 $\kms$ (Sect.~\ref{sect_assym}). 

\item The known $R$-$V_\phi$ ridges discovered with \Gaia DR2, reach larger Galactocentric radius with EDR3 (up to 14 kpc) and there are also new ridges up to about 16-18 kpc, that is much beyond the limits reached in previous studies ({Sect.~\ref{sect_substructure}}).

\item Galactic rotation is detected as far as 18 kpc from the Galactic centre, being this a lower limit on the current thin disc size (Sects.~\ref{sect_vphivprof} and \ref{sect_assym}).

\item {The red sequence of high tangential velocity stars (suggested to be }the ancient disc that was heated after the merger with Gaia-Enceladus-Sausage{) now is seen to} extend {out} to $\sim14$ kpc (Sect.~\ref{sect_halo}).

\item The {blue sequence (assumed to be the} debris of the Gaia-Enceladus-Sausage{)} is much more extended and can be detected at least beyond 17 kpc (Sect.~\ref{sect_halo}). 

\item The far anticentre shows a intricate superposition of structures in the proper motion and photometry diagrams including the leading (in the north) and trailing (in the south) Sagittarius stream, and  known outer disc structures such of Monoceros and ACS in the north (Sect.~\ref{sect_outer}). 

\item There are two structures at latitudes of $-30<b<-20\deg$  approximately at 9 and 14 kpc from the Sun, tentatively related to the Monoceros in the south and an extension of TriAnd in the anticentre direction, respectively (Sect.~\ref{sect_outer}). 

\item The clusters Berkeley~29 and Saurer~1, which are among the oldest open clusters known, are found to be on disc-like orbits despite being located at around 20 kpc from the Galactic centre (Sect.~\ref{sect_clusters}).

\end{enumerate}

\subsection{Discussion (I): MW disc dynamics}\label{sect_discussion1}

As in \Gaia DR2, the disc is found to be rather complex. 
Nearby, the rotation velocities are dominated by the {known} ridges {in the $(R,\Vp)$ plane}, which 
{are now detected up to 14 kpc} from the Galactic centre, that is 3 kpc farther than for DR2, while two additional ridges are discovered that reach 16-18 kpc. The overlap of distinct ridges in $R$ seems to be the cause of some oscillation seen in the rotation curve, as already suggested by \citet{MartinezMedina2019}, although they {could also be} related to the {physical} location of the spiral arms \citep{Sancisi2004,McGaugh2019}, {rather than their resonances {\citep{Barros2013}}}.

The most prominent nearby ridge is  Sirius, followed by the hat, L18, Hyades and Hercules. If indeed the Hercules, Sirius and hat ridges are signatures of the corotation, 4:1, and 2:1 Outer Lindblad Resonance of the bar \citep{Monari2019, Laporte2020}, respectively, there could be the 4:3 and 1:1 resonances beyond that \citep{Kawata2021} but perhaps we would require a different explanation for the new ridges beyond 12 kpc discovered here, 
either spiral structure {(with a lower pattern speed or transient)} or {the} perturbation from Sagittarius (or the two at the same time since perturbations from satellites inevitably induce density spirals and rings,  \citealt{Purcell2011}). 

{The  \Gaia EDR3 now allows for a full characterisation of the velocity ellipsoid and the asymmetric drift as a function of age and radius.} We see clear oscillations in  $\VZ$ with radius {(and angular momentum)} of an amplitude of 1-2 $\kms$ but increasing for younger stars.
  As already noticed before (e.g. {\citealt{Schonrich2018}; \citealt{Huang2018}}; \citealt{Beane2019}; {\citealt{Cheng2020}}) but now seen at a 
 higher precision with \Gaia EDR3, these oscillations could indicate a vertical wave propagating radially and are possibly associated to oscillations in the local mid-plane itself. The vertical velocity dispersions do not show the expected decreasing behaviour with radius but seem to {be flat or} increase and present very prominent oscillations that appear connected to the oscillations in the median velocities. 
  
In the outer disc ($R>12\kpc$), the velocity field is dominated by an upwards motion of about  $5\kms$. {Already seen in} \citet{Katz2018}, \citet{Wang2018},  \citet{Poggio2018}, \citet{RomeroGomez2019},  \citet{Carrillo2019}, \citet{LopezCorredoira2020}  {and \citet{Cheng2020}}, it has been associated to the warp that in the anticentre is near the line-of nodes, a bending wave {due to} Sagittarius, or to a disc that never achieved equilibrium.
Here, however, we go one step beyond and find, coexisting in $R$, {a bimodal distribution of stars moving vertically with opposite directions and with a different amount of rotation. }
 The feature can be observed also as a vertical velocity oscillation in angular momentum space, which can thus have different phases coexisting at the same $R$. This bimodality shows similar phase space correlations {to those of} the phase spiral \citep{Antoja2018}. Each group {of the bimodality} could  be interpreted as different wraps of a phase-mixing feature or a combination of bending waves. Missing data in this study such as {line-of-sight} velocities and chemistry will help in the understanding of this feature. The WEAVE Galactic Archaeology survey \citep{Dalton2016,Famaey2016} has a dedicated science case in the region of the anticentre to obtain {line-of-sight} velocities in complement to \Gaiaf, which will be crucial in this and many other aspects explored in this study. 
{Yet }our exploration reveals that simple 2d projections of phase space often do not capture the full complexity of the disc dynamics: when the vertical velocities are explored alone as a function of radius, only the upward motion (as in previous studies) is seen and adding more coordinates is necessary to observe this bimodality.

{To interpret the complex patterns observed, }a dynamical framework is required that no longer assumes decoupling between the vertical and horizontal movements \citep{DOnghia2016} and is capable of linking the small scale features such as the ridges, the global streaming motions, the phase spiral and perhaps structures such as the warp, {the flare and} the spiral arms. In any case, {we probably} live in a Galaxy with a highly perturbed outer disc {\citep[e.g.][]{Widrow2012,Wang2017,Antoja2018,Beane2019}} as seen in simulations of MW-like galaxies perturbed by a Sagittarius-like galaxy \citep{Purcell2011,Gomez2013,Gomez2017,Laporte2018} {having} rings with {non-null} vertical velocities, qualitatively comparable to what we find here. 

\subsection{Discussion (II): MW constituents}\label{sect_discussion2}

After \Gaia DR2, our understanding of {the} Galactic components has changed, in particular recognising that most of the (local) halo is made of debris from a single accretion event {(forming a blue sequence in the HR diagram of high transverse velocity stars)} and that we here find to be extended beyond the local neighbourhood at least up to distances of 17 kpc 
 from the {Galactic centre}. This is consistent with expectations from, for example, the orbit integrations of \citet{Deason2018}, but also emphasises the global importance of the debris. 
 {The redder component of the HR diagram } does not extend this far, having very few stars already around 14 kpc.
{If this redder component is the heated thick disc after the merger \citep[as claimed in][]{Helmi2018,DiMatteo2019,Gallart2019,Belokurov2020}, this would imply that it is more compact}
 than the canonical thick disc, which can be detected up to this radius. It will be interesting to try to constrain its initial properties, particularly through comparison to simulations of mergers and subsequent disc growth.

{In line with \citet{Carraro2010} and \citet{LopezCorredoira2018} that advocate a disc larger than the previously thought 12-14 kpc \citep[][and references therein]{Minniti2011}, we find evidence of circular rotation up to about} 18 kpc from the Galactic centre.  
{A more precise value for such} an important measurement needs {a} detailed analysis {of the effects of} the adopted zero parallax point, the biases on {any} distance estimate {(see Appendix \ref{app_distances} and \ref{app_zpt})} 
{and }
 aspects such as the flare \citep[e.g.][]{LopezCorredoira2014}. We compared the effects of a constant parallax offset of ${-17}\,\mu$as (the average offset of the quasars) with that of \citet{Lindegren2020} - a more sophisticated prescription as a function of magnitude, colour and ecliptic latitude. We find that the latter gives a more compressed distance scale (that propagates to velocities) but at this point it is not straightforward to claim that one prescription {is} better than the other \citep{Lindegren2020}. In any case, the features observed remain qualitatively the same regardless of the zero point.

\Gaia has also provided us with a window into the structures that dwell at the edge of the disc. We detected the Monoceros and the ACS above the disc plane and other structures in the south. Our southern detections are possibly related to the Monoceros south or south middle structure \citep[e.g.][]{Ibata2003,Xu2015} and TriAnd \citep[][]{Rocha-Pinto2004}, which have not been probed in detail at $\ell\sim180\deg$ so far due to the high extinction \citep[e.g.][]{Slater2014, Xu2015}. If confirmed, this would be the first TriAnd detection with \Gaia data {\citep[but see][]{Ramos2020b}} and the first time it is observed beyond its previously known longitude limit of $\ell\sim160\deg$. Curiously, {the part of the disc bimodality of stars below the plane moving upwards strongly}  at 12 kpc coincides in distance with {this} nearby southern structure, though the latter is at a lower latitude. The connection between these features certainly needs some attention. {Whether these structures are the corresponding northern and southern counterparts of the vertically oscillating disc (bending wave) expected in the scenario proposed by \citet{Widrow2012} and \citet{Yanny2013}, or they are individual rings or feather structures in the outermost parts of the disc as suggested in \citet{Ibata2003},  \citet{Kazantzidis2008}, \citet{Purcell2011}, \citet{Gomez2013} and \citet{Laporte2019a} also remains undetermined. In any case, future studies can benefit from the \Gaia  data that enable the kinematic selection of members of these features, providing a uniform sample of all the stellar types, and the determination of their proper motion. }

 Here we looked at two particular clusters, Berkeley~29 and Saurer~1, that due to their great distances from the Galaxy centre (around 20 kpc, derived photometrically and thus not affected by the parallax offset) and their old age (3-4 Gyr) probe extreme conditions in the Galaxy. 
 {Using an improved membership assignation and the better astrometry of \Gaia EDR3, we ascertain that the two clusters are on disc orbits, unlike what was claimed by previous studies \citep{Frinchaboy06orbit,Wu09,Carraro07,VandePutte10,Carraro09}}. 
Yet, their distant location makes us wonder whether the disc extends to such a distance or whether these clusters were brought there by other means {(radial migration, interaction with a satellite, expelled material from the disc)}. 
In particular Berkeley~29 has been already associated to Monoceros in \citet{Carraro2004} and in \citet{Frinchaboy04}, though in the latter case advocating for a stream origin. Our distances and proper motion of these clusters are compatible with the ones of Monoceros. Similarly, after examining the literature \citep{Rocha-Pinto2004,Li2012,CantatGaudin16,Sheffield2018}, we {note} that their chemistry and {line-of-sight} velocity are also broadly comparable. These clusters thus can be small but relevant pieces of information on the outer disc unknowns.

\subsection{Conclusion}

The quality of the EDR3 \Gaia data together with the advantage of having
astrometry and photometry from the same mission have allowed us to extend the  horizon for exploration towards the very end of the disc, travel to the past to explore its ancient components and detect its small constituents and  phase space features with better resolution. With a simple exploration of the \Gaia  data we find new complex patterns of movement in the outskirts of the Galactic disc, we estimate the extent of the ancient MW disc, show how the anticentre is a crossroad of structures likely both of internal and external origin, and uncover the nature of the orbits of two distant clusters. The anticentre is thus proven to be an excellent testbed region in the quest of deciphering the structure and history of our Galaxy that many astrophysicists are pursuing in the \Gaia era.

\begin{acknowledgements}

{We thank the referee Dr. James Binney for his comments.} This work has made use of data from the European Space Agency (ESA) mission
{\it Gaia} (\url{https://www.cosmos.esa.int/gaia}), processed by the {\it Gaia}
Data Processing and Analysis Consortium (DPAC,
\url{https://www.cosmos.esa.int/web/gaia/dpac/consortium}). Funding for the DPAC
has been provided by national institutions, in particular the institutions
participating in the {\it Gaia} Multilateral Agreement.


This work presents results from the European Space Agency (ESA) space mission \gaia. \gaia\ data are being processed by the \gaia\ Data Processing and Analysis Consortium (DPAC). Funding for the DPAC is provided by national institutions, in particular the institutions participating in the \gaia\ MultiLateral Agreement (MLA). The \gaia\ mission website is \url{https://www.cosmos.esa.int/gaia}. The \gaia\ archive website is \url{https://archives.esac.esa.int/gaia}.

The \gaia\ mission and data processing have financially been supported by, in alphabetical order by country:

the Algerian Centre de Recherche en Astronomie, Astrophysique et G\'{e}ophysique of Bouzareah Observatory;
the Austrian Fonds zur F\"{o}rderung der wissenschaftlichen Forschung (FWF) Hertha Firnberg Programme through grants T359, P20046, and P23737;
the BELgian federal Science Policy Office (BELSPO) through various PROgramme de D\'eveloppement d'Exp\'eriences scientifiques (PRODEX) grants and the Polish Academy of Sciences - Fonds Wetenschappelijk Onderzoek through grant VS.091.16N, and the Fonds de la Recherche Scientifique (FNRS);
the Brazil-France exchange programmes Funda\c{c}\~{a}o de Amparo \`{a} Pesquisa do Estado de S\~{a}o Paulo (FAPESP) and Coordena\c{c}\~{a}o de Aperfeicoamento de Pessoal de N\'{\i}vel Superior (CAPES) - Comit\'{e} Fran\c{c}ais d'Evaluation de la Coop\'{e}ration Universitaire et Scientifique avec le Br\'{e}sil (COFECUB);
the National Science Foundation of China (NSFC) through grants 11573054 and 11703065 and the China Scholarship Council through grant 201806040200;  
the Tenure Track Pilot Programme of the Croatian Science Foundation and the \'{E}cole Polytechnique F\'{e}d\'{e}rale de Lausanne and the project TTP-2018-07-1171 'Mining the Variable Sky', with the funds of the Croatian-Swiss Research Programme;
the Czech-Republic Ministry of Education, Youth, and Sports through grant LG 15010 and INTER-EXCELLENCE grant LTAUSA18093, and the Czech Space Office through ESA PECS contract 98058;
the Danish Ministry of Science;
the Estonian Ministry of Education and Research through grant IUT40-1;
the European Commission’s Sixth Framework Programme through the European Leadership in Space Astrometry (\href{https://www.cosmos.esa.int/web/gaia/elsa-rtn-programme}{ELSA}) Marie Curie Research Training Network (MRTN-CT-2006-033481), through Marie Curie project PIOF-GA-2009-255267 (Space AsteroSeismology \& RR Lyrae stars, SAS-RRL), and through a Marie Curie Transfer-of-Knowledge (ToK) fellowship (MTKD-CT-2004-014188); the European Commission's Seventh Framework Programme through grant FP7-606740 (FP7-SPACE-2013-1) for the \gaia\ European Network for Improved data User Services (\href{https://gaia.ub.edu/twiki/do/view/GENIUS/}{GENIUS}) and through grant 264895 for the \gaia\ Research for European Astronomy Training (\href{https://www.cosmos.esa.int/web/gaia/great-programme}{GREAT-ITN}) network;
the European Research Council (ERC) through grants 320360 and 647208 and through the European Union’s Horizon 2020 research and innovation and excellent science programmes through Marie Sk{\l}odowska-Curie grant 745617 as well as grants 670519 (Mixing and Angular Momentum tranSport of massIvE stars -- MAMSIE), 687378 (Small Bodies: Near and Far), 682115 (Using the Magellanic Clouds to Understand the Interaction of Galaxies), and 695099 (A sub-percent distance scale from binaries and Cepheids -- CepBin);
the European Science Foundation (ESF), in the framework of the \gaia\ Research for European Astronomy Training Research Network Programme (\href{https://www.cosmos.esa.int/web/gaia/great-programme}{GREAT-ESF});
the European Space Agency (ESA) in the framework of the \gaia\ project, through the Plan for European Cooperating States (PECS) programme through grants for Slovenia, through contracts C98090 and 4000106398/12/NL/KML for Hungary, and through contract 4000115263/15/NL/IB for Germany;
the Academy of Finland and the Magnus Ehrnrooth Foundation;
the French Centre National d’Etudes Spatiales (CNES), the Agence Nationale de la Recherche (ANR) through grant ANR-10-IDEX-0001-02 for the 'Investissements d'avenir' programme, through grant ANR-15-CE31-0007 for project 'Modelling the Milky Way in the Gaia era' (MOD4Gaia), through grant ANR-14-CE33-0014-01 for project 'The Milky Way disc formation in the Gaia era' (ARCHEOGAL), and through grant ANR-15-CE31-0012-01 for project 'Unlocking the potential of Cepheids as primary distance calibrators' (UnlockCepheids), the Centre National de la Recherche Scientifique (CNRS) and its SNO Gaia of the Institut des Sciences de l’Univers (INSU), the 'Action F\'{e}d\'{e}ratrice Gaia' of the Observatoire de Paris, the R\'{e}gion de Franche-Comt\'{e}, and the Programme National de Gravitation, R\'{e}f\'{e}rences, Astronomie, et M\'{e}trologie (GRAM) of CNRS/INSU with the Institut National Polytechnique (INP) and the Institut National de Physique nucléaire et de Physique des Particules (IN2P3) co-funded by CNES;
the German Aerospace Agency (Deutsches Zentrum f\"{u}r Luft- und Raumfahrt e.V., DLR) through grants 50QG0501, 50QG0601, 50QG0602, 50QG0701, 50QG0901, 50QG1001, 50QG1101, 50QG1401, 50QG1402, 50QG1403, 50QG1404, and 50QG1904 and the Centre for Information Services and High Performance Computing (ZIH) at the Technische Universit\"{a}t (TU) Dresden for generous allocations of computer time;
the Hungarian Academy of Sciences through the Lend\"{u}let Programme grants LP2014-17 and LP2018-7 and through the Premium Postdoctoral Research Programme (L.~Moln\'{a}r), and the Hungarian National Research, Development, and Innovation Office (NKFIH) through grant KH\_18-130405;
the Science Foundation Ireland (SFI) through a Royal Society - SFI University Research Fellowship (M.~Fraser);
the Israel Science Foundation (ISF) through grant 848/16;
the Agenzia Spaziale Italiana (ASI) through contracts I/037/08/0, I/058/10/0, 2014-025-R.0, 2014-025-R.1.2015, and 2018-24-HH.0 to the Italian Istituto Nazionale di Astrofisica (INAF), contract 2014-049-R.0/1/2 to INAF for the Space Science Data Centre (SSDC, formerly known as the ASI Science Data Center, ASDC), contracts I/008/10/0, 2013/030/I.0, 2013-030-I.0.1-2015, and 2016-17-I.0 to the Aerospace Logistics Technology Engineering Company (ALTEC S.p.A.), INAF, and the Italian Ministry of Education, University, and Research (Ministero dell'Istruzione, dell'Universit\`{a} e della Ricerca) through the Premiale project 'MIning The Cosmos Big Data and Innovative Italian Technology for Frontier Astrophysics and Cosmology' (MITiC);
the Netherlands Organisation for Scientific Research (NWO) through grant NWO-M-614.061.414, through a VICI grant (A.~Helmi), and through a Spinoza prize (A.~Helmi), and the Netherlands Research School for Astronomy (NOVA);
the Polish National Science Centre through HARMONIA grant 2018/30/M/ST9/00311, DAINA grant 2017/27/L/ST9/03221, and PRELUDIUM grant 2017/25/N/ST9/01253, and the Ministry of Science and Higher Education (MNiSW) through grant DIR/WK/2018/12;
the Portugese Funda\c{c}\~ao para a Ci\^{e}ncia e a Tecnologia (FCT) through grants SFRH/BPD/74697/2010 and SFRH/BD/128840/2017 and the Strategic Programme UID/FIS/00099/2019 for CENTRA;
the Slovenian Research Agency through grant P1-0188;
the Spanish Ministry of Economy (MINECO/FEDER, UE) through grants ESP2016-80079-C2-1-R, ESP2016-80079-C2-2-R, RTI2018-095076-B-C21, RTI2018-095076-B-C22, BES-2016-078499, and BES-2017-083126 and the Juan de la Cierva formaci\'{o}n 2015 grant FJCI-2015-2671, the Spanish Ministry of Education, Culture, and Sports through grant FPU16/03827, the Spanish Ministry of Science and Innovation (MICINN) through grant AYA2017-89841P for project 'Estudio de las propiedades de los f\'{o}siles estelares en el entorno del Grupo Local' and through grant TIN2015-65316-P for project 'Computaci\'{o}n de Altas Prestaciones VII', the Severo Ochoa Centre of Excellence Programme of the Spanish Government through grant SEV2015-0493, the Institute of Cosmos Sciences University of Barcelona (ICCUB, Unidad de Excelencia ’Mar\'{\i}a de Maeztu’) through grants MDM-2014-0369 and CEX2019-000918-M, the University of Barcelona's official doctoral programme for the development of an R+D+i project through an Ajuts de Personal Investigador en Formaci\'{o} (APIF) grant, the Spanish Virtual Observatory through project AyA2017-84089, the Galician Regional Government, Xunta de Galicia, through grants ED431B-2018/42 and ED481A-2019/155, support received from the Centro de Investigaci\'{o}n en Tecnolog\'{\i}as de la Informaci\'{o}n y las Comunicaciones (CITIC) funded by the Xunta de Galicia, the Xunta de Galicia and the Centros Singulares de Investigaci\'{o}n de Galicia for the period 2016-2019 through CITIC, the European Union through the European Regional Development Fund (ERDF) / Fondo Europeo de Desenvolvemento Rexional (FEDER) for the Galicia 2014-2020 Programme through grant ED431G-2019/01, the Red Espa\~{n}ola de Supercomputaci\'{o}n (RES) computer resources at MareNostrum, the Barcelona Supercomputing Centre - Centro Nacional de Supercomputaci\'{o}n (BSC-CNS) through activities AECT-2016-1-0006, AECT-2016-2-0013, AECT-2016-3-0011, and AECT-2017-1-0020, the Departament d'Innovaci\'{o}, Universitats i Empresa de la Generalitat de Catalunya through grant 2014-SGR-1051 for project 'Models de Programaci\'{o} i Entorns d'Execuci\'{o} Parallels' (MPEXPAR), and Ramon y Cajal Fellowship RYC2018-025968-I;
the Swedish National Space Agency (SNSA/Rymdstyrelsen);
the Swiss State Secretariat for Education, Research, and Innovation through
 the ESA PRODEX programme,
the Mesures d’Accompagnement, the Swiss Activit\'es Nationales Compl\'ementaires, and the Swiss National Science Foundation;
the United Kingdom Particle Physics and Astronomy Research Council (PPARC), the United Kingdom Science and Technology Facilities Council (STFC), and the United Kingdom Space Agency (UKSA) through the following grants to the University of Bristol, the University of Cambridge, the University of Edinburgh, the University of Leicester, the Mullard Space Sciences Laboratory of University College London, and the United Kingdom Rutherford Appleton Laboratory (RAL): PP/D006511/1, PP/D006546/1, PP/D006570/1, ST/I000852/1, ST/J005045/1, ST/K00056X/1, ST/K000209/1, ST/K000756/1, ST/L006561/1, ST/N000595/1, ST/N000641/1, ST/N000978/1, ST/N001117/1, ST/S000089/1, ST/S000976/1, ST/S001123/1, ST/S001948/1, ST/S002103/1, and ST/V000969/1.
 \end{acknowledgements}

%
%

\bibliographystyle{aa}
\bibliography{mybib}

\begin{appendix}

\section{Queries to the \Gaia Archive}\label{queries}

In this appendix we show a few examples of queries to the \Gaia Archive \url{https://gea.esac.esa.int/archive/} to retrieve the data:

\begin{lstlisting}[title=A,label={q1},caption=An example of query to retrieve stars in the rectangular sky patch of the AC20 sample.]
SELECT * from gaiaedr3.gaia_source WHERE l<190 and l>170 and b>-10 and b<10 } 
\end{lstlisting}

\begin{lstlisting}[title=B,label={q2},caption=An example of query to retrieve the number of stars and average quantities in all healpix of level 8 inside a rectangular patch in the sky.]
SELECT sub.healpix_8,COUNT(*) as N,AVG(phot_g_mean_mag) as avg_g,  AVG(visibility_periods_used) as avg_vp FROM (SELECT gaia_healpix_index(8, source_id) AS healpix_8,phot_g_mean_mag,visibility_periods_used FROM user_edr3int4.gaia_source WHERE l<240 AND l>120 AND b<60 and b>-60 AND ruwe < 1.4) AS sub  GROUP BY sub.healpix_8
\end{lstlisting}

\section{Selection of red clump stars}\label{app_RCsel}

In this appendix we describe the selection of the RC sub-sample. 
 First, in order to compute the absolute magnitude, we need good estimates of the extinction A$_{\lambda}$ in band ${\lambda}$. For each star, one could in principle use the 2D ($l,b$) maps of reddening, $E(B-V)$,  from \citet{Schlegel1998} which estimates the extinction at infinity. However, these 2D extinction values will overestimate the reddening. Since, we have parallax information for our sample, we can use this as a prior for distance and estimate the 3D extinction. For this, we made use of the 3D dust-reddening maps from \bayestar{} \citep{Green2019}. These are derived using a Bayesian scheme that combines \Gaia parallaxes with photometry from the \twomass{} and \panstars{} surveys, and covers the sky north of declination of $-30^\circ$. Only 3 stars in our AC20 sample are missing from \bayestar. The multiplicative factor ($f_{\lambda}$) between reddening and extinction that we use is listed in \autoref{tab:tab_extinct} for various bands. 

For the RC selection, we first apply the following photometric cuts:
\begin{equation}
BP -G > 0.6, BP-RP_{0} > 0.91 .       
\end{equation}Then, for each star, we compute the absolute magnitude (\absmag) in each of the \twomass{} bands, and in Gaia $G$:
\begin{equation}
\label{eqn:absmag_cal}
M_{\lambda} = m_{\lambda} - A_{\lambda} - {\rm dmod} ,   
\end{equation}using dmod = 5 $log_{10} (100 /\varpi^{'} [\rm mas])$. Here, $\varpi^{'}$ is the parallax corrected for the offset of {-17} $\mu as$. In \autoref{tab:tab_extinct} we list the  literature absolute magnitude ($\Bar{M_{\lambda}}$) and dispersion in various photometric bands for the RC population. Using this, for each star we can write down a likelihood function per bandpass i, and take their product

\begin{equation}
P_{RC}(m_{\lambda}, A_{\lambda},\varpi) = 
\prod\limits_{i} \sqrt{2\pi}\sigma_{\Bar{M_{\lambda}}} \mathcal{N}(M_{\lambda}||\Bar{M_{\lambda}},\sigma_{M_{\lambda}}^{*}),
\end{equation}

where  $\sigma_{M_{\lambda}}^{*} =  \sqrt{\sigma_{er, M_{\lambda}}^{2} +  \sigma_{\Bar{M_{\lambda}}}^{2}} $ combines the propagated error in the absolute magnitude from \autoref{eqn:absmag_cal},  $\sigma_{er, M_{\lambda}}$,  and the dispersion in the true absolute magnitude, $\sigma_{\Bar{M_{\lambda}}}$.

\begin{table}
\centering
\caption{Median absolute magnitude  $\Bar{M_{\lambda}}$, and dispersion in absolute magnitude $\sigma_{\Bar{M_{\lambda}}}$  for RC stars selected from \citet{2017arXiv170508988H}. Also listed are the extinction factors ($f_{\lambda}$) for the four passbands used, with the \twomass{} values taken from \citet{Green2019} and \esagaia{} from  \citet{2018MNRAS.479L.102C}.\label{tab:tab_extinct}}
\begin{tabular}{llll}
\hline
Passband ($\lambda$) & $\Bar{M_{\lambda}}$ & $\sigma_{\Bar{M_{\lambda}}}$ & $f_{\lambda} = \frac{A_{\lambda}}{E(B-V)}$ \\
\hline
$J$        & $-0.93 \pm 0.01  $ & $0.20 \pm 0.02  $ &  0.7927 \\	
$H$        & $-1.46 \pm 0.01  $ & $0.17 \pm 0.02 $ &  0.469 \\
$K$        & $-1.61 \pm 0.01  $ & $0.17 \pm 0.02  $ &  0.3026 \\
$G$ & $+0.44 \pm 0.01 $ & $0.20 \pm 0.02 $ & 2.74  \\
$G_{BP}$ & - & -  & 3.374  \\
$G_{RP}$ & - & -  & 2.035  \\
\hline
\end{tabular}
\end{table}

\begin{figure}
\includegraphics[width=1.\columnwidth]{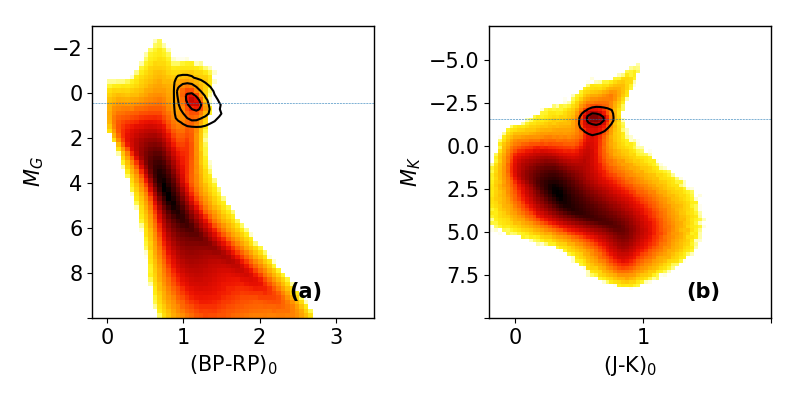} 
\caption{HR diagrams for the AC20 sample, with contours marking the RC selection.  \label{fig:absmag_rc_sel_gmag}}
\end{figure}

\begin{figure}
\includegraphics[width=1.\columnwidth]{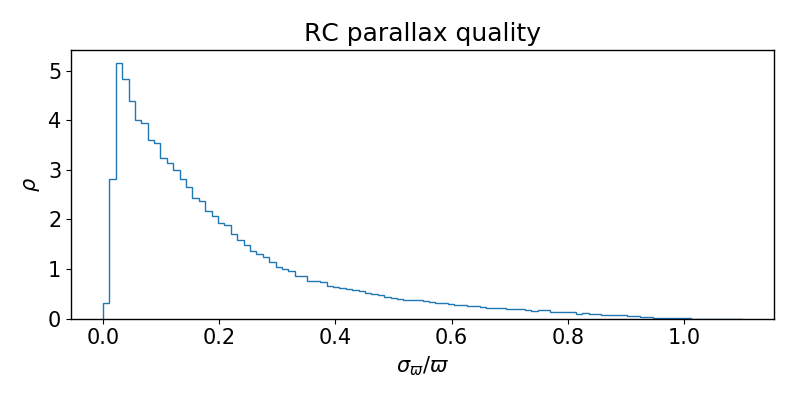} \caption{Parallax error quality of the RC sample. The tail of the distribution extends down to $\sigma_{\varpi}/\varpi \approx 0.8$. \label{fig:perror_compare}}
\end{figure}

For any distribution, the distance between the centroid ($x_{0}$) and a point of interest ($x_{1}$) can be given in terms of its Mahalanobis distance ($ML$) \begin{eqnarray}
ML^{2} = (x_{1} - x_{0})^{T} \Sigma^{-1} (x_{1} - x_{0}),  
\end{eqnarray} that respects the combined covariance of $x_{0}$ and $x_{1}$, which we have written as $\Sigma$. Essentially, $ML$ is a measure of the distance from the centroid in units of the standard deviation. Then, we can define a p-value, that is the probability of finding a value of  $ML^{2}$ or more extreme under the null-hypothesis of the star not being part of the RC, from a chi-square distribution, and select those stars for which the probability ($P_{RC}$) is greater than the p-value:
\begin{eqnarray}
P_{RC} > 1 - P[\chi^{2} \leq ML^{2}].
\end{eqnarray} In this work we limited our analysis to a maximum of two bands, namely, \gaia $G$ and \twomass{} $K$. So, we used a chi-square distribution with 2 degrees of freedom, and $ML$ is essentially the confidence level used to set a minimum probability threshold. The tolerance parameters used in our selection is shown in \autoref{tab:tab_parameters}, and we obtain a high quality RC sample of $N_{RC}=${121857}. The HR diagram with our RC selection is shown in \autoref{fig:absmag_rc_sel_gmag}. The parallax quality for the selection is shown in \autoref{fig:perror_compare}, with the tail of the distribution extending down to $\sigma_{\varpi}/\varpi \approx 0.8$.

\begin{table}
\centering
\caption{Parameters used for selecting the $RC$ population. $ML$ is essentially the confidence level used to set a minimum probability threshold ($P_{\rm RC} >$). Finally, $N_{\rm RC}$ gives the resulting number of stars classed as RC that lie between $170^\circ<l<190^\circ$ and $|b|<10^\circ$.}\label{tab:tab_parameters}
\begin{tabular}{llll}
\hline
$ML$ & $N_{\rm RC}$ & $P_{\rm RC} >$ & band(s)\\
\hline
 3 & {121857}  & 0.01 & \twomass{} $J$, \esagaia{} $G$ 	\\
\hline
\end{tabular}
\end{table}

In \autoref{fig:absmag_hist_gmag}, we further inspect the RC selection. Panels(a-b) show the absolute magnitude distribution in $G$ and $K$ bands. We find that the median absolute magnitudes for our sample is offset from their literature values by -0.05 ($G$) and 0.05 ($K$) in the two bands. The yellow curves use the \bayestar{} reddening, but we also show the distribution for absolute magnitudes computed with $A_{\lambda}=0$, just to illustrate that our extinction correction shifts the distribution in the correct direction.
In Panels (c-d), we compare the distances computed as in \autoref{app_RCdist}, against inverse parallax. It is encouraging to see that the running median for nearby stars lies on the 1:1 line. This is further shown in panels (e-f), where we look at the relative difference between the two distance estimates. Compared to inverse parallax method, our distances are slightly under (over) estimated in $G$ ($K$) beyond 5 kpc from the Sun. This is likely due to the fact that distance modulus $\propto$ -$M_{\lambda}$. Since the literature absolute magnitudes are slightly offset, this would result in smaller distances, but the effect is minor given the small offset, especially in the $K$ band. 
Finally, in panels (g-h), is shown the distance error as a function of distance. The errors in the inverse parallax distances are quadratic with $d$, while the trend is linear for the {RC} distances. Beyond, $d>5$ kpc, the errors in inverse parallax grow significantly, while for RC distances, the prediction is $\sigma_d{} ~ 1.5 $ kpc at 10 kpc. The distribution in Galactocentric cylindrical radius $R$ is shown in \autoref{fig:nbstarsperradialbin}. Our sample extends out to $R
\sim$17 kpc, consisting of about 1000 stars at that distance.

\begin{figure}
\includegraphics[width=1.\columnwidth]{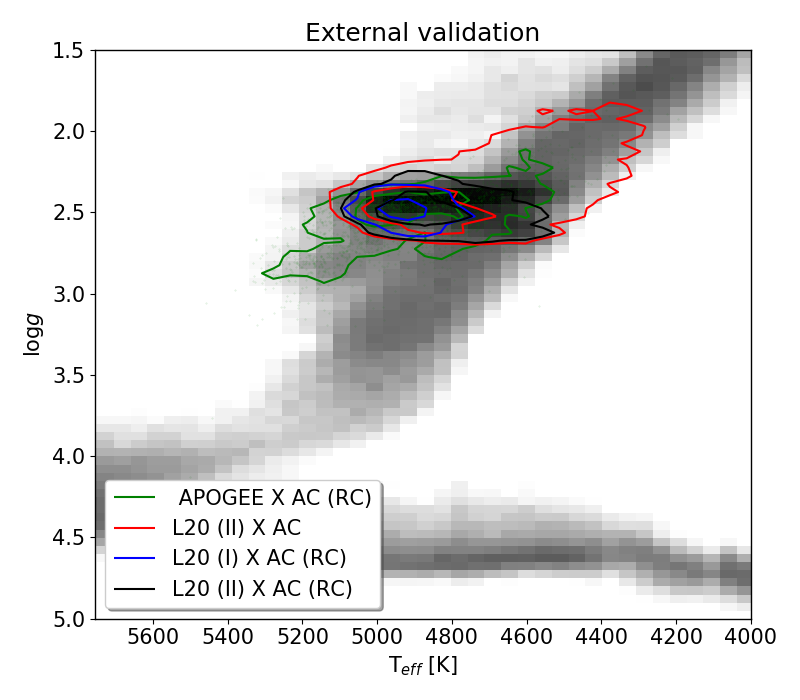}
\caption{External validation comparing the distribution of the AC20 RC stars on a \kiel{} diagram. In grey is the full distribution from {\sl APOGEE}-DR16, and the green contours show common stars between our RC sample and {\sl APOGEE}-DR16. Red contours show common stars between \lucey{} and the entire anticentre sample used here. The blue contours show common stars between \lucey{}-`Tier I' (i.e. 20\% contamination) and our RC sample. The black contours show common stars between \lucey{}-`Tier II' (i.e. 33\% contamination) and our RC sample.\label{fig:validate_galah_gmag}}
\end{figure}

\citet{Lucey2020} recently put out a catalogue of 2.6 million RC stars. Their method involves predicting asteroseismic parameters ($\Delta P, \Delta \nu$) and stellar parameters (\logg{},$T_{\rm eff}$) from spectral energy distributions (SED). They combined photometry from \panstars{}, \wise{}, \twomass{} and \Gaiaf. In their catalogue (hereafter \lucey) they classified RC stars with contamination rate of $\approx33\%$ as `Tier II', and a superior subset with contamination rate of $\approx20\%$ as `Tier I'. In \autoref{fig:validate_galah_gmag} we show the distribution of our sample on a $Kiel$ diagram by cross-matching with the \lucey{} catalogue. We notice that their `Tier I' sample does not have too many cooler stars. Conversely, their less stringent `Tier II' sample, extends out to \logg$\approx1.8$, which is typically the lower limit of the RC range, and thus prone to contamination from regular giants. 

Finally, we used {\sl APOGEE}-DR16 \citep{apogee_dr16}, to construct the background $Kiel$ diagram. This shows that our RC sample is largely concentrated around the horizontal branch (blue contour), thus missing several common stars with L20, but at the same time is likely a `purer' sample for the purpose of distance estimation.

\section{Distances to stars}\label{app_distances}

\subsection{Distance estimates}\label{sect_bayesiandistances}

   \begin{figure}
   \centering
   \includegraphics[width=\hsize]{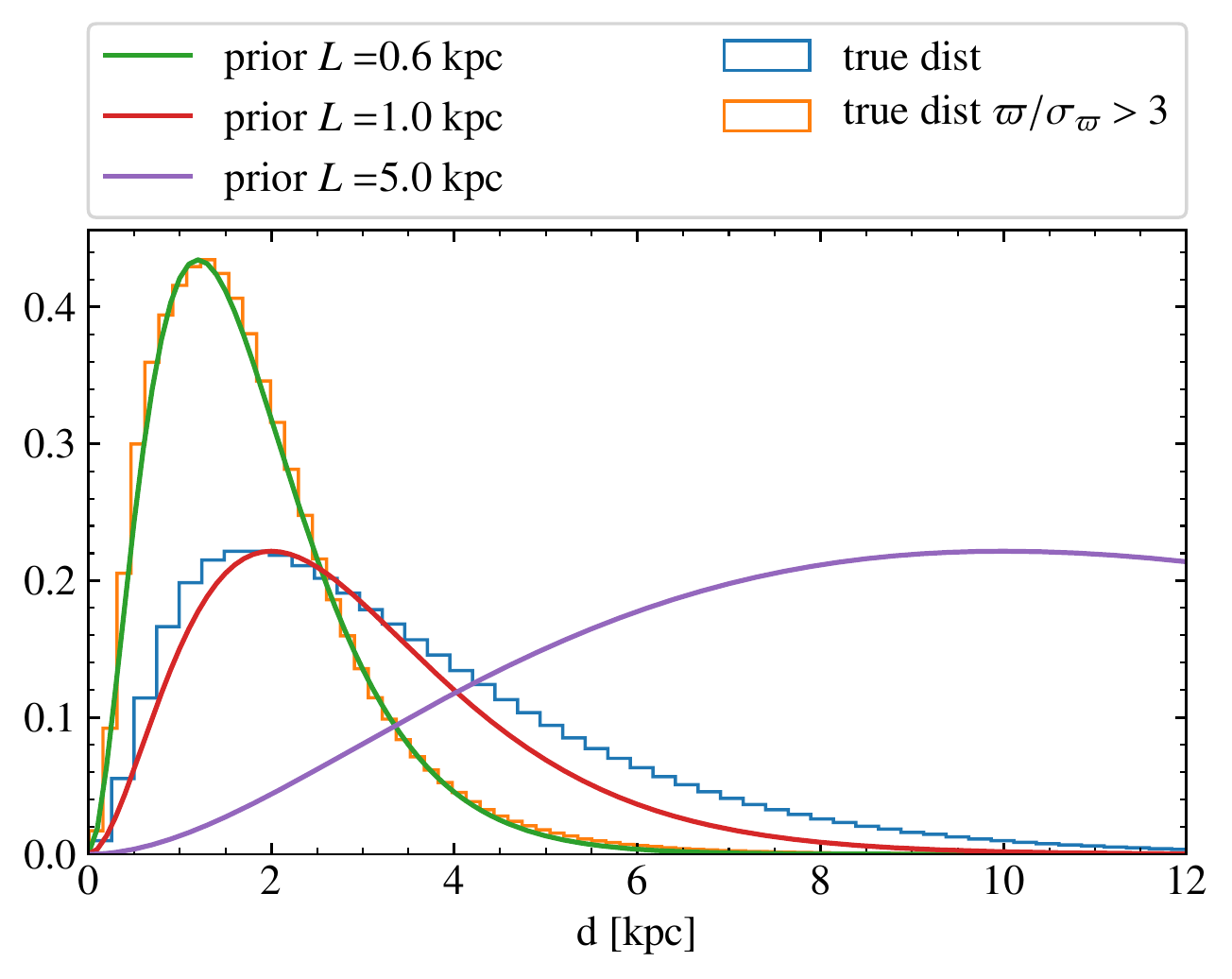}
      \caption{Distribution of true distances of GOG in the anticentre. We show all stars in GOG (blue histogram) and stars with $\piepi>3$ (orange histogram). The different solid lines are the exponential decreasing prior (\equref{eqprior1}) with scalelength of 0.6, 1 and 5 kpc.}
         \label{prior}
   \end{figure}

As discussed in \secref{sect_distances} there is no existing perfect recipe for estimating distances from a measured parallax. In this work we approach this problem by testing how robust  our conclusions are to the use of different distance estimators. We used three different  methods, which we tested with the mock \Gaia data from GOG (described in \secref{sect_selection}):
   
\paragraph{1. $d_\varpi$: simple inversion of parallax $1/\varpi$.}

\paragraph{2. $d_{\rm{PM}}$: Bayesian distances with an iterative prior.}
This approach is closely related to that used by \citet{Schonrich2017}.
In general, the Bayesian approach relies on the statement that for an observed parallax, $\varpi$, and uncertainty, $\sigma_\varpi$, the probability of a given distance $d$ is 
\[
P(d|\varpi,\sigma_\varpi)  \propto P(\varpi | d,\sigma_\varpi)\, P(d),
\]
where $P(d)$ is the prior on distance. This prior takes three factors into account. Firstly, 
the volume at distances between $d$ and $d+\delta d$ increases like $d^2$. Secondly, the true spatial distribution of stars is not uniform. Thirdly, there are selection effects:  the probability of a star at distance $d$ entering the catalogue varies with $d$ because there is a magnitude limit to the survey (because for example intrinsically faint stars become too faint to enter the catalogue). 

For the distances $d_{\rm{PM}}$, these considerations lead us to a prior $P(d) \propto d^2 P({r}(d)) S(d)$, where $S(d)$ is the selection function, and ${r}(d)$ is the position in a galaxy of an object at distance $d$ along a given line-of-sight, so $P({r}(d))$ is proportional to the density of a model Galaxy. The distance estimate, $\tilde{d}$ and uncertainty $\sigma_d$ is then found as the expectation value and standard deviation of $d$ given this prior and the measured parallax (with uncertainty). 

The model from which we take $P({r}(d))$ is taken from \citet{McMillan2018}, and has two exponential discs (thin and thick) and a power-law halo. It has no warp. We approximated the selection function as $S(d)\propto\exp(-d/L_s)$ where $L_s$ is a value we determine. Experiments with GOG (see below) and investigation with the \gaia data both suggest that this is a reasonable approximation.

Following \citet{Schonrich2017}, we derived the selection function from the data by recognising that $S(d)\propto N(d)/(d^2\int\rho(d,l,b)\cos b\mathrm{d}l\,\mathrm{d}b)$ where $N(d)$ is the number of stars in the catalogue at a distance $d$ and the integral over $\ell,b$ is taken over the field we consider. We don't know $N(d)$, but we can make the approximation that $N(d)\approx N(\tilde{d})$ for some range of distances and subset of the more accurate parallaxes. We used this to find the scale length $L_s$ which enters into $S(d)$. We then iterated this process -- using this estimate of the selection function to find new distance estimates, $\tilde{d}$, then using these to make a new estimate of $S(d)$. Experiments with GOG indicate that fitting $S(d)$ for distances $1<d/\mathrm{kpc}<3$ and for stars with $\varpi/\sigma_\varpi>3$ give a reasonable approximation. The value of $L_s$ we find converges after a few iterations and we find $L_s=0.963$ for our sample and $L_s=1.16$ for GOG. 

\paragraph{3. $d_{L}$: Bayesian distances with exponentially decreasing prior with scale length of $L$ following \citet{BailerJones2018}.} These distances were computed following \citet{BailerJones2018} with a simpler prior, in particular an exponentially decreasing prior with distance $d$:
\begin{equation}
P(d | L)  \ = \  
\begin{dcases}
  \ \frac{1}{2 L ^3}\,d^2e^{-d/ L }  & \:{\rm if}~~ d >0 \\
  \ 0                          & \:{\rm otherwise.}
\end{dcases}
\label{eqprior1}
\end{equation}
\figrefalt{prior} shows the true distribution of distances of GOG (blue histogram) and the same for a selection of sources with $\piepi>3$ (orange histogram). As explained in \citet{BailerJones2018}, a good approximation for the maximum likelihood estimate for the scale-length $L$  of the prior \equref{eqprior1} of a given distance distribution is  
$MED(d)/3$, where $MED(d)$ is the median of the distribution. For GOG in the anticentre this turns out to be 0.977 kpc  and 0.527 kpc for stars with $\piepi>3$. 
The red and green solid lines show the shape of the prior with $L=1\kpc$ and $L=0.6\kpc$, respectively, which fairly reproduce the true distribution of distances in each case. We also show the prior for $L=5\kpc$. Hereafter, we choose two different scale-length $L$ of $1\kpc$ (that we name $d_{L_1}$) and $5\kpc$ ($d_{L_5}$), motivated by the tests shown below. While \citet{BailerJones2018} uses a scale-length that depends on the sky coordinates, here for simplicity we use a single value for the whole field of $20\times20\deg$. 

\subsection{Tests with GOG}

   \begin{figure}
   \centering
   \includegraphics[width=0.5\textwidth]{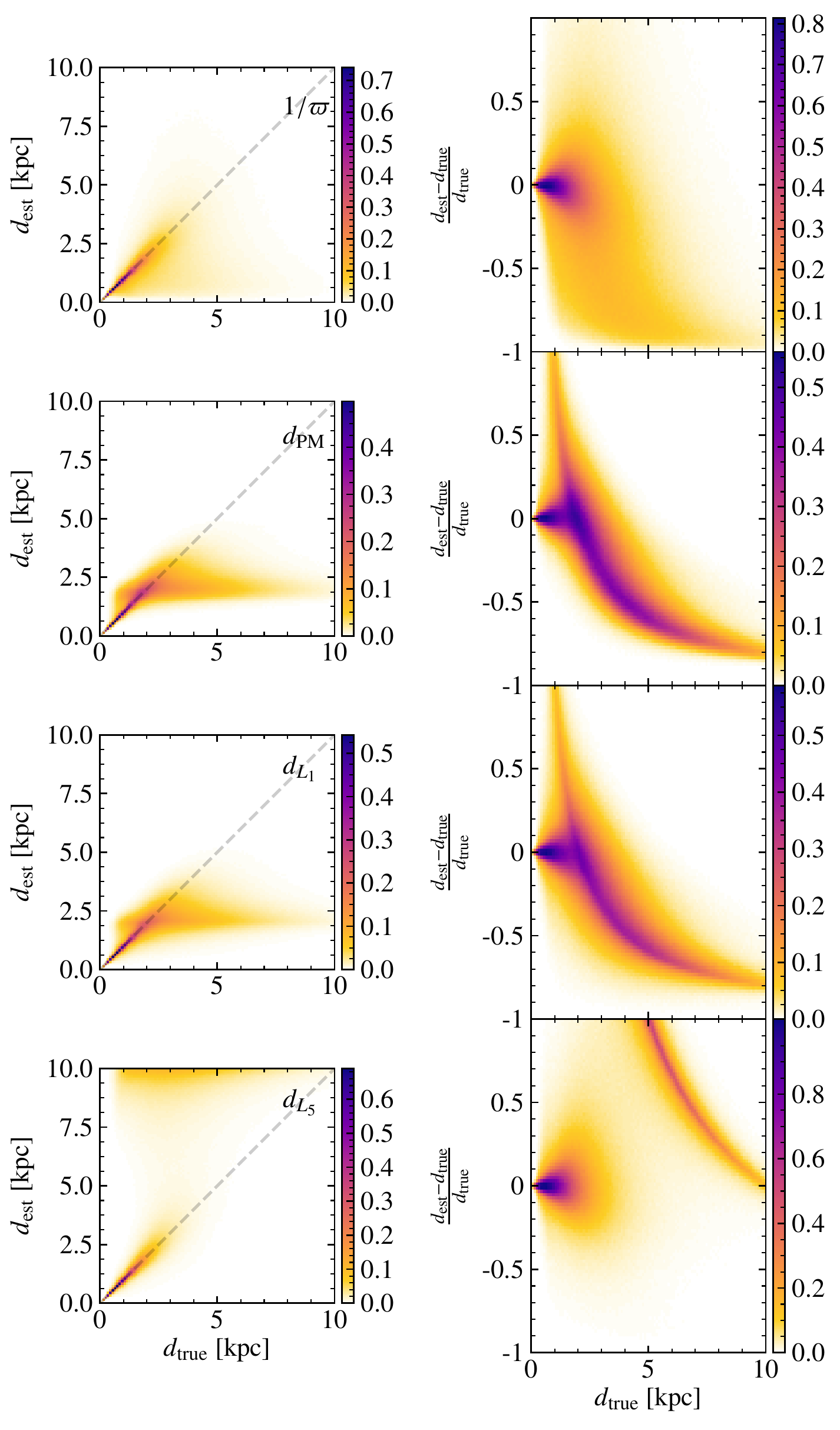}
     \caption{True versus estimated distances for different methods. {\it Left:} Direct comparison between true and estimated distances of GOG for the different distance estimations as indicated in the legends.  {\it Right:} Fractional error in the estimated distance as a function of true distance for the same estimators. }
         \label{trueversusobserved}
   \end{figure}

 Here we test the different distances estimations with GOG. First we note that due to deficiencies in the \Gaia error model, the uncertainties in the astrometric values in GOG somewhat disagree with the values for EDR3. In particular, we see an overestimation of the parallax errors as a function of magnitude $G$, actually more similar to the DR2 scenario than to EDR3 (Fig.~\ref{fig_astrometricerrors} top). The exercises presented here thus show a worse case scenario.
  
  \figrefalt{trueversusobserved} shows the comparison between true distance and estimated distance for the different methods presented above applied to the whole GOG sample, for which the true distances are known. We see in the left panels a good fraction of stars with properly determined distances falling on the 1:1 line (those with small parallax uncertainties). However, we also see large fraction of stars with badly estimated distances corresponding to parallaxes with large uncertainty (including negative parallaxes). For the $d_\varpi$ case (top panels), most of these problematic cases appear scattered in the underestimated region. For Bayesian estimations $d_{PM}$, $d_{L_1}$ and $d_{L_5}$ (three bottom panels), they appear concentrated at the nearly horizontal line at $d_{\rm{est}}=2L$ (coinciding with two times the mode of the prior, that is $\sim2\kpc$ in the two middle rows, and 10 kpc for the bottom row), completely dominated by the choice of the prior as explained in \citet{BailerJones2018}.
These numerous uninformative parallaxes forces us to perform a cut in fractional parallax uncertainty, which, unfortunately, may introduce biases in our samples as discussed for instance in \citet{Luri2018}. 

   \begin{figure}
   \centering
   \includegraphics[width=1\textwidth]{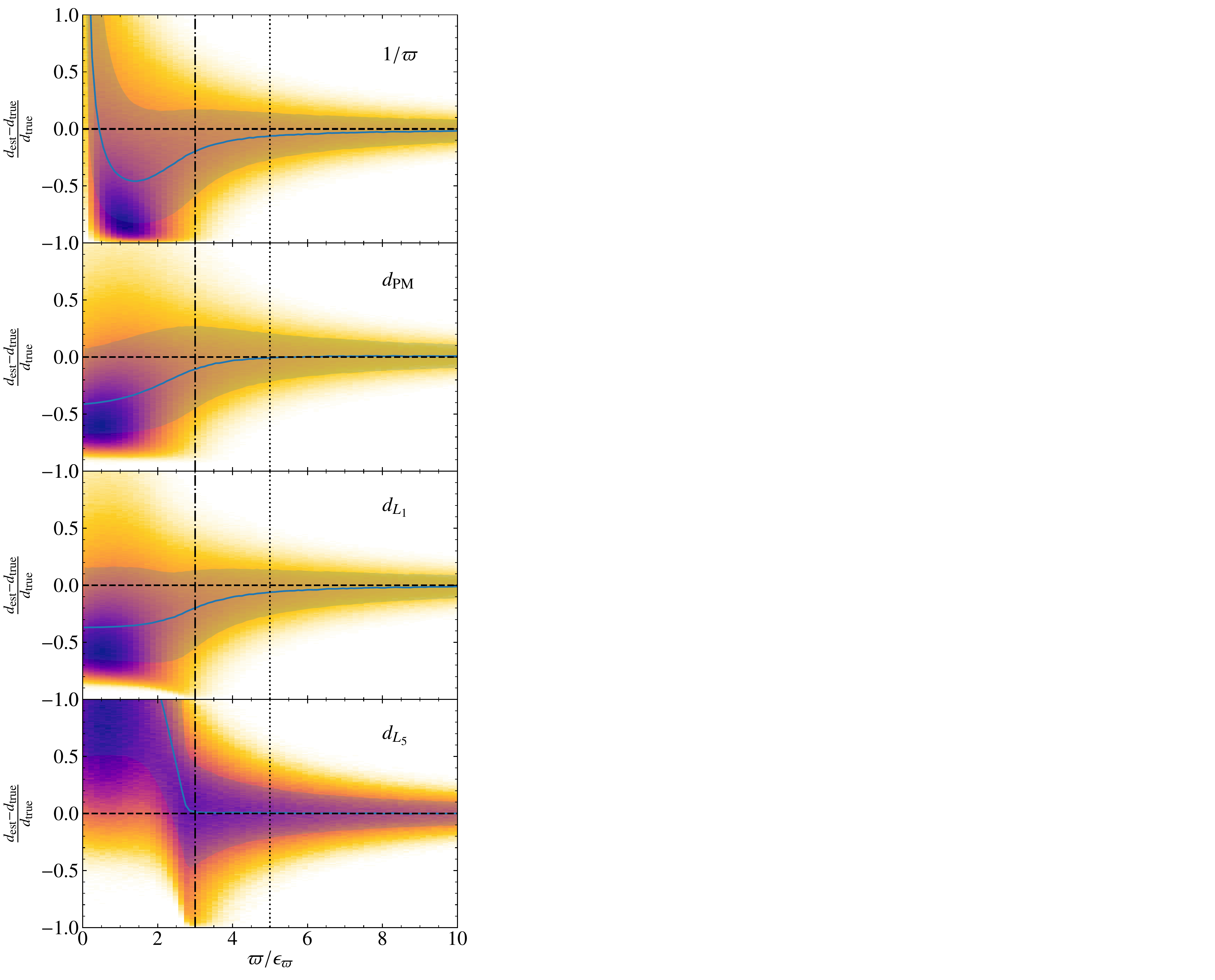}
     \caption{Fractional error in the estimated distance as a function of fractional error in parallax. We show the errors for the different distance estimations as indicated in the labels for the GOG sample. Solid lines indicate the median fractional error and shaded areas show the 25 and 75\% quartiles.}
         \label{disterrpiepi}
   \end{figure}

\figrefalt{disterrpiepi} shows the fractional error in the estimated distance as a function of fractional error in parallax with different panels for different distance estimations. All panels show large errors for large parallax uncertainties (smaller $\piepi$) as expected and highlight the need to use a certain criteria to select good parallaxes while finding a proper balance with the final number of sources kept and trying not to bias the sample as a result of eliminating specific populations. Depending on the particular analysis, these considerations might lead to different choices. Here we choose to select sources with $\piepi>3$ in the case of the AC20 sample (dash-dotted vertical line) while a more restrictive cut at $\piepi>5$ is used {in Sect.~\ref{sect_halo}} (dotted line). 

From \figref{disterrpiepi} we also note two important aspects. First, the performance of the 4 different methods is quite similar when one chooses cuts in $\piepi$ as the ones mentioned above, with only a slight underestimation of the distances in the case of $d_\varpi$ compared  for instance to $d_{PM}$. Second, we also want to emphasise that, even if the median differences between estimated and true distances are small, at $\piepi>5$ (dotted vertical line) 50\% of the sources have errors in the derived distances $\gtrapprox 20\%$ (sources outside the shaded areas which enclose the other 50\%) independently of the method used. 

   \begin{figure}
   \centering
   \includegraphics[width=0.45\textwidth]{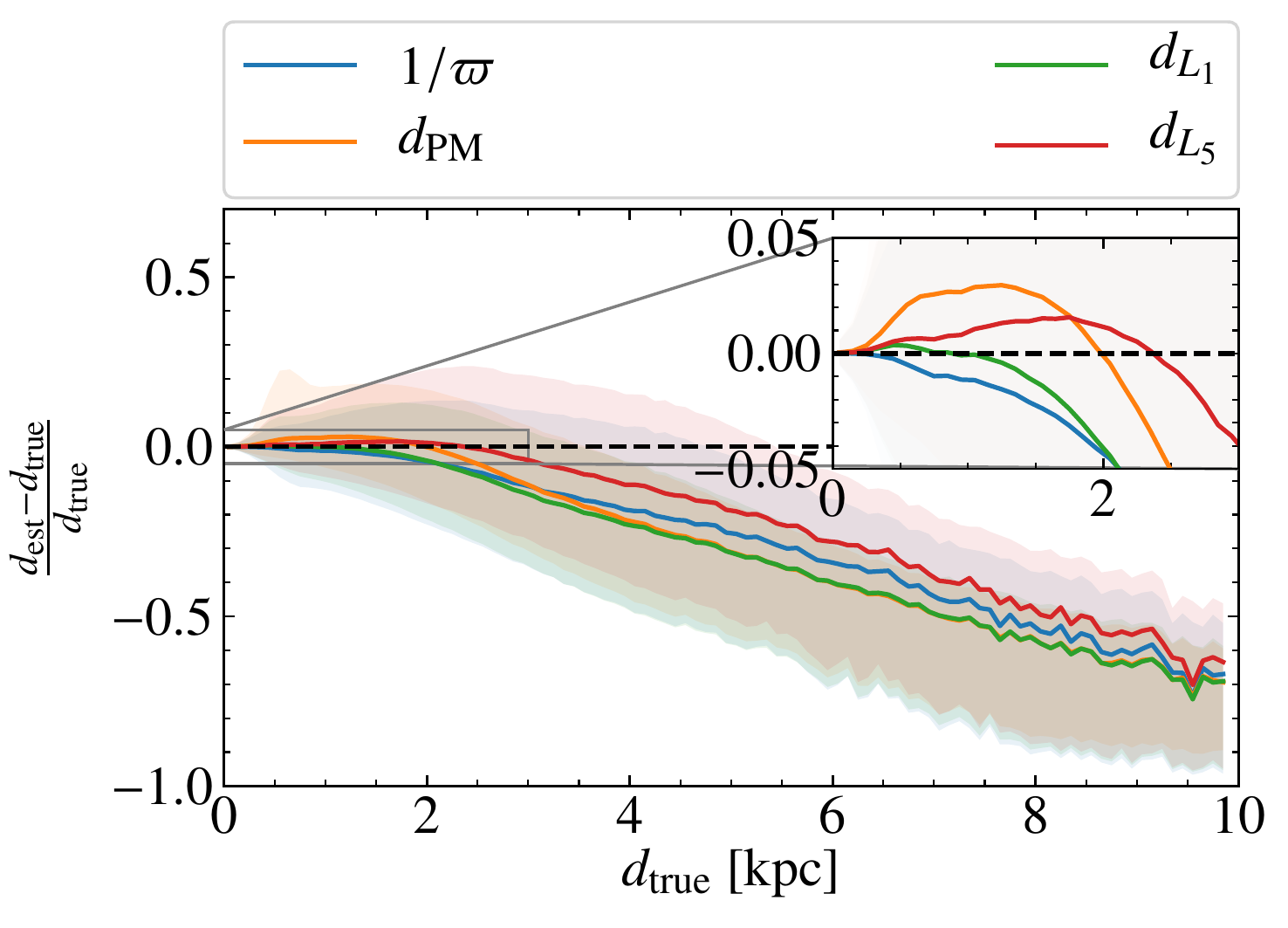}
 \caption{Median fractional error in the estimated distance for good quality parallaxes. We show the errors as a function of true distance for different distance estimations as indicated in the labels for stars with $\piepi>3$ in the GOG sample. Solid lines indicate the median fractional error and shaded areas show the 25 and 75\% quartiles. }
         \label{disterrdistrueL}
   \end{figure}

Now focusing on the selection of sources with $\piepi>3$, the distance error of these different estimators as a function of true distances is shown in \figref{disterrdistrueL}. We see a slightly better performance of the $d_{L_5}$ at larger distances but a better one for $d_{L_1}$ at nearby distances. The $d_\varpi$ is  underestimated in median for all distances while $d_{PM}$ shows overestimated distances at nearby distances, but the contrary beyond 2 kpc. Apart from these little differences, we note that non of the estimators is completely free of bias even with the selection of $\piepi>3$, as already mentioned above. We see underestimations of the distance that start to be important (20\%) at around 4 kpc and biases larger than $\sim$40\% for 25\% of the sources at this same distance. Again we emphasise that the parallax errors in GOG are overestimated with respect to \Gaia EDR3, and therefore the expected biases as a function of distance in EDR3 are possibly smaller than shown here.

All these tests show that different priors might work better in different regimes and that there can be multiple criteria to choose which method provides a better estimate (e.g. minimising the median distance error at small versus large distances). We also need to keep in mind that these conclusions are somewhat model dependent, influenced by the particular MW density model and selection function imposed in GOG.
Our  approach of exploring varied distances estimations wants to mitigate this model-dependency and the appropriateness of different methods and priors in different cases. We highlight that it is necessary to evaluate the impact of these biases and the effects of the parallax quality cut on the different analysis.

So far what we showed regarded only the estimation of the distance. This estimation and a single value for its uncertainty is then used, together with the proper motions, to calculate velocities and their uncertainties. However, we know this is not strictly correct. On one hand, because the proper motion errors are correlated with the parallax errors and, on the other, because the distribution of uncertainties in the estimated distance in general is not Gaussian and asymmetric. Ideally, then, one would use a method to estimate simultaneously the distance and the tangential velocity of each star. 
The \Gaia technical note GAIA-C8-TN-MPIA-CBJ-081 described a way to infer velocities and distances at the same time, from the proper motions and parallax, using a Markov chain Monte Carlo (MCMC) method. This approach is mathematically more accurate and allows us to deal properly with the correlations between velocities and distances. We have tested it with a random subset of the GOG sample and conclude, firstly, that the resulting velocities are similar to the ones obtained with the usual and simpler approach; secondly, that the correlation between velocities and distances is dominated by the transformation rather than by the correlation in the uncertainties; thirdly, that a cut in parallax quality is still necessary; and, finally, that the high computation cost renders it unfeasible to use for even modest-sized samples. For all these reasons, we do not use it here.


\subsubsection{Red clump distances}
\label{app_RCdist}

For each star classified as RC, we can invert \autoref{eqn:absmag_cal} to calculate the distance modulus. For this we used the literature absolute magnitudes in each band pass, $\Bar{M_{\lambda}}$. The errors in the computed distances using the RC and parallax only is given by, 
\begin{eqnarray}
\sigma_{d_{RC, \lambda}} = 0.2 \ln(10) \sigma_{M_{\lambda}}  d \\
\sigma_{d_{\varpi}} = \sigma_{\varpi}d^{2}, 
\end{eqnarray} where  $\sigma_{M_{\lambda}}$ is the dispersion in the computed absolute magnitudes of the RC selection, and $\sigma_{\varpi}$ is the parallax error.
The parameters in \autoref{tab:tab_parameters} are fine tuned in order to maximise the number of RC stars and minimise the dispersion and thus the errors in distances. 

As mentioned earlier, we did not apply the 'qfl' quality flag on 
\twomass{} photometry, but instead use the photometric errors to decide if the distances will be estimated using the $K$ band or $G$ band. In general, the $K$ band suffers from lower extinction than the broader $G$ band, so we prefer to use distances estimated using $K$. However, if for a given star the photometric errors, $(e\_jmag$| $e\_kmag) >0.025$, the typical value above which photometry in \twomass{} becomes unreliable, then we estimate distances for these using the $G$ band. This is illustrated in \autoref{fig:drc_luc_kcorr}, where we compare our distance estimates to the external catalogue of  \lucey{}. Essentially, for stars with poor \twomass{} photometry we overestimate the distances if the $K$ band is used. Replacing these with $G$ band estimates results in a much better agreement with the external catalogue.

\begin{figure}
\includegraphics[width=1.\columnwidth]{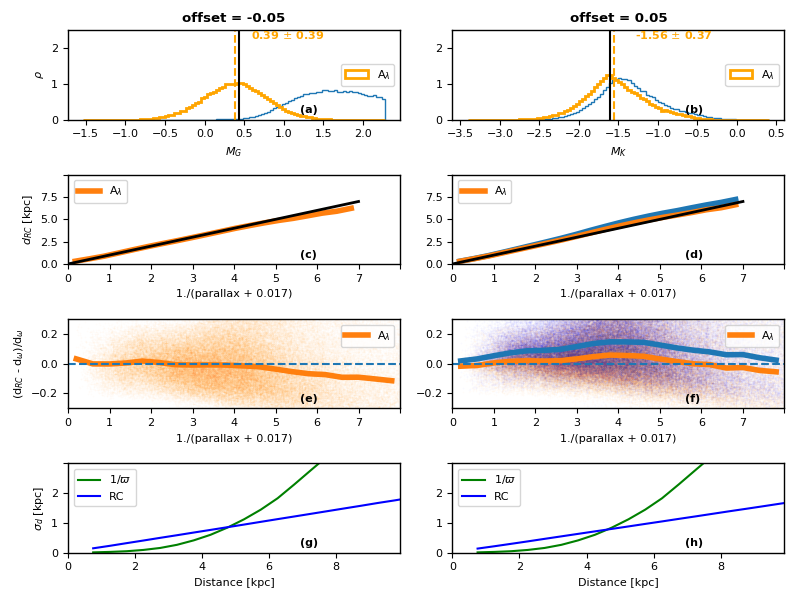} 
\caption{Red clump sample inspection. (a-b) Absolute magnitudes in $G$ and $K$ for the selected sample. The yellow curves use the 3D extinctions from \bayestar{}, while the blue curves are for zero extinction shown just for illustration of shift towards the correct literature value upon reddening correction. (c-f) compare the RC distances to inverse parallax, while panels (g-h) show the error in distances as a function of d for the two methods. Beyond 5 kpc, RC distances become more reliable than inverse parallax. \label{fig:absmag_hist_gmag}}
\end{figure}

\begin{figure}
\includegraphics[width=1.\columnwidth]{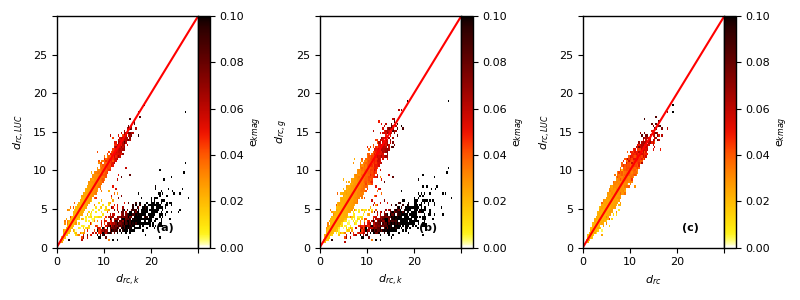} 
\caption{Red clump distance validation with the external catalogue L20. To enhance the illustration we use a larger RC sample here ($147^\circ<l<219^\circ$ and $|b|<30^\circ$). Panel a) Comparison to L20 shows the presence of a population for which distances are overestimated using the $K$ band. This is due to very high photometric errors (i.e. $(e\_jmag$| $e\_kmag) >0.025$). b) Comparison between $G$ and $K$ band derived distances also highlights the same trend: $K$ band distances are overestimated for poor photometry stars. c) Replacing $K$ band estimates with $G$ where $(e\_jmag$| $e\_kmag) >0.025$ improves agreement with L20. \label{fig:drc_luc_kcorr}}
\end{figure}

\subsubsection{Comparison of the different distances for EDR3}
\label{sec_distcomp}

\begin{figure}
   \centering
   \includegraphics[width=0.3\textwidth]{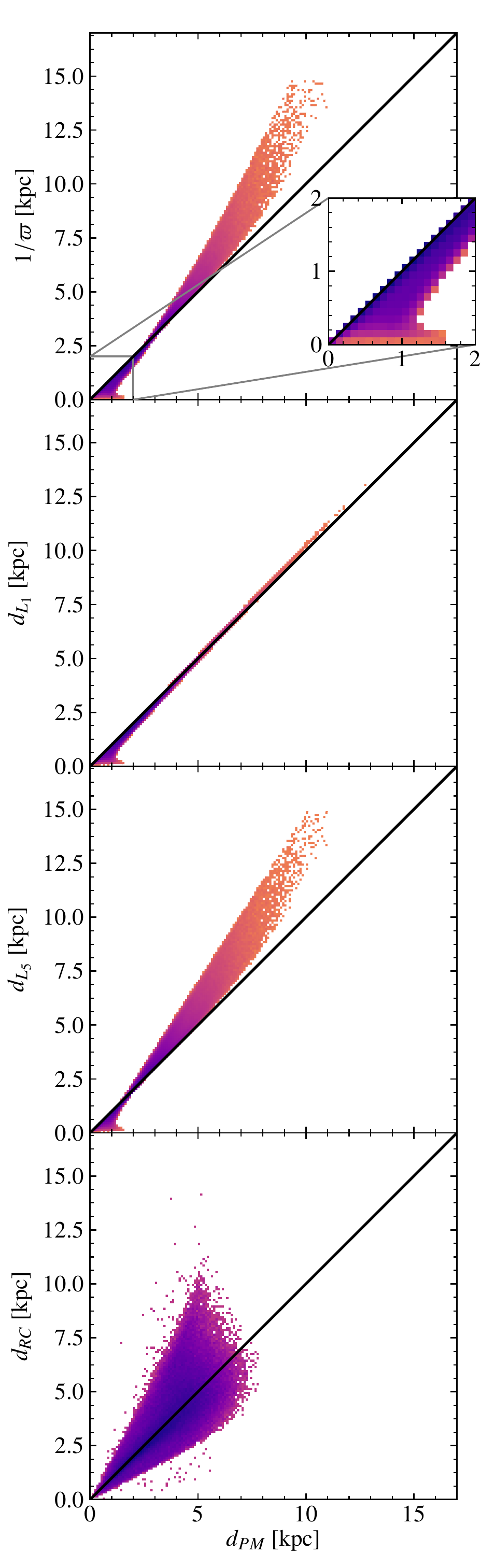}
      \caption{Comparison of the different distances used in this study. The comparison is done with respect to the $d_{PM}$ distances. The discrepancies are small for small distances. For the case of the RC, we only compare stars with $\piepi>3$, since the rest of RC sources are not included in our set of $d_{PM}$. This is then misleading since for these stars the parallax retrieves better distances, but the real gain for the RC occurs exactly for the stars missing in this panel, in the regime where the photometric distances might be better than the ones from parallax alone. The peculiar shape shown in the inset of the top panel and present in the three top panels is composed of stars with large parallax error, for which the expectation values used in the $d_{PM}$ estimation are larger than for instance the medians used in $d_{L}$.}
         \label{distcomp}
   \end{figure}

Finally, \figref{distcomp} compares all sets of distances derived in this work using the $d_{PM}$ case as a baseline (see caption for more details).

\section{Parallax zero point}\label{app_zpt}

\begin{figure}
   \centering

      \includegraphics[width=0.8\hsize]{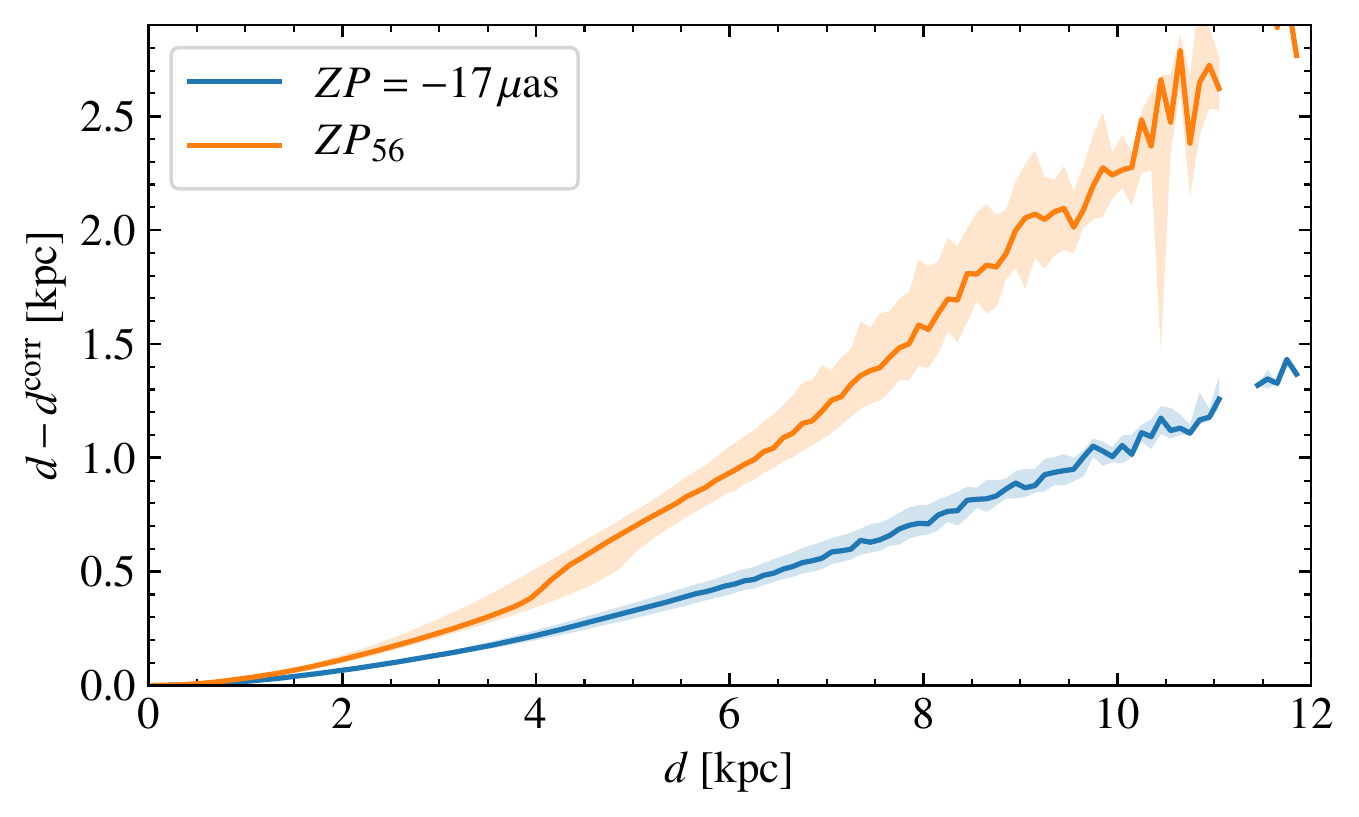}
   
   \includegraphics[width=0.8\hsize]{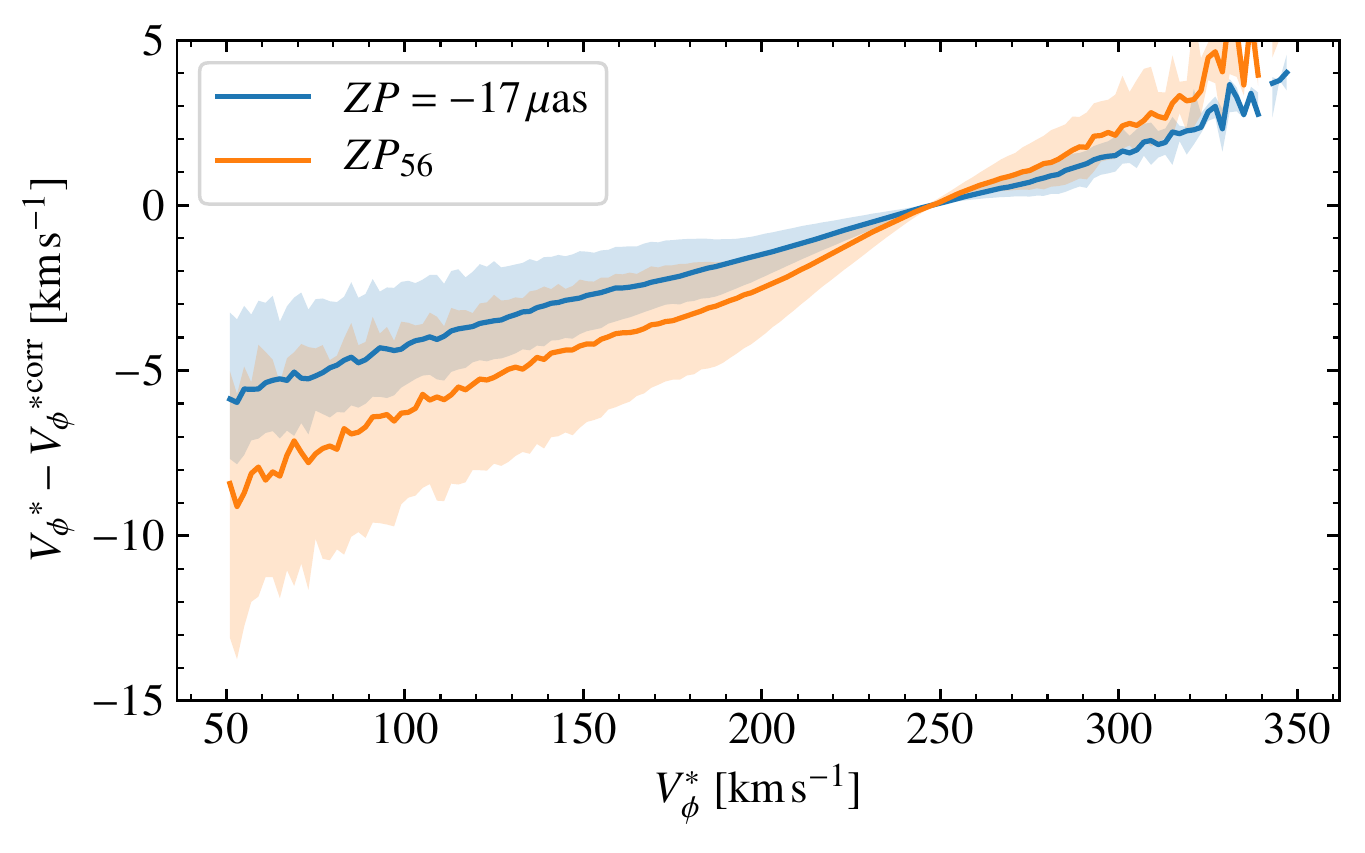}
   
   \includegraphics[width=0.8\hsize]{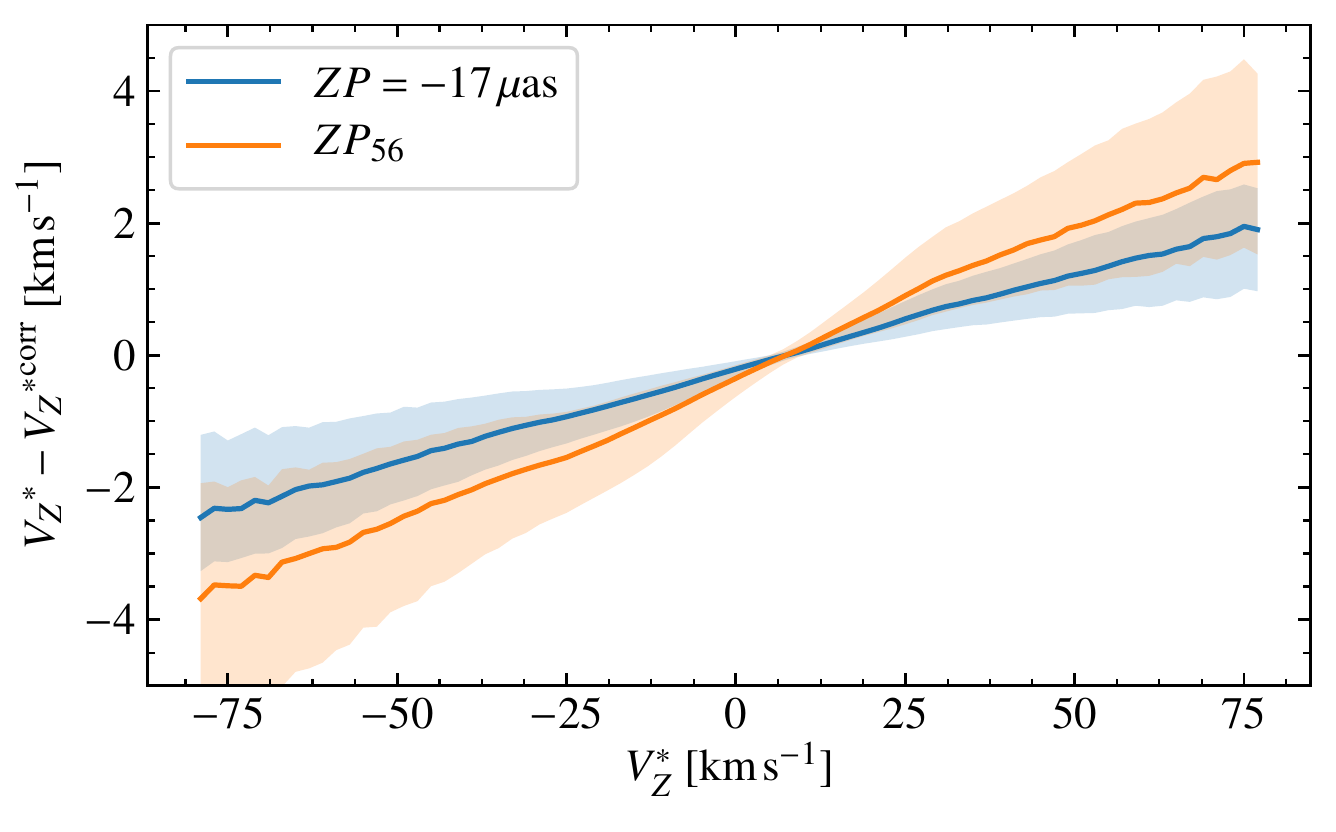}
   
      \caption{Effects of the zero point in parallax on distances and velocities. The comparison is done with respect to the case {with the Bayesian distances $d_{PM}$} where the zero point is not considered (x axis) and the shaded areas show the percentiles 10 and 90 (i.e. they enclose 80\% of stars). In the top panel, we see how not correcting for the zero point produces overestimated distances. The zero point prescription $ZP_{56}$ reduces even more the distances compared to the case of a fixed zero point $ZP={-17}\,\mu$as. The velocities (middle and bottom) scale linearly with the distance and thus we see the absolute magnitude of the velocities being larger when the zero point is not considered. We see null differences in the case of null proper motion, that is when the velocities equal that of the Local Standard of Rest. }
         \label{zpt} 
   \end{figure}

\begin{figure*}
   \centering

   \includegraphics[width=0.45\hsize]{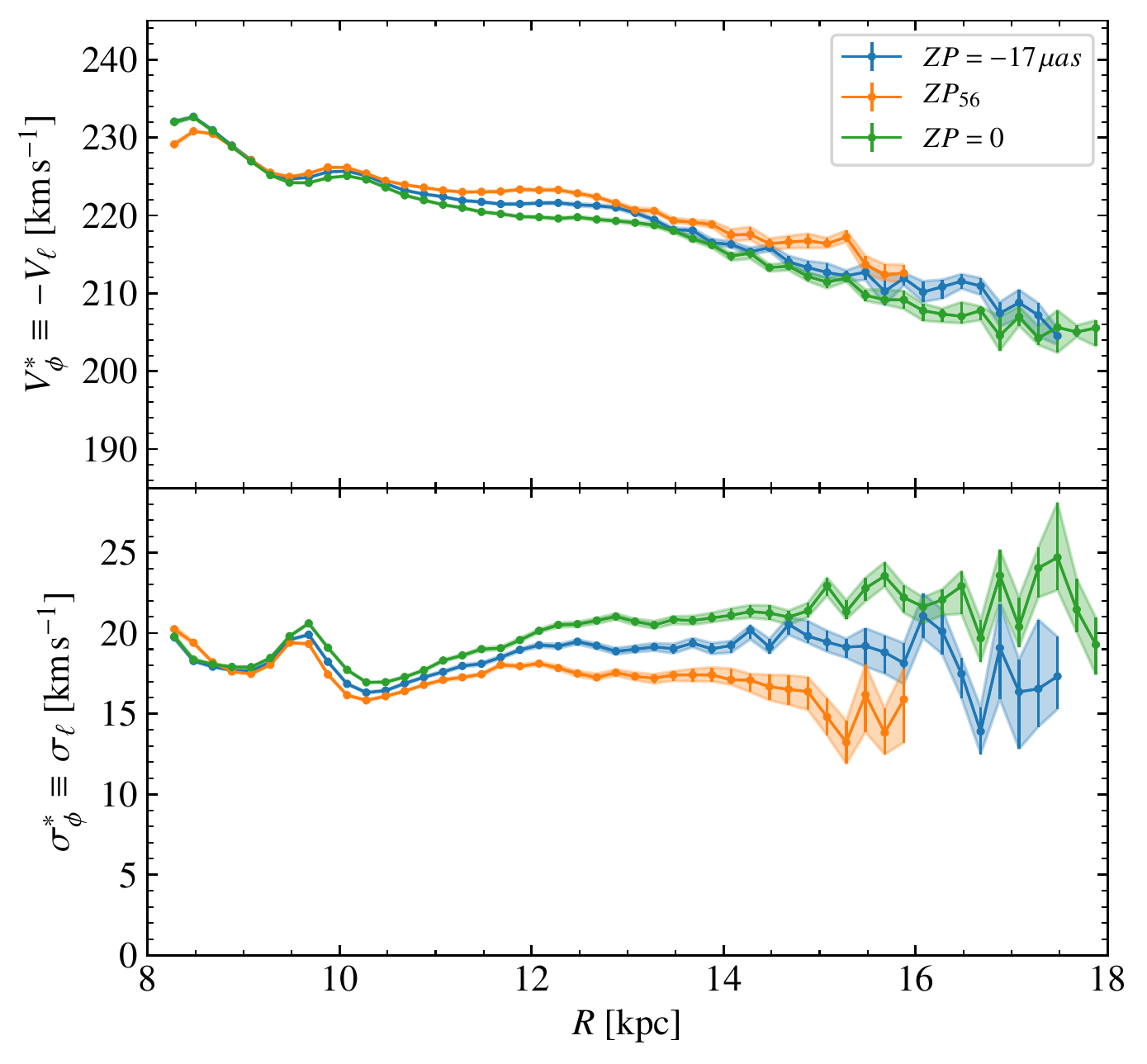}
\includegraphics[width=0.45\hsize]{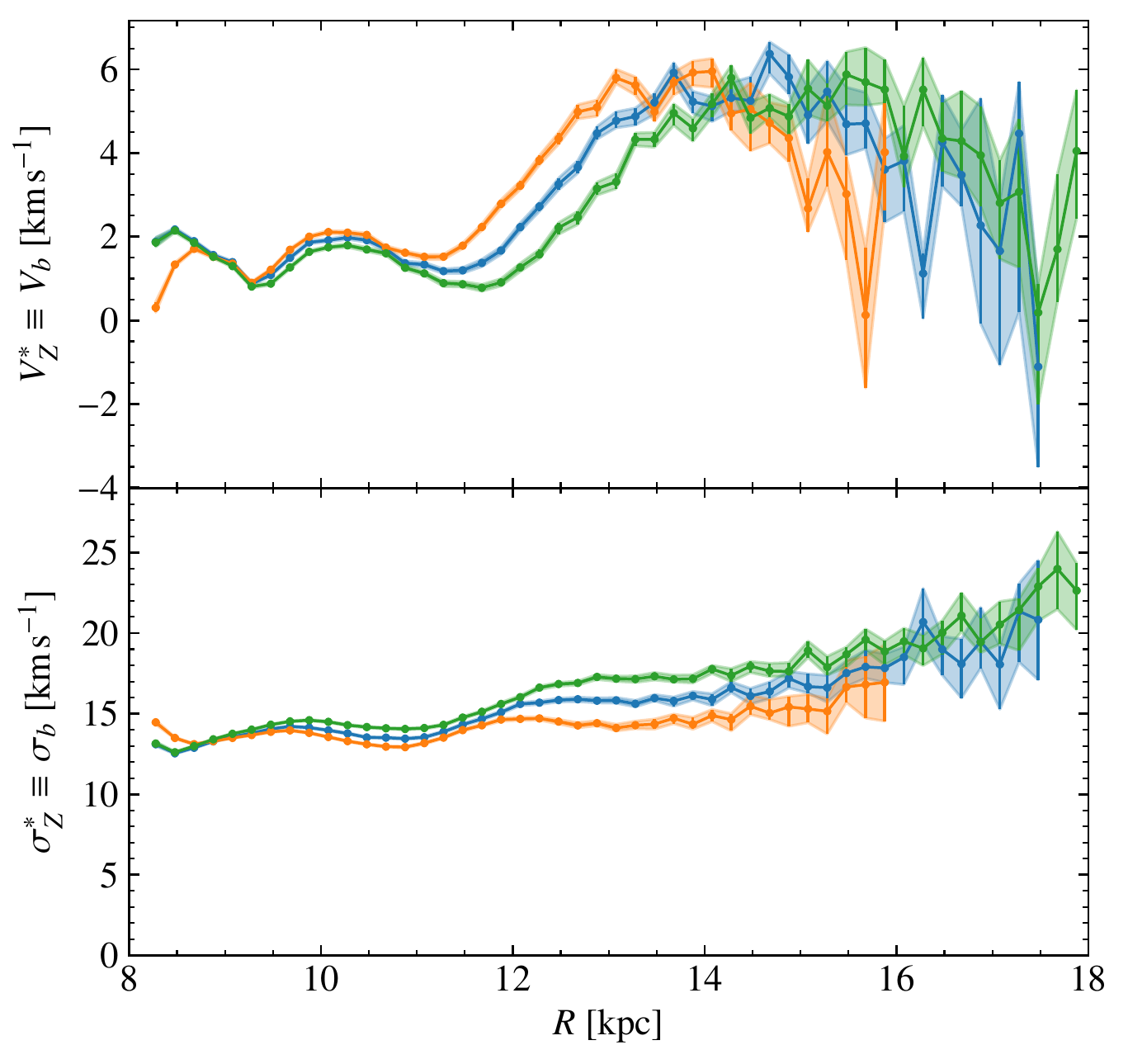}

         \caption{Velocity profiles for different parallax zero points. In the rotation curve (top left), as expected,  the rotation curve computed using $ZP_{56}$ (orange curve) is slightly shifted to the left ($R$ decreases by about 0.5 kpc at $R=14$ kpc) and $\Vp$ also decreases, but always in amounts smaller than $\sim2 \kms$. In the vertical velocity plot (top right), we observe similar effects, though a notable effect is seen in the first kpc. The velocity dispersions ({computed as the $mad$ values, }bottom) appear also slightly different, with  $ZP_{56}$ yielding smaller dispersions but without changing the overall shape.}
         \label{figprofiles_zpts}
   \end{figure*}

\begin{figure}
   \centering
   \includegraphics[width=1.\hsize]{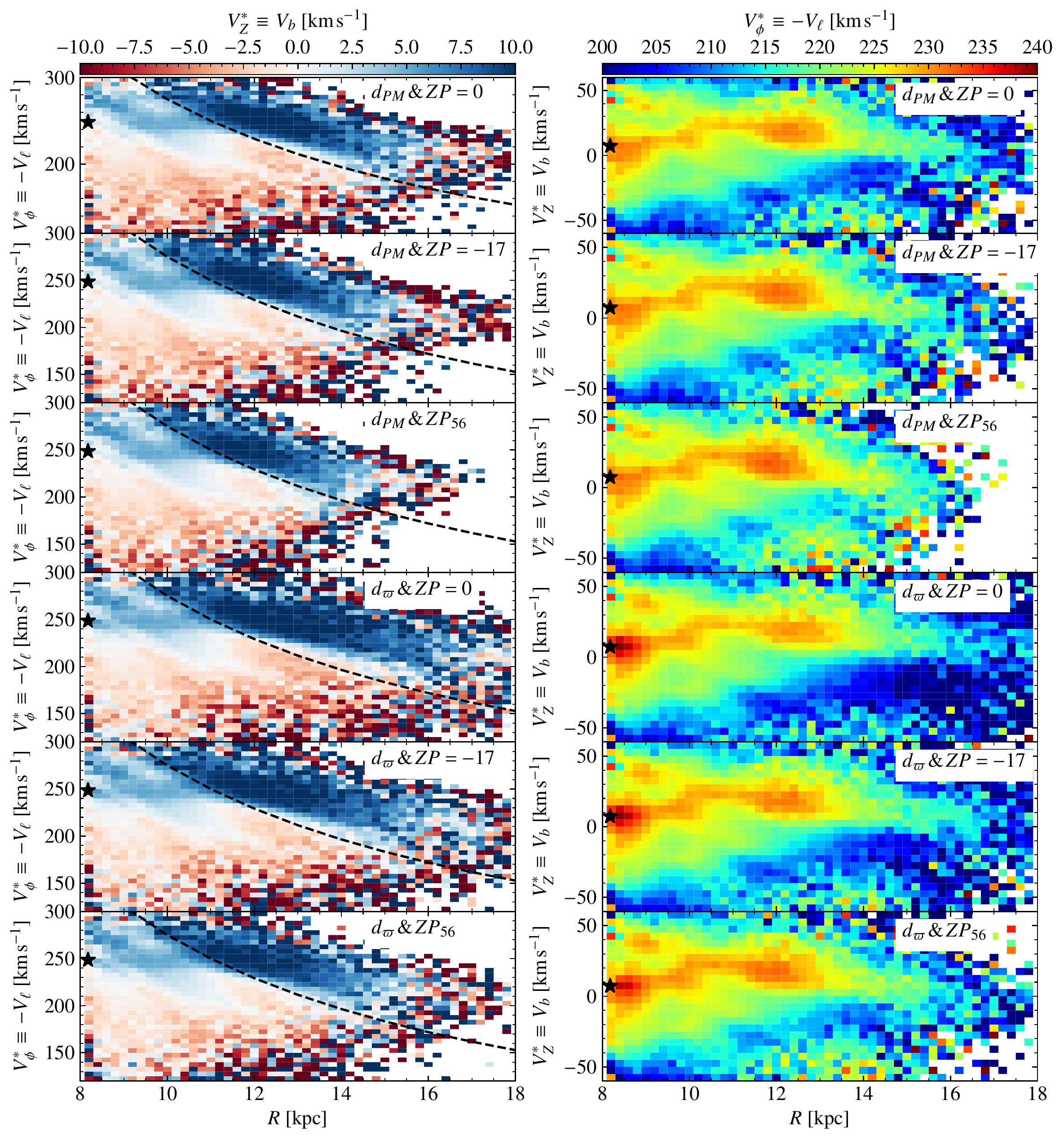}
         \caption{Phase space projections for different parallax zero point. The plot reproduces panels b and e of Fig.~\ref{figRVb} using different distance estimations and parallax zero point as indicated in the legends. As explained in other parts of the article, the correction of the zero point combined with the different distance estimators used produce a change in the distance scale but in any case induces or removes the phase space substructure such as the one observed in this panels. The smallest distances are found when the Bayesian distances $d_{PM}$ and the zero point $ZP_{56}$ are used.}
        \label{figRVlVb_zpt}
   \end{figure}

\begin{figure}
\centering
\includegraphics[width=\hsize]{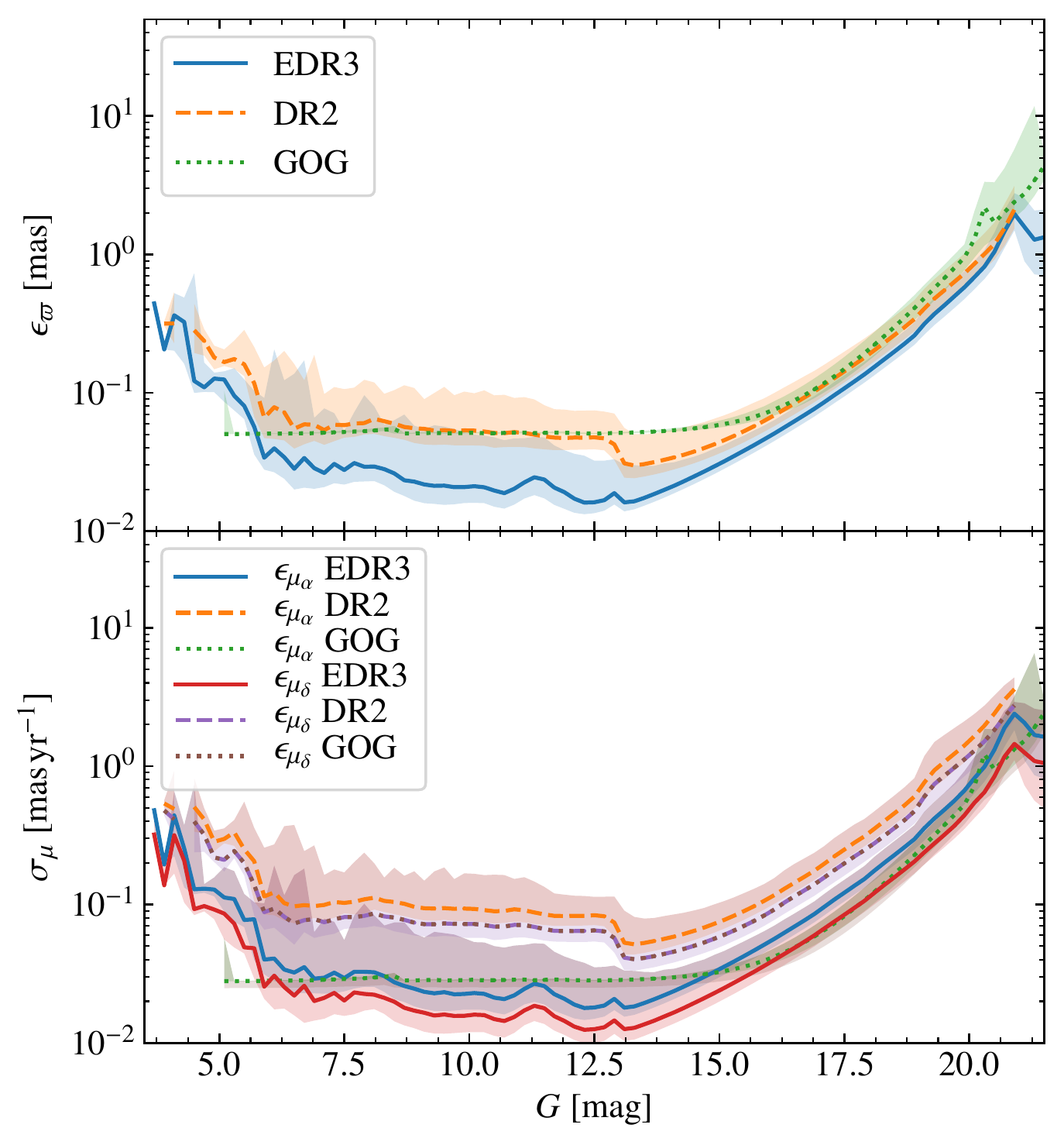}
\caption{Comparison of the astrometric uncertainties for DR2, EDR3 and GOG. Due to deficiencies in the GOG \Gaia error model, the astrometric uncertainties in GOG do not match perfectly those for EDR3. The error model retrieves unique values of the formal uncertainties as a function of $G$, while a large range is obtained for the data (shaded areas showing the 10\% and 90\% percentiles). We also see an overestimation of the parallax errors (top), which actually look more similar to the DR2 scenario than to EDR3. {By definition, the errors in $\mu^*_\alpha$ and $\mu_\delta$ are the same for GOG.} The errors of the proper motions are closer to the true uncertainties although no distinction between the different components is made for this mock data.}
\label{fig_astrometricerrors}
\end{figure}

In this appendix we illustrate the differences in distance and velocities when different parallax zero points are used (Fig.~\ref{zpt}), and we reproduce several figures of the main part done with and without different parallax zero points (Figs.~\ref{figprofiles_zpts} and \ref{figRVlVb_zpt}).

\section{Additional material}\label{app_material}

In this appendix we present a miscellanies set of plots that serve as supporting material to the rest of the sections. A describing text can be found in each of the figures.

\begin{figure}
   \centering
   \includegraphics[width=\hsize]{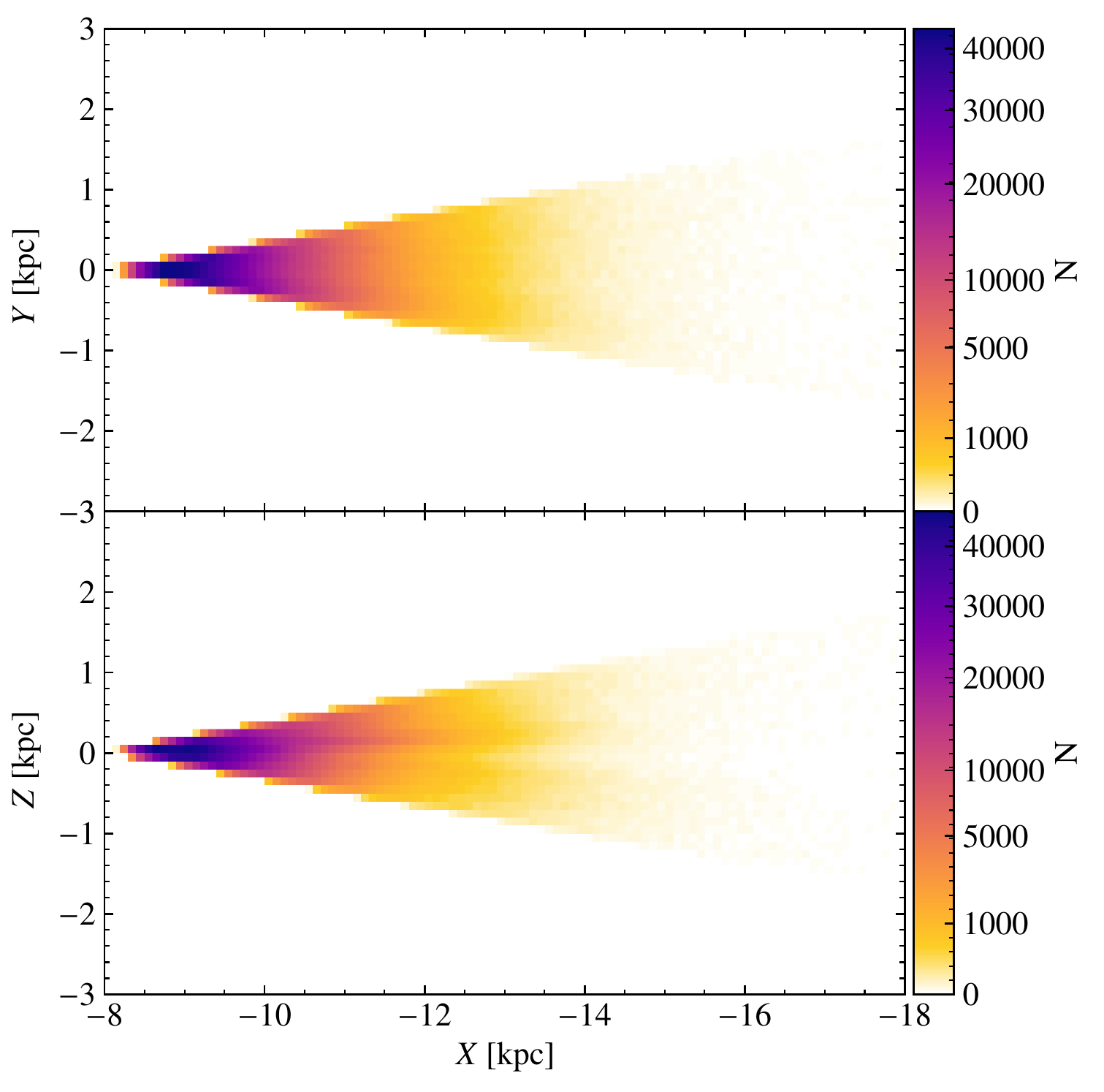}
      \caption{Spatial distribution of the \AC. a) Counts in bins of 0.1 kpc in the $X$-$Y$ projection. b) Same but for the $X$-$Z$ projection. }
         \label{XYZ}
   \end{figure}

   \begin{figure}
   \centering
   \includegraphics[width=0.9\hsize]{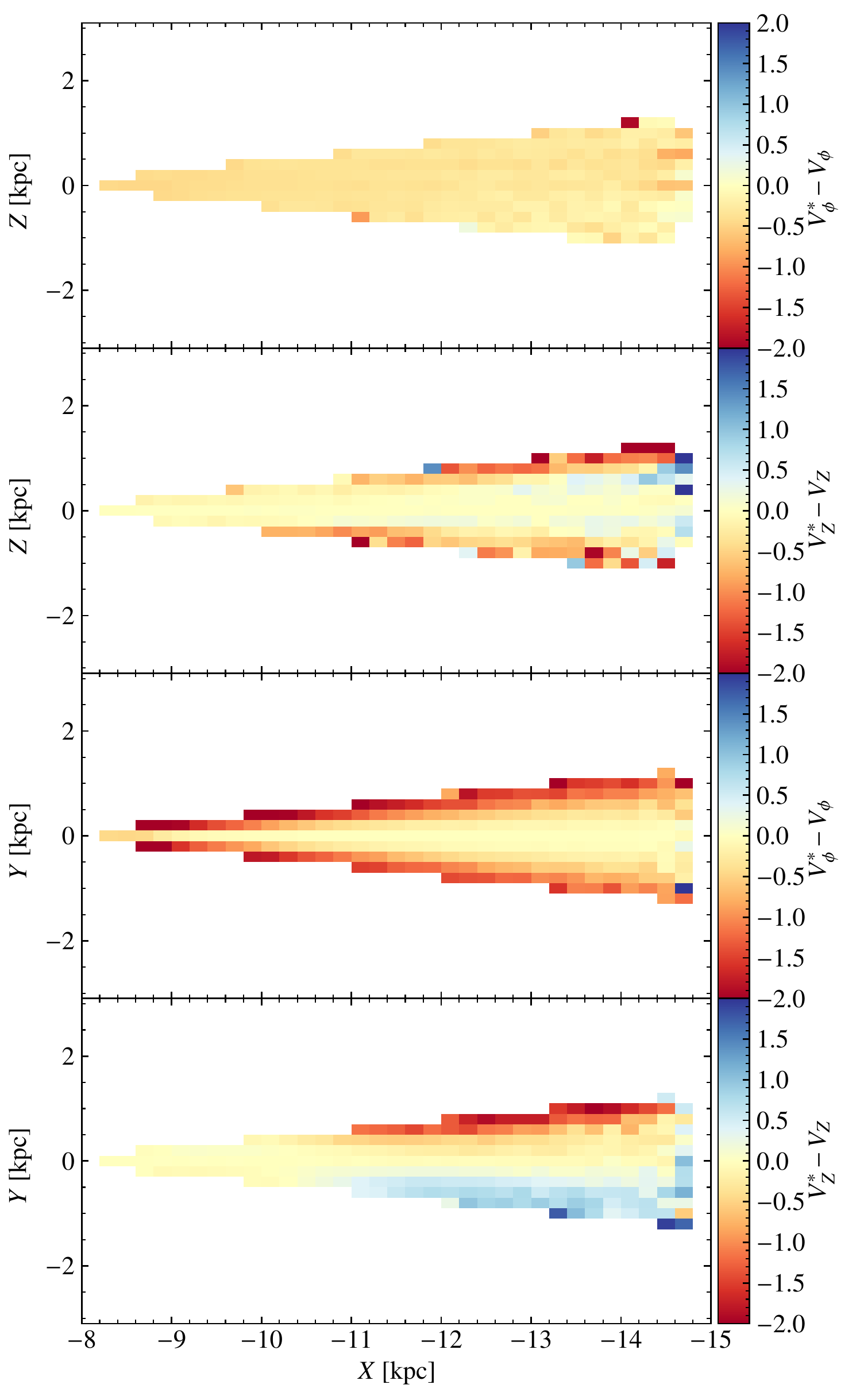}
      \caption{Error in the velocities for GOG when using our approximations. Median differences between $\Vp$ and $V_\phi$, and between $\VZ$ and $V_Z$ (Eq.~\ref{eq:vl} and \ref{eq:vb}) in bins in the $X$-$Y$ and $X$-$Z$ projections.}
         \label{5d6d_3}
   \end{figure}

   \begin{figure}
   \centering
   \includegraphics[width=0.9\hsize]{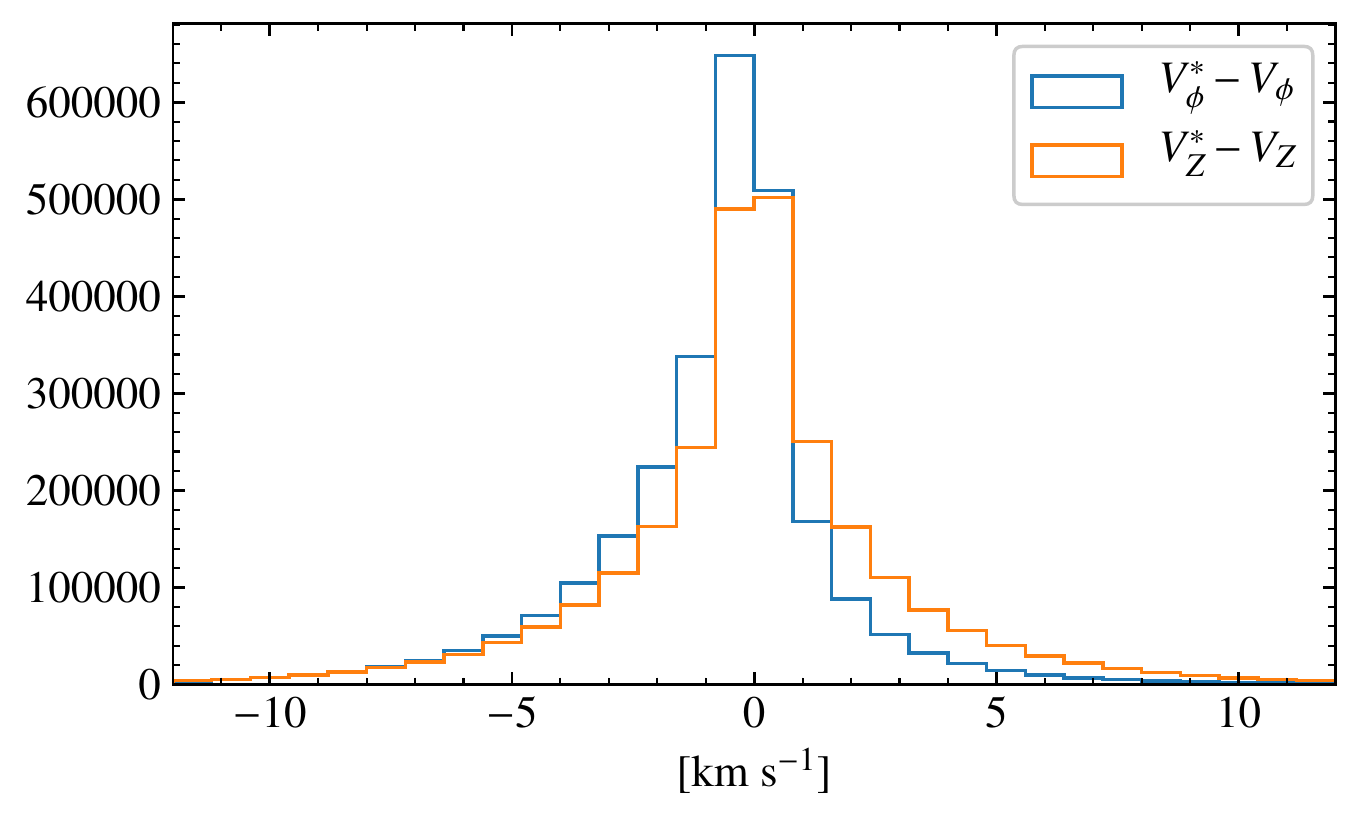}
      \caption{Error in the velocities for GOG when using our approximations (part 2). The histogram of the differences between $\Vp$ and $V_\phi$, and between $\VZ$ and $V_Z$ is shown. The 10\% and 90\% percentiles of the differences are -1.5 and 3.9 $\kms$ and -3.6 and 3.4 for $V_\phi*$ and for $V_Z*$, respectively.}
         \label{5d6d_1}
   \end{figure}

   \begin{figure}
   \centering
   \includegraphics[width=\hsize]{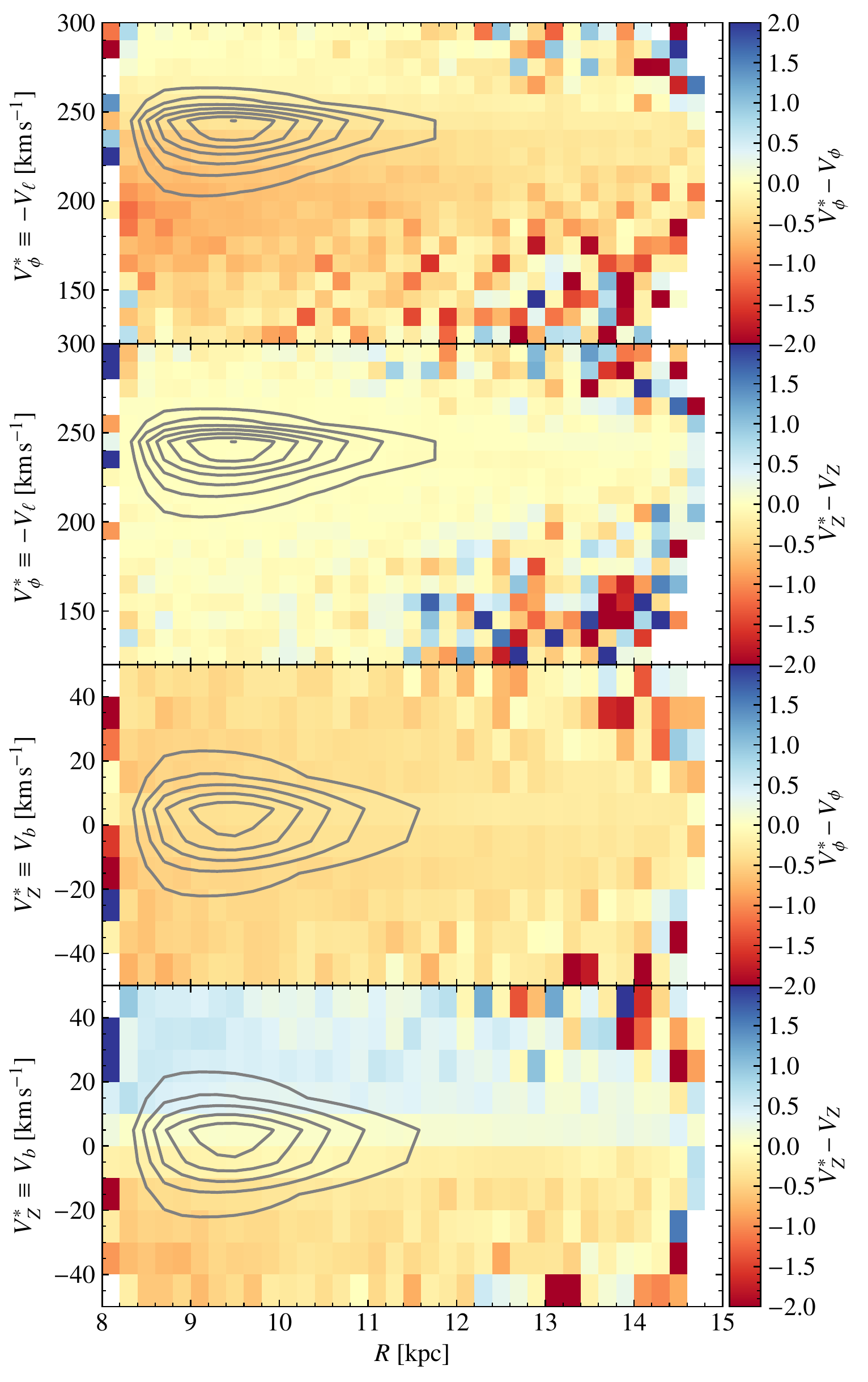}
      \caption{Error in the velocities for GOG when using our approximations (part 3). Median differences  between $\Vp$ and $V_\phi$, and between $\VZ$ and $V_Z$ in bins in the $R$-$\Vp$ and $R$-$\VZ$ projections.}
         \label{5d6d_2}
   \end{figure}

   \begin{figure*}
   \centering
   \includegraphics[width=0.8\hsize]{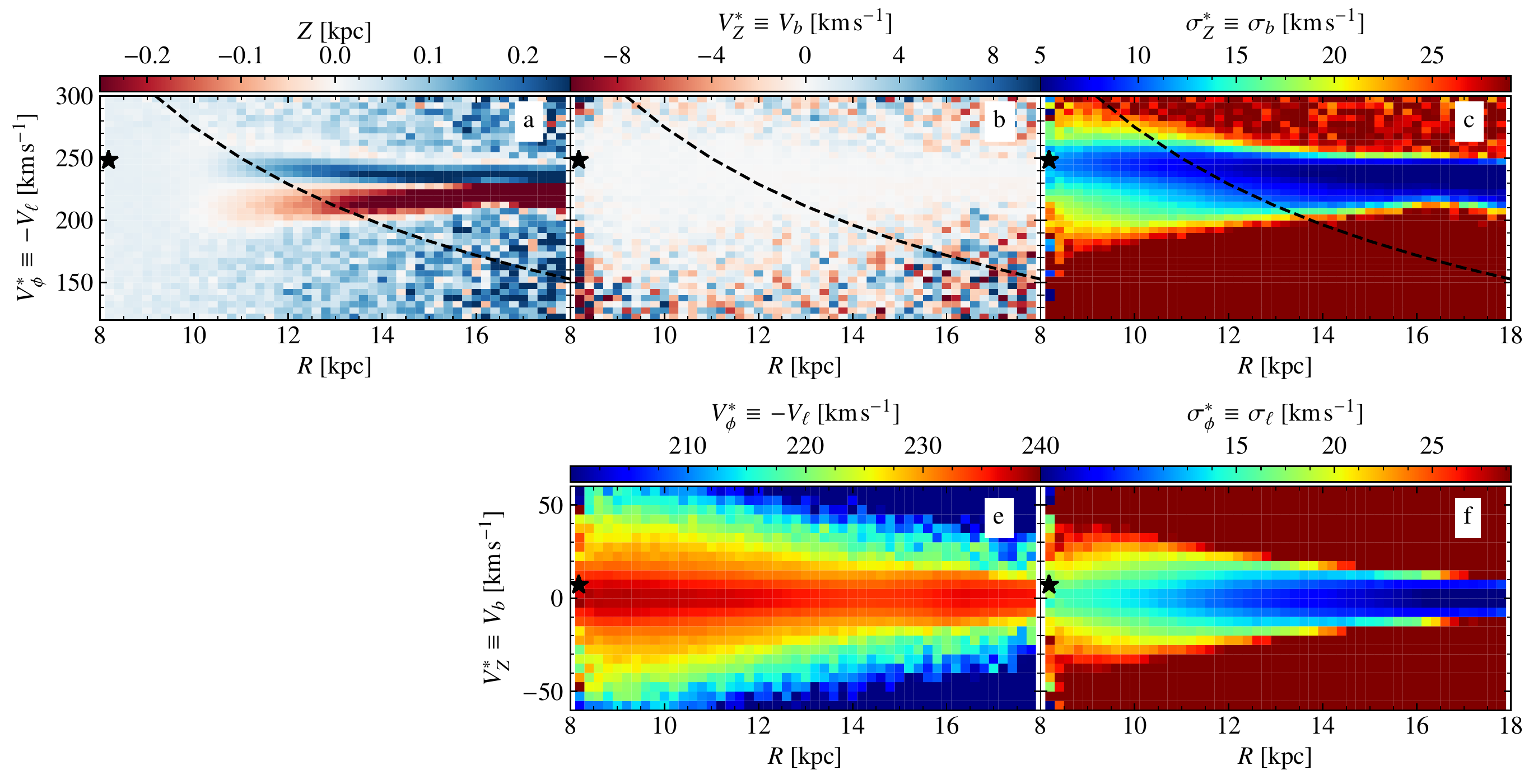}
   
   \includegraphics[width=0.8\hsize]{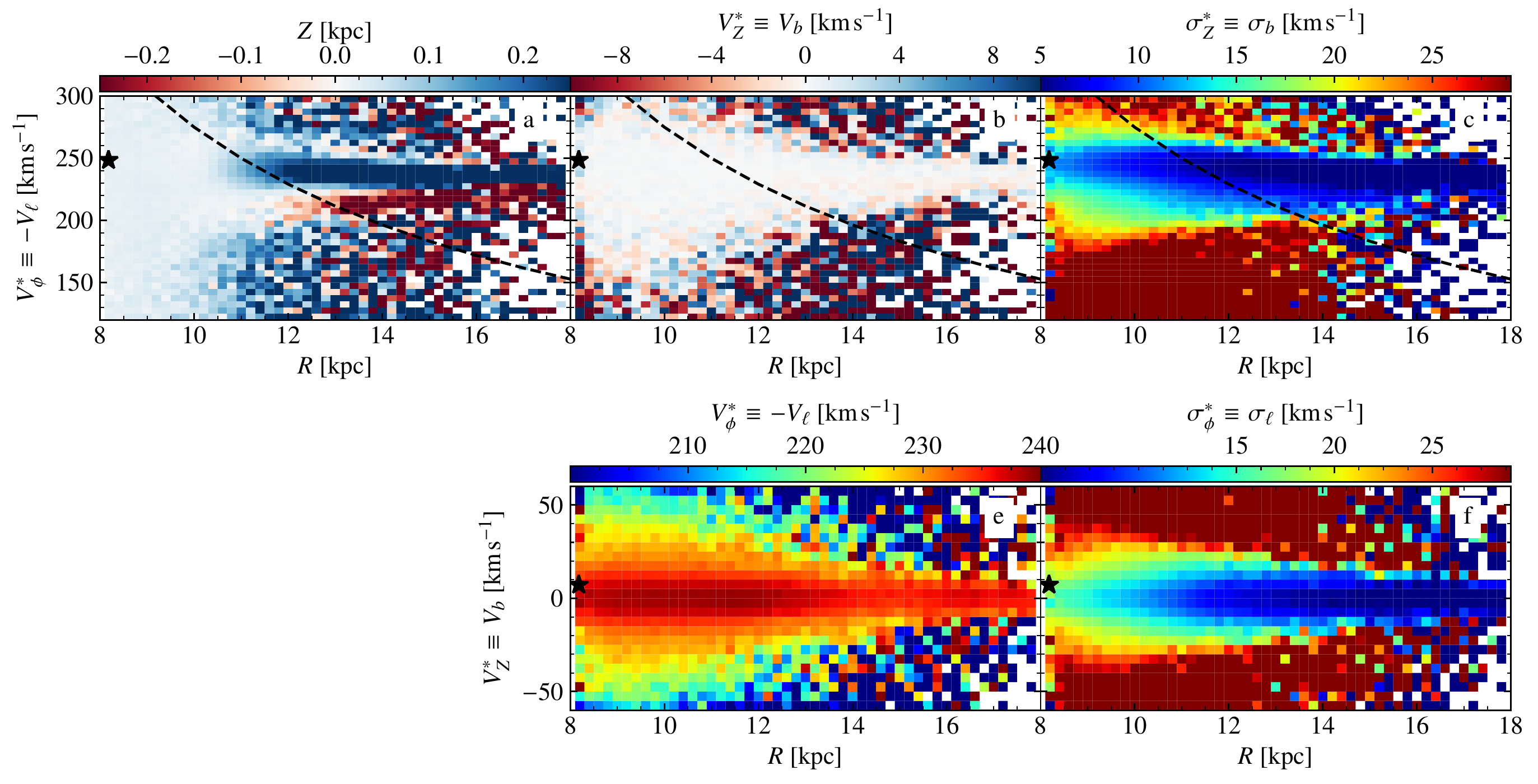}
   
   \includegraphics[width=0.8\hsize]{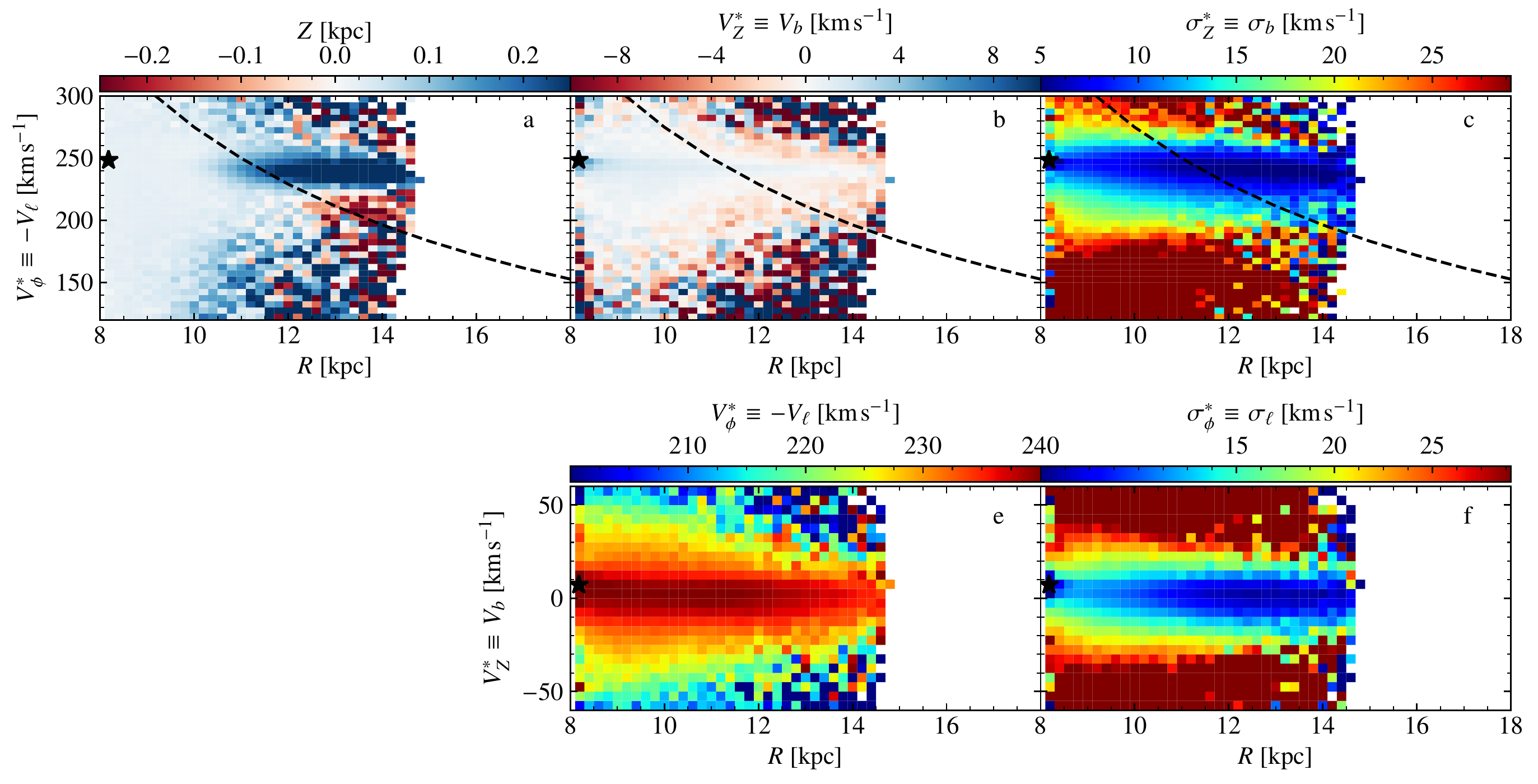}
         \caption{Phase space projections for model and mock data. Same as Fig.~\ref{figRVb} but for the UM (top), the UM with the sources that in GOG have $\piepi>3$ (middle), and for GOG with the selection $\piepi>3$ (bottom). The phase space spiral does not exist in GOG and panels d {are not shown}.}
         \label{figRVlVb_GOGUM}
   \end{figure*}

  \begin{figure*}
   \centering
   \includegraphics[width=0.6\hsize]{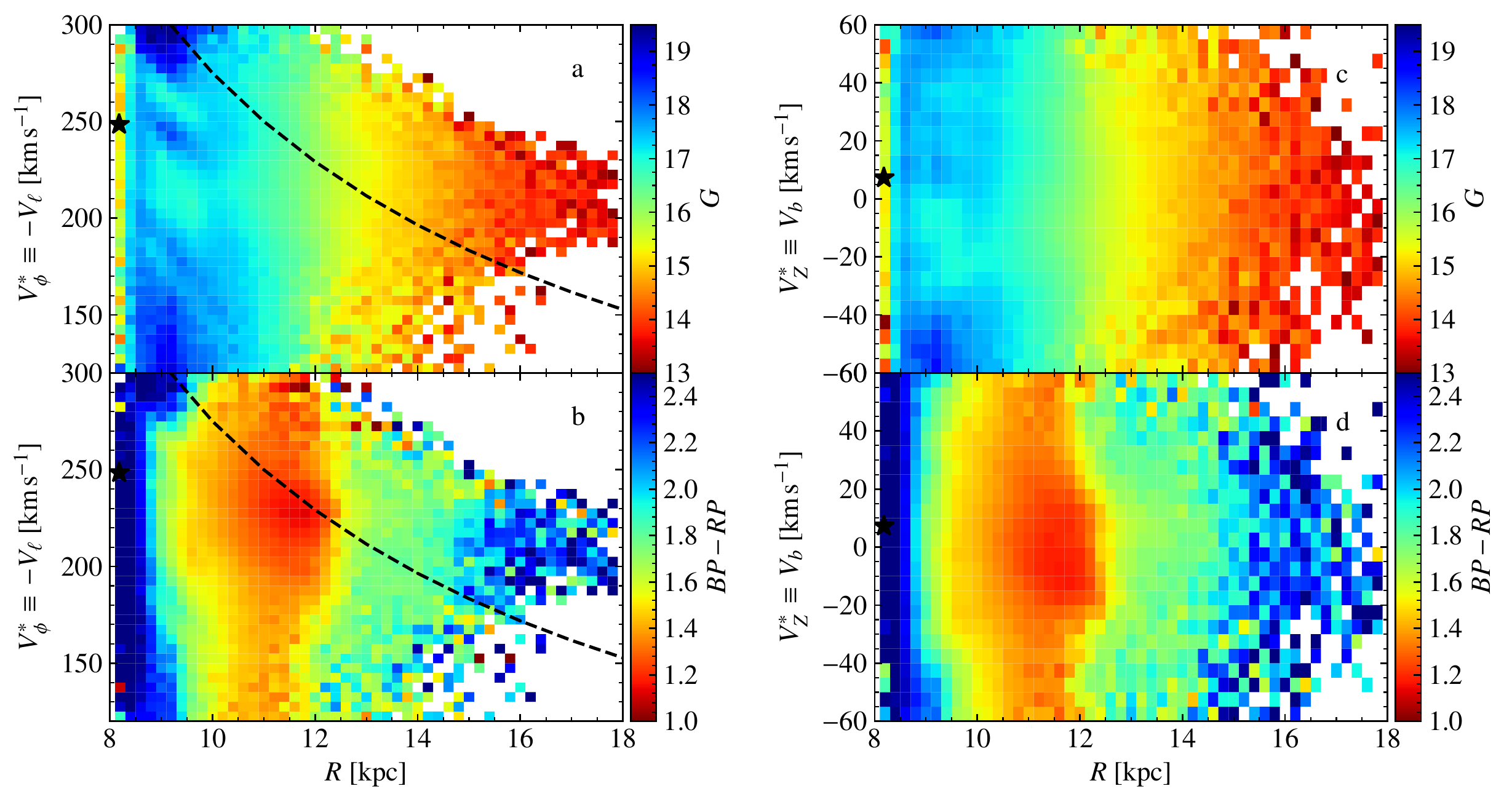}
         \caption{Photometry in different phase space projections. Median magnitudes $G$ (top) and colour $BP-RP$ (bottom) in the $R$-$\Vp$ plane (left) and $R$-$\VZ$ plane (right) for the \AC sample.}
         \label{figRVlVb_phot}
   \end{figure*}

 \begin{figure*}
  \centering
 \includegraphics[width=\hsize]{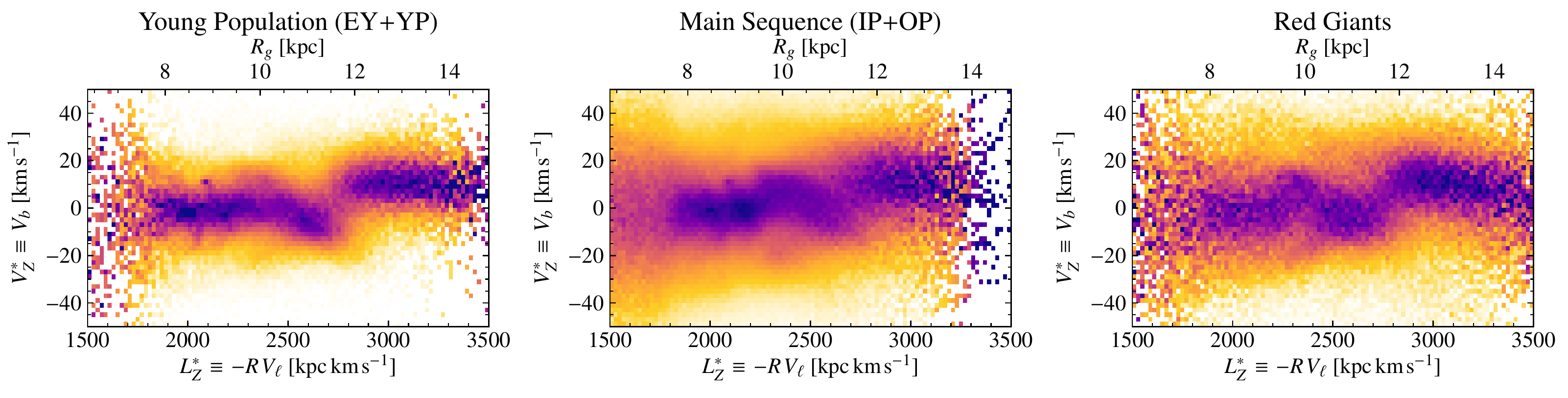}
        \caption{Structures in the vertical velocity and angular momentum space for different populations. As in the top panel of \figref{fig:LzVz}, these show a column normalised histogram of star numbers in the $\Lz,\VZ$ plane but for a given population (as in Sect.~\ref{sect_selection}). In all cases the feature at $\sim2750\kms\kpc$ is clearly visible. The young population has the lowest velocity dispersion, and therefore shows the feature most cleanly.}
        \label{fig:LzVzPop}
  \end{figure*}

\end{appendix}


\end{document}